\documentclass[pdftex,twocolumn,epjc3]{svjour3}  

\input{./Main/packages.tex}

\newcommand{\ednote}[1]{}

\newcommand{\tabt}[1]{\multicolumn{1}{c}{#1}}
\newcommand{\tabtt}[1]{\multicolumn{2}{c}{#1}}
\newcommand{\tabttt}[1]{\multicolumn{3}{c}{#1}}

\newcommand{\clicsid}{\textsc{CLIC}\_\text{SiD}\xspace}
\newcommand{\clicild}{\textsc{CLIC}\_\text{ILD}\xspace}

\newcommand{\mokka}{\textsc{Mokka}\xspace}
\newcommand{\marlin}{\textsc{Marlin}\xspace}
\newcommand{\geant}{\textsc{Geant4}\xspace}
\newcommand{\slic}{\textsc{Slic}\xspace}
\newcommand{\lcsim}{\texttt{org.lcsim}\xspace}
\newcommand{\pythia}{\textsc{Pythia}\xspace}
\newcommand{\whizard}{\textsc{Whizard}\xspace}
\newcommand{\pandora}{\textsc{PandoraPFA}\xspace}
\newcommand{\fastjet}{\textsc{FastJet}\xspace}
\newcommand{\ilcdirac}{\textsc{iLCDirac}\xspace}
\newcommand{\lcfiplus}{\textsc{LcfiPlus}\xspace}

\newcommand{\tauola}{\textsc{Tauola}\xspace}
\newcommand{\physsim}{\textsc{PhysSim}\xspace}
\newcommand{\tmva}{\textsc{Tmva}\xspace}
\newcommand{\taufinder}{\textsc{TauFinder}\xspace}
\newcommand{\minuit}{\textsc{Minuit}\xspace}

\newcommand{\higgsstrahlung}{\textnormal{Higgs\-strah\-lung}\xspace}
\newcommand{\gamgam}{\ensuremath{\PGg\PGg}\xspace}
\newcommand{\gghadrons}{\mbox{\ensuremath{\gamgam \to \text{hadrons}}}\xspace}
\newcommand{\gghadron}{\mbox{\ensuremath{\gamgam \to \text{hadron}}}\xspace}
\newcommand{\roots }{\ensuremath{\sqrt{s}}\xspace}
\newcommand{\mZ }{\ensuremath{m_{\PZ}}\xspace}
\newcommand{\mW }{\ensuremath{m_{\PW}}\xspace}
\newcommand{\mH }{\ensuremath{m_{\PH}}\xspace}
\newcommand{\GH }{\ensuremath{\Gamma_{\PH}}\xspace}

\newcommand{\GeV}{\ensuremath{\text{GeV}}\xspace}
\newcommand{\MeV}{\ensuremath{\text{MeV}}\xspace}
\newcommand{\TeV}{\ensuremath{\text{TeV}}\xspace}
\newcommand{\fb}{\ensuremath{\text{fb}}\xspace}
\newcommand{\ab}{\ensuremath{\text{ab}}\xspace}
\newcommand{\abinv}{\ensuremath{\text{ab}^{-1}}\xspace}
\newcommand{\fbinv}{\ensuremath{\text{fb}^{-1}}\xspace}

\newcommand{\micron}{\ensuremath{\upmu\text{m}}\xspace}

\newcommand{\mrad}{\ensuremath{\text{mrad}}\xspace}

\newcommand{\gHZZ}{\ensuremath{g_{\PH\PZ\PZ}}\xspace}
\newcommand{\gHWW}{\ensuremath{g_{\PH\PW\PW}}\xspace}
\newcommand{\gHtt}{\ensuremath{g_{\PH\PQt\PQt}}\xspace}
\newcommand{\gHbb}{\ensuremath{g_{\PH\PQb\PQb}}\xspace}
\newcommand{\gHcc}{\ensuremath{g_{\PH\PQc\PQc}}\xspace}
\newcommand{\gHTauTau}{\ensuremath{g_{\PH\PGt\PGt}}\xspace}
\newcommand{\gHMuMu}{\ensuremath{g_{\PH\PGm\PGm}}\xspace}

\newcommand{\BR}{\ensuremath{BR}\xspace}
\newcommand{\rootsprime }{\ensuremath{\sqrt{s^\prime}}\xspace}
\newcommand{\pT}{\ensuremath{p_\mathrm{T}}\xspace}
\newcommand{\kT}{\ensuremath{k_{\mathrm{t}}}\xspace}

\newcommand{\mrec}{\ensuremath{m_\mathrm{rec}}\xspace}
\newcommand{\mvis}{\ensuremath{m_\mathrm{vis}}\xspace}
\newcommand{\mll}{\ensuremath{m_{\Pl\Pl}}\xspace}
\newcommand{\Evis}{\ensuremath{E_\mathrm{vis}}\xspace}
\newcommand{\mqq}{\ensuremath{m_{\PQq\PAQq}}\xspace}
\newcommand{\cosQ}{\ensuremath{|\cos\theta_{\PQq}|}}
\newcommand{\costhetamis}{\ensuremath{|\cos\theta_{\text{miss}}|}}
\newcommand{\LumiIntDiff}{\ensuremath{\int\frac{d \mathcal{L}}{d s'} ds'}}

\newcommand{\egam}{\ensuremath{\Pepm\PGg}}
\newcommand{\epem}{\ensuremath{\Pep\Pem}\xspace}
\newcommand{\mpmm}{\ensuremath{\PGmp\PGmm}\xspace}  %mu+mu-
\newcommand{\tptm}{\ensuremath{\PGtp\PGtm}\xspace} %tau+tau-
\newcommand{\qq}{\ensuremath{\PQq \PAQq}\xspace}
\newcommand{\bb}{\ensuremath{\PQb \PAQb}\xspace}
\newcommand{\cc}{\ensuremath{\PQc \PAQc}\xspace}
\newcommand{\ttbar}{\ensuremath{\PQt \PAQt}\xspace}

\newcommand{\qqbar}{\ensuremath{\PQq \PAQq}\xspace} 
\def \qq {\qqbar}
 
\newcommand{\nuenuebar}{\ensuremath{\PGne\PAGne}\xspace} %nu_e nue_e
\newcommand{\nunubar}{\ensuremath{\PGn\PAGn}\xspace}   %nu nu 

\newcommand{\lplm}{\ensuremath{\Plp\Plm}\xspace}
\newcommand{\qqqq}{\ensuremath{\PQq\PAQq\PQq\PAQq}\xspace}

\newcommand{\ww}{\ensuremath{\PWp\PWm}\xspace} %W+W-
\newcommand{\WW}{\ensuremath{\PW\PW}\xspace} %WW
\newcommand{\ZZ}{\ensuremath{\PZ\PZ}\xspace} % ZZ

\newcommand{\ghThree}{\ensuremath{\lambda\xspace}}

\DeclareRobustCommand{\PepR}{\HepParticle{\Pe}{R}{+}\Xspace} 
\DeclareRobustCommand{\PemR}{\HepParticle{\Pe}{R}{-}\Xspace}

\DeclareRobustCommand{\PepL}{\HepParticle{\Pe}{L}{+}\Xspace} 
\DeclareRobustCommand{\PemL}{\HepParticle{\Pe}{L}{-}\Xspace}

\journalname{Eur. Phys. J. C}

\begin{document}

\begin{NoHyper}

% The main title
\title{\boldmath Higgs Physics at the CLIC Electron-Positron Linear Collider}

% The subtitle
\subtitle{}

\author{H.~Abramowicz\thanksref{inst:TelAviv}
        \and
        A.~Abusleme\thanksref{inst:Santiago}
        \and
        K.~Afanaciev\thanksref{inst:Minsk}
        \and
        N.~Alipour~Tehrani\thanksref{inst:CERN}
        \and
        C.~Bal\'{a}zs\thanksref{inst:Monash}
        \and
        Y.~Benhammou\thanksref{inst:TelAviv}
        \and
        M.~Benoit\thanksref{inst:Geneva}
        \and
        B.~Bilki\thanksref{inst:Argonne}
        \and
        J.-J.~Blaising\thanksref{inst:Annecy}
        \and
        M.J.~Boland\thanksref{inst:Melbourne}
% ...
}

\thankstext[$\star$]{t1}{Corresponding Editors: \href{mailto:clicdp-higgs-paper-editors@cern.ch}{clicdp-higgs-paper-editors@cern.ch}}
\thankstext[$\dagger$]{deceased}{Deceased}
\thankstext[a]{now:Bonn}{Now at University of Bonn, Bonn, Germany}
\thankstext[b]{also:Bonn}{Also at University of Bonn, Bonn, Germany}
\thankstext[c]{now:XFEL}{Now at European XFEL GmbH, Hamburg, Germany}
\thankstext[d]{also:Vienna}{Also at Vienna University of Technology, Vienna, Austria}
\thankstext[e]{now:PSI}{Now at Paul Scherrer Institute, Villigen, Switzerland}
\thankstext[f]{now:PNNL}{Now at Pacific Northwest National Laboratory, Richland, Washington, USA}
\thankstext[g]{now:Wuppertal}{Now at University of Wuppertal, Wuppertal, Germany}

\institute{Raymond \& Beverly Sackler School of Physics \& Astronomy, Tel Aviv University, Tel Aviv, Israel\label{inst:TelAviv}
          \and
          Pontificia Universidad Cat\'{o}lica de Chile, Santiago, Chile\label{inst:Santiago}
          \and
          National Scientific and Educational Centre of Particle and High Energy Physics, Belarusian State University, Minsk, Belarus\label{inst:Minsk}
          \and
          CERN, Geneva, Switzerland\label{inst:CERN}
          \and
          Monash University, Melbourne, Australia\label{inst:Monash}
          \and
          D\'{e}partement de Physique Nucl\'{e}aire et Corpusculaire (DPNC), Universit\'{e} de Gen\`{e}ve, Geneva, Switzerland\label{inst:Geneva}
          \and
          Argonne National Laboratory, Argonne, USA\label{inst:Argonne}
          \and
          Laboratoire d'Annecy-le-Vieux de Physique des Particules, Annecy-le-Vieux, France\label{inst:Annecy}
          \and
          University of Melbourne, Melbourne, Australia\label{inst:Melbourne}
          \and
          IFIC, CSIC-University of Valencia, Valencia, Spain\label{inst:Valencia}
          \and
          Vin\v{c}a Institute of Nuclear Sciences, University of Belgrade, Belgrade, Serbia\label{inst:Belgrade}
          \and
          University of Liverpool, Liverpool, United Kingdom\label{inst:Liverpool}
          \and
          Faculty of Physics and Applied Computer Science, AGH University of Science and Technology, Cracow, Poland\label{inst:AGH-UST-Cracow}
          \and
          Oxford University, Oxford, United Kingdom\label{inst:Oxford}
          \and
          The Henryk Niewodnicza\'{n}ski Institute of Nuclear Physics Polish Academy of Sciences , Cracow, Poland\label{inst:IFJPAN-Cracow}
          \and
          Department of Physics and Technology, University of Bergen, Bergen, Norway\label{inst:Bergen}
          \and
          Institute of Space Science, Bucharest, Romania\label{inst:Bucharest}
          \and
          Max-Planck-Institut f\"{u}r Physik, Munich, Germany\label{inst:Munich}
          \and
          DESY, Hamburg, Germany\label{inst:DESY}
          \and
          University of Bristol, Bristol, United Kingdom\label{inst:Bristol}
          \and
          Cavendish Laboratory, University of Cambridge, Cambridge, United Kingdom\label{inst:Cambridge}
          \and
          University of Edinburgh, Edinburgh, United Kingdom\label{inst:Edinburgh}
          \and   
          School of Physics and Astronomy, University of Birmingham, Birmingham, United Kingdom\label{inst:Birmingham}
          \and
          Faculty of Physics, University of Warsaw, Poland\label{inst:Warsaw}
          \and
          Institute of Physics of the Academy of Sciences of the Czech Republic, Prague, Czech Republic\label{inst:Prague}
          \and
          IFCA, CSIC-University of Cantabria, Santander, Spain\label{inst:Santander}
          \and
          Karlsruher Institut f\"{u}r Technologie (KIT), Institut f\"{u}r Prozessdatenverarbeitung und Elektronik (IPE), Karlsruhe, Germany\label{inst:Karlsruhe}
          \and
          University of Glasgow, Glasgow, United Kingdom\label{inst:Glasgow}
          \and
          University of Barcelona, Barcelona, Spain\label{inst:Barcelona}
          \and
          Aarhus University, Aarhus, Denmark\label{inst:Aarhus}
          \and
          Physics Department, University of Michigan, Ann Arbor, Michigan, USA\label{inst:Michigan}
}

\date{Received: date / Accepted: date}

\maketitle

\end{NoHyper}

\begin{abstract}
  The Compact Linear Collider (CLIC) is an option for a future $\epem$
  collider operating at centre-of-mass energies up to $3\,\TeV$,
  providing sensitivity to a wide range of new physics phenomena and
  precision physics measurements at the energy frontier. This paper is 
  the first comprehensive presentation of 
  the Higgs physics reach of CLIC operating at three energy
  stages: $\roots = 350\,\GeV$, $1.4\,\TeV$, and $3\,\TeV$. The initial
  stage of operation allows the study of Higgs boson production in \higgsstrahlung 
  ($\epem\to\PZ\PH$) and $\PW\PW$-fusion ($\epem\to\PH\nuenuebar$),
  resulting in precise measurements of the production cross sections, the Higgs
  total decay width $\Gamma_{\PH}$, and model-independent
  determinations of the Higgs couplings. Operation at $\roots >
  1\,\TeV$ provides high-statistics samples of Higgs bosons produced
  through $\PW\PW$-fusion, enabling tight constraints on the Higgs
  boson couplings. Studies of the rarer processes
  $\epem\to\PQt\PAQt\PH$ and $\epem\to\PH\PH\nuenuebar$ allow
  measurements of the top Yukawa coupling and the Higgs boson
  self-coupling. This paper presents detailed studies of the precision
  achievable with Higgs measurements at CLIC and describes the
  interpretation of these measurements in a global fit.
\end{abstract}

\tableofcontents

\section{Introduction}
\label{sec:introduction}

The discovery of a Higgs 
boson~\cite{Aad:2012tfa, Chatrchyan:2012xdj} at the Large Hadron Collider (LHC) provided confirmation of 
the electroweak symmetry breaking
mechanism~\cite{higgs:englert,higgs:HiggsA,higgs:HiggsB,higgs:KibbleA,higgs:HiggsC,higgs:KibbleB}
of the Standard Model (SM). However, it is not yet known if the
observed Higgs boson is the fundamental scalar of the SM or is
either a more complex object or part of an extended Higgs
sector. Precise studies of the properties of the Higgs boson at the
LHC and future colliders are essential to understand its true nature.

The Compact Linear Collider (CLIC) is a mature option for a future multi-\TeV high-luminosity
linear \epem collider that is currently under development at CERN. It
is based on a novel two-beam acceleration technique providing
accelerating gradients of 100\,MV/m. Recent implementation studies for
CLIC have converged towards a staged approach.
% offering a unique physics programme spanning several decades. 
In this scheme, CLIC 
provides high-luminosity \epem collisions at centre-of-mass energies from a few hundred \GeV up to
3\,TeV. 
The ability of CLIC to collide \epem up to  multi-TeV energy 
scales is unique.  
For the current study, the nominal centre-of-mass energy of the first energy stage
is $\roots=350\,\GeV$. At this centre-of-mass energy, the
\higgsstrahlung and $\PW\PW$-fusion processes have significant cross
sections, providing access to precise measurement of the absolute values of the
Higgs boson couplings to both fermions and bosons. Another advantage of
operating CLIC at $\roots \approx 350\,\GeV$ is that
it enables a programme of precision top quark physics, including a
scan of the $\PQt\PAQt$ cross section close to the production
threshold. In practice, the centre-of-mass energy of the second stage
of CLIC operation will be motivated by both the machine design and
results from the LHC. In this paper, it is assumed that the second CLIC
energy stage has $\roots=1.4\,\TeV$ and that the ultimate CLIC
centre-of-mass energy is $3\,\TeV$. In addition to direct and indirect
searches for Beyond the Standard Model (BSM) phenomena, these higher
energy stages of operation provide a rich potential for Higgs physics
beyond that accessible at lower energies, such as the direct
measurement of the top Yukawa coupling and a direct probe of the Higgs
potential through the measurement of the Higgs self-coupling. 
Furthermore, rare Higgs boson decays become accessible due
to the higher integrated luminosities at higher energies and the
increasing cross section for Higgs production in $\PW\PW$-fusion.
The proposed staged approach spans around twenty years of running.

The following sections describe the experimental conditions at CLIC,
an overview of Higgs production at CLIC, and the Monte Carlo samples,
detector simulation, and event reconstruction used for the subsequent
studies.  Thereafter, Higgs production at $\roots = 350\,\GeV$, 
Higgs production in $\PW\PW$-fusion at $\roots > 1\,\TeV$, 
Higgs production in $\PZ\PZ$-fusion, the measurement of the top Yukawa 
coupling, double Higgs production, and measurements of the Higgs boson 
mass are presented.  The paper concludes
with a discussion of the measurement precisions on the Higgs
couplings obtained in a combined fit to the expected CLIC results, 
and the systematic uncertainties associated with the measurements.

The detailed study of the CLIC potential for Higgs physics 
presented here supersedes earlier preliminary estimates~\cite{CLIC_snowmass13}.
The work is carried out by the CLIC Detector and Physics (CLICdp) 
collaboration.

\section{Experimental Environment at CLIC}
\label{sec:ExperimentalEnvironment}

The experimental environment at CLIC is characterised by challenging
conditions imposed by the CLIC accelerator technology, by detector
concepts optimised for the precise reconstruction of complex final
states in the multi-TeV energy range, 
and by the operation in several energy
stages to maximise the physics potential.

\subsection{Accelerator and Beam Conditions}

The CLIC accelerator design is based on a two-beam acceleration
scheme. It uses a high-intensity drive beam to efficiently generate
radio frequency (RF) power at 12\,GHz. The RF power is used to
accelerate the main particle beam that runs parallel to the drive
beam. CLIC uses normal-conducting accelerator structures, operated at
room temperature. These structures permit high accelerating gradients,
while the short pulse duration discussed below limits ohmic losses to
tolerable levels. The initial drive beams and the main
electron/positron beams are generated in the central complex and are
then injected at the ends of the two linac arms. The feasibility of
the CLIC accelerator has been demonstrated through prototyping,
simulations and large-scale tests, as described in the Conceptual
Design Report~\cite{CLICCDR_vol1}. In particular, the two-beam
acceleration at gradients exceeding 100\,MV/m has been demonstrated in
the CLIC test facility, CTF3. High luminosities are achievable by very
small beam emittances, which are generated in the injector complex and
maintained during transport to the interaction point.

\begin{figure}
\begin{center}
\includegraphics[width=\columnwidth, trim=8cm 4cm 1cm 1cm,clip]{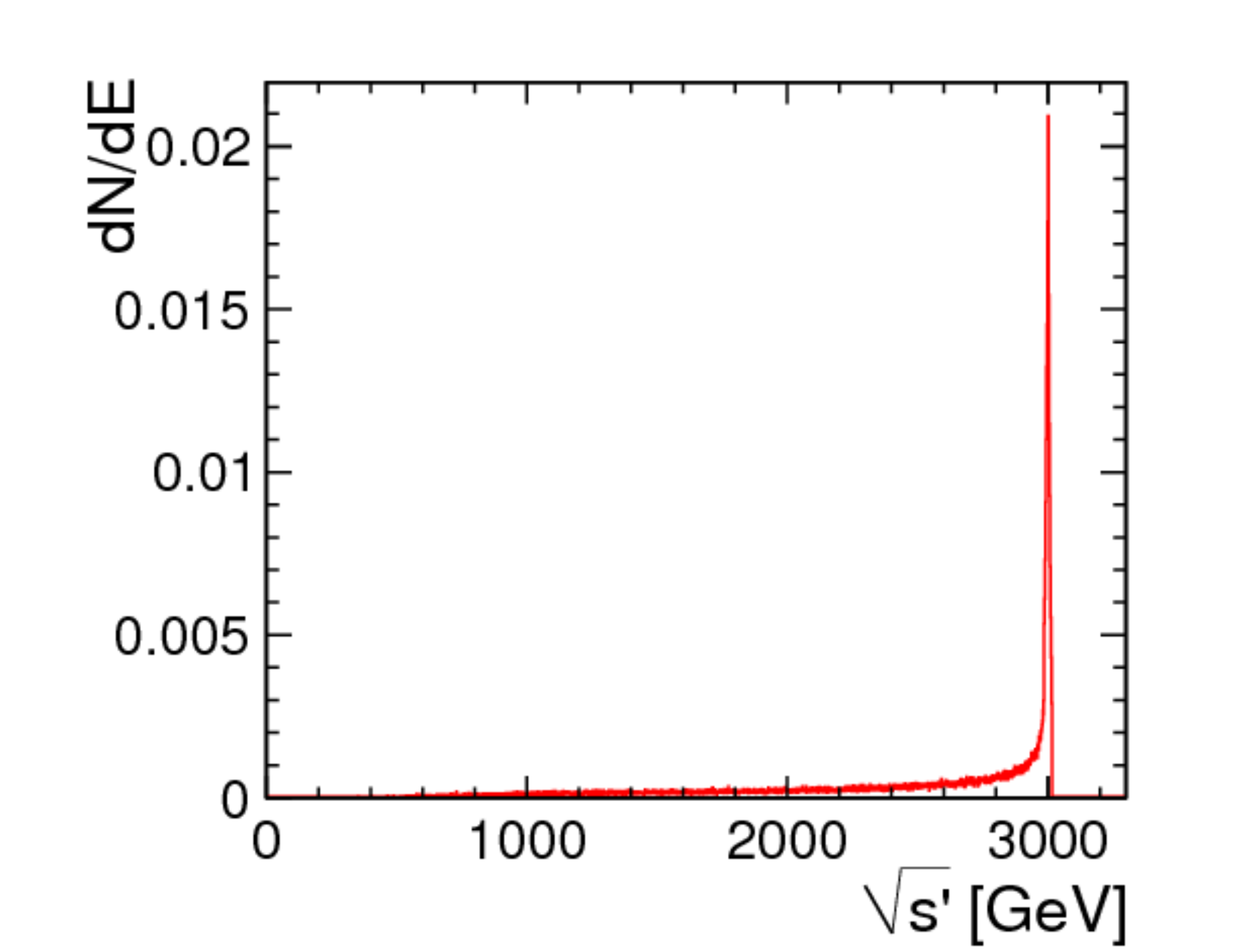}
\caption{The luminosity spectrum for CLIC operating at $\roots = 3\,\TeV$, where $\rootsprime$ is the effective centre-of-mass energy after beamstrahlung and initial state radiation \cite{CLIC_PhysDet_CDR}.\label{fig:luminosityspectrum}}
\end{center}
\end{figure}

CLIC will be operated with a bunch train repetition rate of 
50\,Hz. Each bunch train consists of 312 individual bunches, with
0.5\,ns between bunch crossings at the interaction point. The average
number of hard $\epem$ interactions in a single bunch
train crossing is much less than one.
However, for CLIC operation at $\roots >
1\,\TeV$, the highly-focussed intense beams lead to significant
beamstrahlung (radiation of photons from electrons/positrons in the
electric field of the other beam). Beamstrahlung results in high rates
of incoherent electron--positron pairs and low-$Q^2$ $t$-channel
multi-peripheral \gghadron events, where $Q^2$ is the negative of the
four-momentum squared of the virtual space-like photon. In addition, the
energy loss through beamstrahlung generates a long lower-energy tail
to the luminosity spectrum that extends well below the nominal
centre-of-mass energy, as shown in \autoref{fig:luminosityspectrum}.
Both the CLIC detector design and the event
reconstruction techniques employed are optimised to mitigate the
influence of these backgrounds, which are most severe at the higher
CLIC energies; this is discussed further in \autoref{sec:simreco}.

The baseline machine design allows for up to $\pm$80\,\% longitudinal 
electron spin-polarisation by using GaAs-type cathodes \cite{CLICCDR_vol1}; 
and provisions have been made to allow positron
polarisation as an upgrade option. Most studies presented in this
paper are performed for zero beam polarisation and are subsequently
scaled to account for the increased cross sections with left-handed
polarisation for the electron beam.

\subsection{Detectors at CLIC}

\begin{figure*}[tb]
  \centering
  \includegraphics[scale=1,clip]{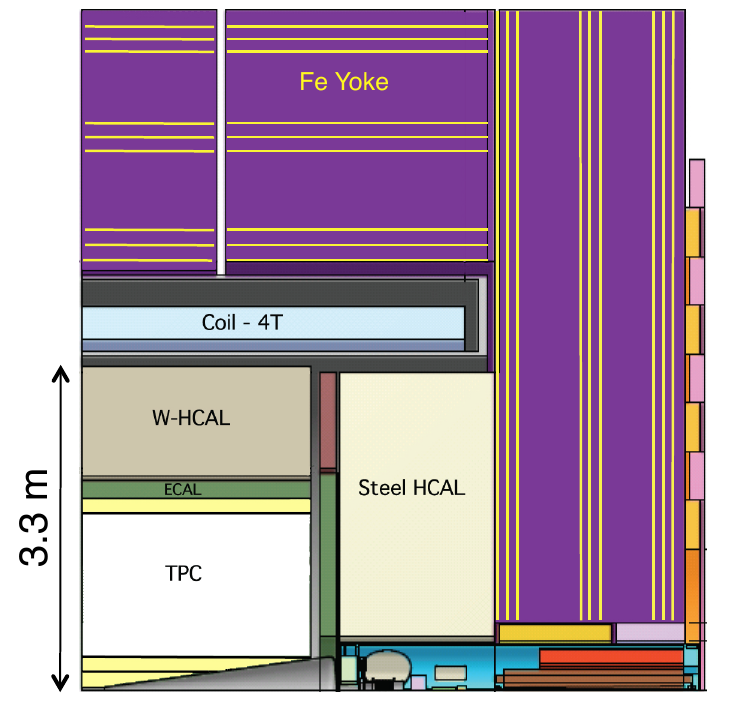}\put(-210,190){\large{\bf a)}}
  \includegraphics[scale=1,clip]{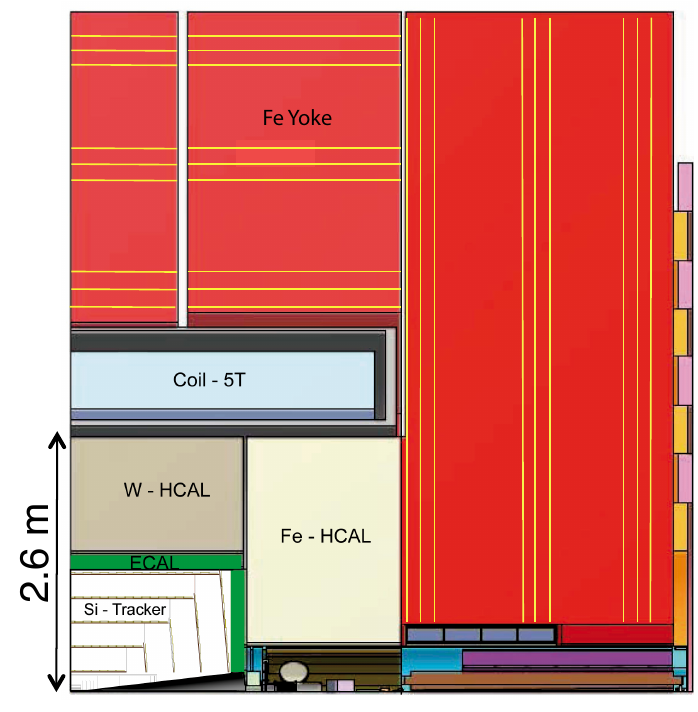}\put(-198,190){\large{\bf b)}}
  \caption{Longitudinal cross section of the top right quadrant of the
    \clicild (a) and \clicsid (b) detector concepts.}
  \label{fig:detectors}
\end{figure*}

The detector concepts used for the CLIC physics studies, described
here and elsewhere~\cite{CLIC_PhysDet_CDR}, are based on the
SiD~\cite{Aihara:2009ad,ilctdrvol4:2013} and
ILD~\cite{ildloi:2009,ilctdrvol4:2013} detector concepts for the
International Linear Collider (ILC). They were initially adapted for
the CLIC $3\,\TeV$ operation, which constitutes the most challenging
environment for the detectors in view of the high beam-induced background levels. For most sub-detector systems, the
$3\,\TeV$ detector design is suitable at all energy stages, the only
exception being the inner tracking detectors and the vertex detector,
where the lower backgrounds at $\roots=350\,\GeV$ enable detectors to
be deployed with a smaller inner radius.

The key performance parameters of the CLIC detector concepts with
respect to the Higgs programme are:
\begin{itemize}
\item excellent track-momentum resolution of $\sigma_{p_T}/p_T^2 \lesssim 2 \cdot 10^{-5}$
  $\GeV^{-1}$, required for a precise
  reconstruction of leptonic $\PZ$ decays in $\PZ\PH$ events;
\item precise impact parameter resolution, defined by $a \lesssim 5\,\micron$ and $b \lesssim 15\,\micron\,\GeV$ 
  in $\sigma_{d_0}^2 = a^2 + b^2/(p^2\sin^3\theta)$ to provide accurate vertex
  reconstruction, enabling flavour tagging with clean $\PQb$-, $\PQc$-
  and light-quark jet separation;
\item jet-energy resolution $\sigma_E/E \lesssim 3.5\,\%$ for light-quark jet
  energies in the range $100\,\GeV$ to $1\,\TeV$, required for the
  reconstruction of hadronic $\PZ$ decays in $\PZ\PH$ events and the
  separation of $\PZ\to\qq$ and $\PH\to\qq$ based on the
  reconstructed di-jet invariant mass;
\item detector coverage for electrons extending to very low angles
  with respect to the beam axis, to maximise background rejection for
  $\PW\PW$-fusion events.
\end{itemize}

The main design driver for the CLIC (and ILC) detector concepts is the
required jet-energy resolution. As a result, the CLIC detector
concepts~\cite{CLIC_PhysDet_CDR}, \clicsid\ and \clicild, are based on
fine-grained electromagnetic and hadronic calo\-rimeters (ECAL and
HCAL), optimised for particle-flow reconstruction techniques. In the
particle-flow approach, the aim is to reconstruct the individual
final-state particles within a jet using information from the
tracking detectors combined with that from the highly granular
calo\-rimeters~\cite{thomson:pandora,Marshall2013153,ALEPH:pflow,CMS:pflow}. 
In addition,
particle-flow event reconstruction provides a powerful tool for the
rejection of beam-induced backgrounds~\cite{CLIC_PhysDet_CDR}. The
CLIC detector concepts employ strong central solenoid magnets, located
outside the HCAL, providing an axial magnetic field of 5\,T in
\clicsid\ and 4\,T in \clicild. The \clicsid\ concept employs central
silicon-strip tracking detectors, whereas \clicild\ assumes a large
central gaseous Time Projection Chamber. In both concepts, the central
tracking system is augmented with silicon-based inner tracking detectors.  The two detector concepts are shown
schematically in \autoref{fig:detectors} and are described in detail
in~\cite{CLIC_PhysDet_CDR}.

\subsection{Assumed Staged Running Scenario}

The studies presented in this paper are based on a scenario 
in which CLIC runs at three energy stages.  
The first stage is at $\roots = 350\,\GeV$, around the 
top-pair production threshold. The second stage is at
$\roots = 1.4\,\TeV$; this energy is chosen because it 
can be reached with a single CLIC drive-beam complex. The third stage
is at $\roots = 3\,\TeV$; the ultimate energy of CLIC. At each
stage, four to five years of running with a fully commissioned accelerator 
is foreseen, providing integrated luminosities of $500\,\fbinv$,
$1.5\,\abinv$ and $2\,\abinv$ at $350\,\GeV$, $1.4\,\TeV$ and
$3\,\TeV$, respectively\footnote{As a result of this paper and other studies, 
a slightly different staging scenario for CLIC, with a first stage at
$\roots = 380\,\GeV$ to include precise measurements 
of top quark properties as a probe for BSM physics, and the next stage at 1.5\,TeV, has recently been adopted and will be used for future studies~\cite{staging_baseline_yellow_report}.}.  Cross sections and integrated luminosities for the three stages are summarised in \autoref{tab:higgs:events}.

\section{Overview of Higgs Production at CLIC}
\label{sec:higgs_production}

\begin{figure}[t]
  \centering
  \includegraphics[width=0.9\columnwidth]{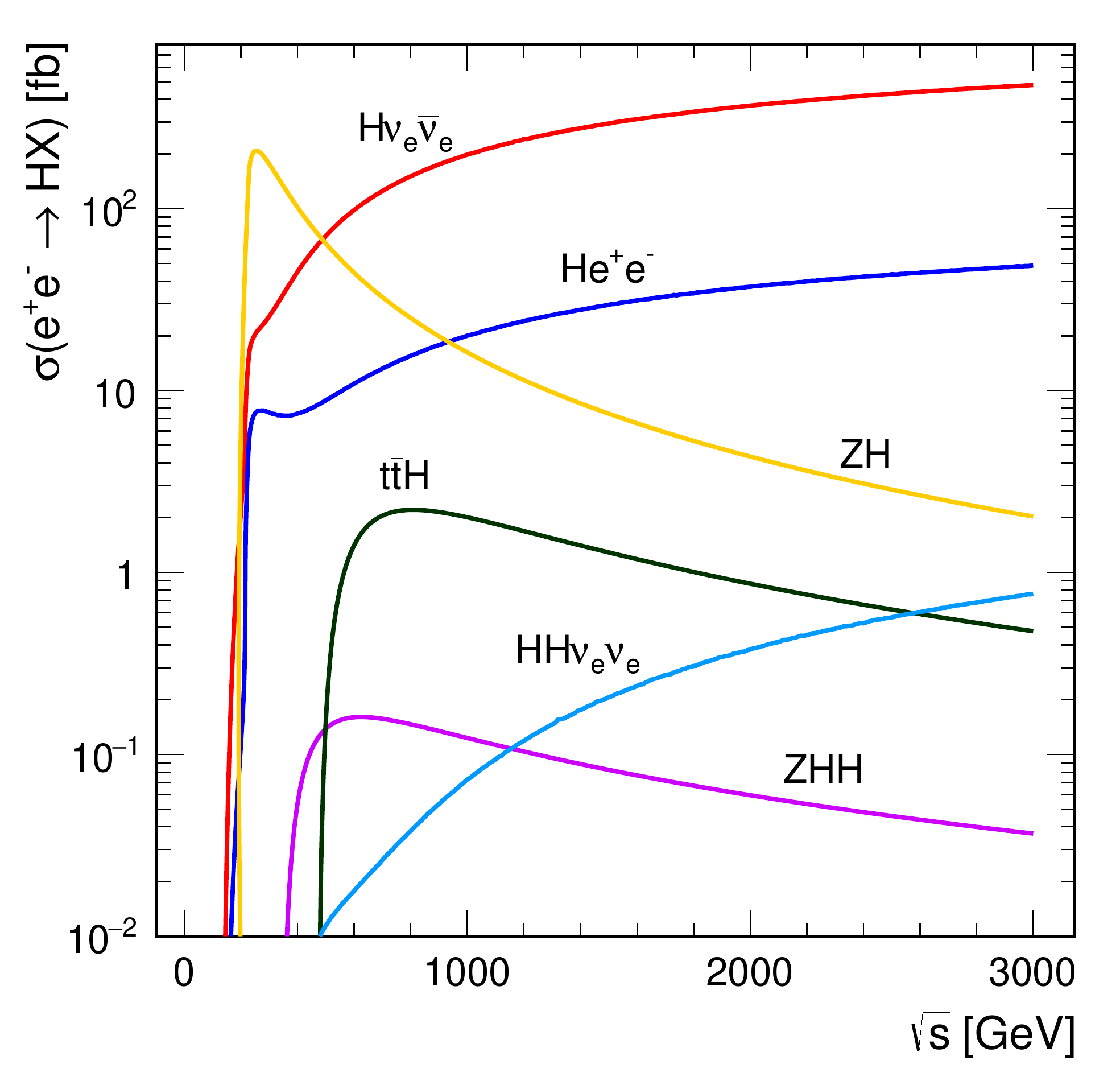}
  \caption{Cross section as a function of centre-of-mass energy for
    the main Higgs production processes at an $\epem$ collider for
    a Higgs mass of $\mH=126\,\GeV$. The values shown correspond to unpolarised beams
    and do not include the effect of beamstrahlung.
    \label{fig:higgs:cross}}
\end{figure}

A high-energy $\epem$ collider such as CLIC provides an
experimental environment that allows the study of Higgs boson properties 
with high precision. The evolution of the leading-order $\epem$ Higgs
production cross sections with centre-of-mass energy,
as computed using the \whizard 1.95~\cite{Kilian:2007gr} program, is shown in
\autoref{fig:higgs:cross} for a Higgs boson mass of $126\,\GeV$~\cite{Agashe:2014kda}.

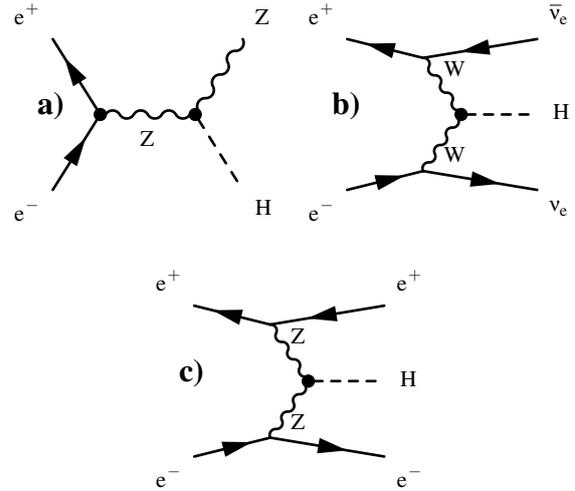
\begin{figure}[htb]
\unitlength = 1mm
\vspace{6mm}
\centering
\begin{fmffile}{higgs_production/eezh}
\begin{fmfgraph*}(25,20)
\put(-2,10){\large{\bf a)}}
\fmfstraight
\fmfleft{i1,i2}
\fmfright{o1,o2}
\fmflabel{$\Pem$}{i1}
\fmflabel{$\Pep$}{i2}
\fmflabel{$\PZ$}{o2}
\fmflabel{$\PH$}{o1}
\fmf{photon,tension=1.0,label=$\PZ$}{v1,v2}
\fmf{fermion,tension=1.0}{i1,v1,i2}
\fmf{photon,tension=1.0}{o2,v2}
\fmf{dashes,tension=1.0}{v2,o1}
\fmfdot{v1}
\fmfdot{v2}
\end{fmfgraph*}
\end{fmffile}
\hspace{10mm}
\begin{fmffile}{higgs_production/eevvh}
\begin{fmfgraph*}(25,20)
\put(-2,10){\large{\bf b)}}
\fmfstraight
\fmfleft{i1,i2}
\fmfright{o1,oh,o2}
\fmflabel{$\Pem$}{i1}
\fmflabel{$\Pep$}{i2}
\fmflabel{$\PAGne$}{o2}
\fmflabel{$\PH$}{oh}
\fmflabel{$\PGne$}{o1}
\fmf{fermion, tension=2.0}{i1,v1}
\fmf{fermion, tension=1.0}{v1,o1}
\fmf{fermion, tension=1.0}{o2,v2}
\fmf{fermion, tension=2.0}{v2,i2}
\fmf{photon, lab.side=right,lab.dist=1.5,label=$\PW$,tension=1.0}{v1,vh}
\fmf{photon, lab.side=right, lab.dist=1.5,label=$\PW$,tension=1.0}{vh,v2}
\fmf{dashes, tension=1.0}{vh,oh}
\fmfdot{vh}
\end{fmfgraph*}
\end{fmffile} \\
\vspace{15mm}
\begin{fmffile}{higgs_production/eeeeh}
\begin{fmfgraph*}(25,20)
\put(-2,10){\large{\bf c)}}
\fmfstraight
\fmfleft{i1,i2}
\fmfright{o1,oh,o2}
\fmflabel{$\Pem$}{i1}
\fmflabel{$\Pep$}{i2}
\fmflabel{$\Pep$}{o2}
\fmflabel{$\PH$}{oh}
\fmflabel{$\Pem$}{o1}
\fmf{fermion, tension=2.0}{i1,v1}
\fmf{fermion, tension=1.0}{v1,o1}
\fmf{fermion, tension=1.0}{o2,v2}
\fmf{fermion, tension=2.0}{v2,i2}
\fmf{photon, lab.side=right,lab.dist=1.5,label=$\PZ$,tension=1.0}{v1,vh}
\fmf{photon, lab.side=right, lab.dist=1.5,label=$\PZ$,tension=1.0}{vh,v2}
\fmf{dashes, tension=1.0}{vh,oh}
\fmfdot{vh}
\end{fmfgraph*}
\end{fmffile}
\vspace{5mm}
\caption{Leading-order Feynman diagrams of the highest cross section Higgs
  production processes at CLIC; Higgsstrahlung (a),
  $\PW\PW$-fusion (b) and $\PZ\PZ$-fusion
  (c). \label{fig:higgs:eezh}}
\end{figure}

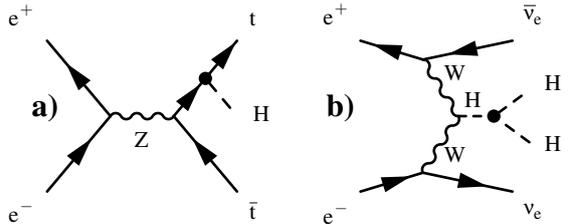
\begin{figure}[hbt]
\unitlength = 1mm
\vspace{5mm}
\centering
     \begin{fmffile}{higgs_production/eetth} 
     \begin{fmfgraph*}(25,20)
\put(-2,10){\large{\bf a)}}
        \fmfstraight
         \fmfleft{i1,i2}  
         \fmfright{o1,oh,o2}
         \fmflabel{$\Pem$}{i1}
        \fmflabel{$\Pep$}{i2}  
         \fmflabel{$\PQt$}{o2} 
         \fmflabel{$\PH$}{oh}
         \fmflabel{$\PAQt$}{o1}
          \fmf{photon,label=$\PZ$,tension=2.0}{v1,v2}
          \fmf{fermion,tension=1.0}{i1,v1,i2}
           \fmf{phantom,tension=1.0}{o1,v2,o2}
            \fmffreeze
           \fmf{fermion,tension=1.0}{o1,v2}
           \fmf{fermion,tension=1.0}{v2,vh,o2}
           \fmf{dashes,tension=0.0}{vh,oh}
           \fmfdot{vh}
	   \end{fmfgraph*}
	   \end{fmffile}
	   \hspace{10mm} 
	 \begin{fmffile}{higgs_production/eevvhh}
        \begin{fmfgraph*}(25,20)
\put(-2,10){\large{\bf b)}}
            \fmfleft{i1,i2}  
            \fmfright{o1,oh1,oh2,o2}
            \fmflabel{$\Pem$}{i1}
            \fmflabel{$\Pep$}{i2}  
            \fmflabel{$\PAGne$}{o2} 
            \fmflabel{$\PH$}{oh1}            
            \fmflabel{$\PH$}{oh2}
            \fmflabel{$\PGne$}{o1}
             \fmf{fermion, tension=2.0}{i1,v1}
              \fmf{fermion, tension=1.0}{v1,o1}
             \fmf{fermion, tension=1.0}{o2,v2}
                          \fmf{fermion, tension=2.0}{v2,i2}
             \fmf{photon,  label=$\PW$,label.dist=1.5,tension=1.0}{v1,vh}
             \fmf{photon,  label=$\PW$,label.dist=1.5,tension=1.0}{v2,vh}
             \fmf{dashes, label=$\PH$, label.dist=1.5,tension=2.0}{vh,vh1}
             \fmf{dashes,  tension=1.0}{oh1,vh1,oh2}
             \fmfdot{vh1}
        \end{fmfgraph*}
    \end{fmffile}
    %\vspace{15mm}
    %    \begin{fmffile}{higgs_production/ghhww}
    %    \begin{fmfgraph*}(25,20)
    %\put(-2,10){\large{\bf c)}}
    %        \fmfleft{i1,i2}  
    %        \fmfright{o1,oh1,oh2,o2}
    %        \fmflabel{$\Pem$}{i1}
    %        \fmflabel{$\Pep$}{i2}  
    %        \fmflabel{$\PAGne$}{o2} 
    %        \fmflabel{$\PH$}{oh1}            
    %        \fmflabel{$\PH$}{oh2}
    %        \fmflabel{$\PGne$}{o1}
    %         \fmf{fermion, tension=2.0}{i1,v1}
    %          \fmf{fermion, tension=1.0}{v1,o1}
    %         \fmf{fermion, tension=1.0}{o2,v2}
    %          \fmf{fermion, tension=2.0}{v2,i2}
    %         \fmf{photon, label=$\PW$,label.side=right,label.dist=1.5,  tension=1.0}{vh,v1}
    %          \fmf{photon, label=$\PW$,label.sde=right,label.dist=1.5, tension=1.0}{v2,vh}
    %         \fmf{dashes,  tension=1.0}{oh1,vh,oh2}
    %         \fmfdot{vh}
    %    \end{fmfgraph*}
    %\end{fmffile}
    \vspace{5mm}
    \caption{Feynman diagrams of the leading-order processes at CLIC
      involving (a) the top Yukawa coupling $g_{\PH\PQt\PQt}$, and (b) 
      the Higgs boson trilinear self-coupling $\lambda$.
      \label{fig:higgs:lambda}}
\end{figure}

The Feynman diagrams for the three highest cross section Higgs
production processes at CLIC are shown in \autoref{fig:higgs:eezh}.
At $\roots \approx 350\,\GeV$, the
\higgsstrahlung process ($\epem\to\PZ\PH$) has the largest cross
section, but the $\PW\PW$-fusion process ($\epem\to\PH\PGne\PAGne$) is
also significant. The combined study of these two processes probes the
Higgs boson properties (width and branching ratios) in a
model-independent manner. In the higher energy stages of CLIC
operation ($\roots = 1.4\,\TeV$ and $3\,\TeV$), Higgs production is
dominated by the $\PW\PW$-fusion process, with the $\PZ\PZ$-fusion
process ($\epem\to\PH\epem$) also becoming significant. Here the
increased $\PW\PW$-fusion cross section, combined with the high
luminosity of CLIC, results in large data samples, allowing precise
${\cal{O}}(1\,\%)$ measurements of the couplings of the Higgs boson to
both fermions and gauge bosons. In addition to the main Higgs
production channels, rarer processes such as $\epem\to\PQt\PAQt\PH$
and $\epem\to\PH\PH\PGne\PAGne$, provide access to the top Yukawa coupling and the Higgs trilinear self-coupling. 
%governed by the parameter $\lambda$ in the Higgs potential. 
Feynman diagrams for these processes are shown in \autoref{fig:higgs:lambda}.
In all cases, the Higgs production cross sections can be
increased with polarised electron (and positron) beams 
as discussed in \autoref{sec:higgs:polarisation}.

\autoref{tab:higgs:events} lists the expected numbers of $\PZ\PH$,
$\PH\PGne\PAGne$ and $\PH\epem$ events for the three main CLIC
centre-of-mass energy stages. These numbers account for the effect of
beamstrahlung and initial state radiation (ISR), which result in a
tail in the distribution of the effective centre-of-mass energy
$\rootsprime$. The impact of beamstrahlung on the expected numbers of
events is mostly small. For example, it results in an
approximately $10\,\%$ reduction in the numbers of $\PH\PGne\PAGne$
events at $\roots > 1\,\TeV$ (compared to the beam spectrum with ISR
alone), because the cross section rises relatively slowly with
$\roots$. The reduction of the effective centre-of-mass energies due to 
ISR and beamstrahlung increases the $\PZ\PH$ cross section 
at $\roots = 1.4\,\TeV$ and $3\,\TeV$.

\begin{figure}[t]
  \centering
  \includegraphics[width=0.9\columnwidth]{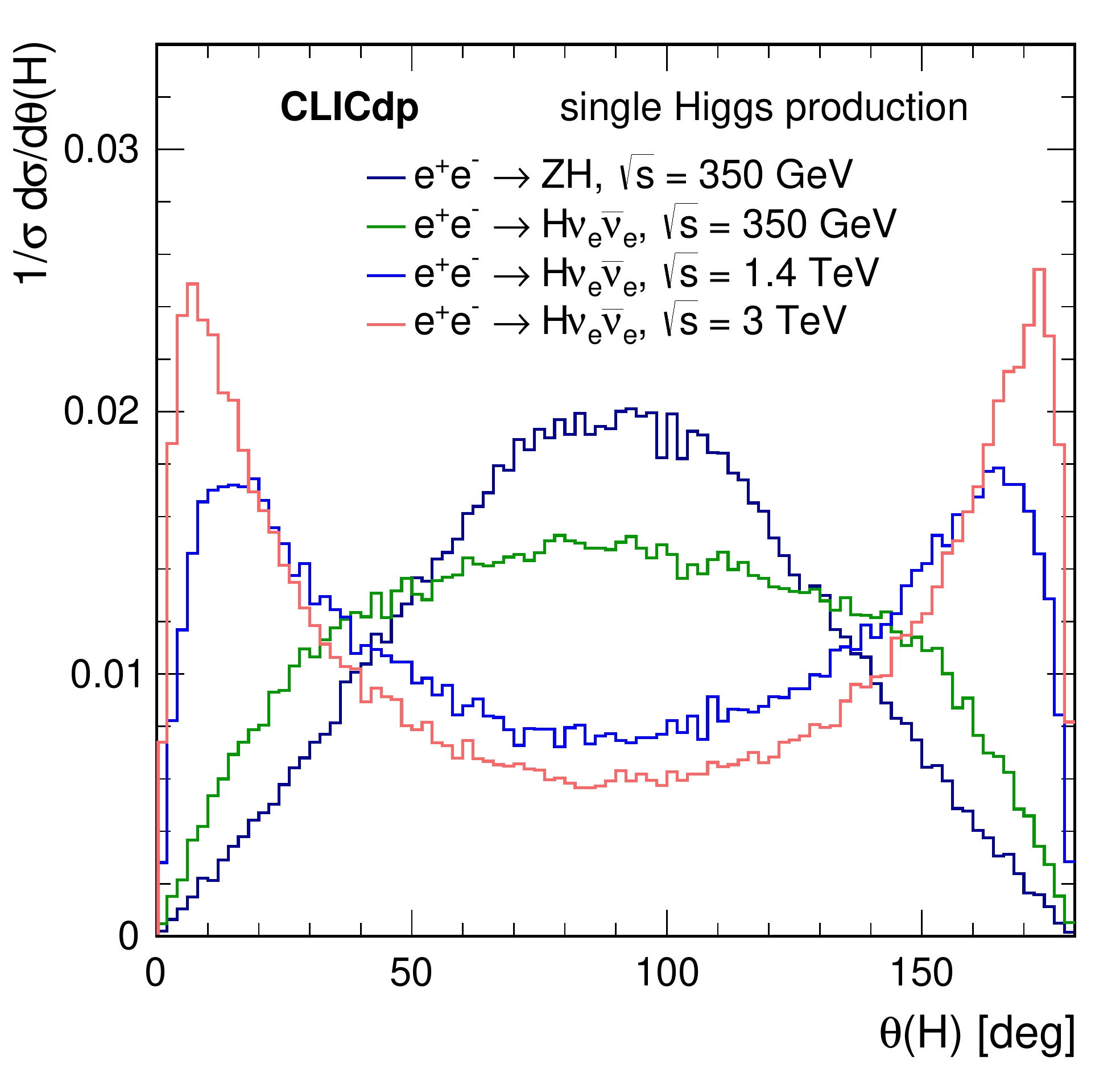}
  \caption{Polar angle distributions for single Higgs
    events at $\roots = 350\,\GeV$, $1.4\,\TeV$ and $3\,\TeV$,
    including the effects of the CLIC beamstrahlung spectrum and
    ISR. The distributions are normalised to unity.
    \label{fig:higgs:theta}}
\end{figure}

The polar angle distributions for single Higgs production obtained using \whizard 1.95~\cite{Kilian:2007gr} for the CLIC
centre-of-mass energies are shown in \autoref{fig:higgs:theta}. Most
Higgs bosons produced at $\roots = 350\,\GeV$ can be reconstructed in
the central parts of the detectors while  Higgs bosons produced in 
the $\PW\PW$-fusion process and their decay products tend towards the beam 
axis with increasing energy. Hence good 
detectors capabilities in the forward regions are crucial at $\roots = 1.4\,\TeV$
and $3\,\TeV$.

\begin{table}[tb]\centering
 {\renewcommand{\arraystretch}{1.2}%
 \begin{tabular}{lrrr}
   \toprule 
                  $\roots=$                     & \tabt{350\,GeV} & \tabt{1.4\,TeV} & \tabt{3\,TeV} \\ \midrule

    \LumiIntDiff                             & 500\,\fbinv     & 1.5\,\abinv    & 2\,\abinv  \\
    $\sigma(\epem\to\PZ\PH)$          & 133\,fb         & 8\,fb           & 2\,fb         \\
    $\sigma(\epem\to\PH\PGne\PAGne)$  & 34\,fb          & 276\,fb         & 477\,fb       \\
    $\sigma(\epem\to\PH\epem)$     & 7\,fb           & 28\,fb          & 48\,fb        \\
    No.\@ $\PZ\PH$ events                   & 68,000         & 20,000         & 11,000       \\
    No.\@ $\PH\PGne\PAGne$ events           & 17,000         & 370,000        & 830,000      \\
    No.\@ $\PH\epem$ events              & 3,700          & 37,000         & 84,000       \\
    \bottomrule
  \end{tabular}
  }%end of arraystretch
    \caption{Leading-order, unpolarised cross sections for
       \higgsstrahlung, $\PW\PW$-fusion, and $\PZ\PZ$-fusion
      processes for $\mH=126\,\GeV$ at the three centre-of-mass
      energies discussed in this paper. $\rootsprime$ is the 
      effective centre-of-mass energy of the $\epem$ collision.  
      The presented cross sections include the effects of ISR 
      but exclude the effects of beamstrahlung. 
      Also given are numbers of expected events,
      including the effects of ISR and the CLIC beamstrahlung spectrum. 
      The presented cross sections and event numbers do not include 
      possible enhancements from polarised beams.
    \label{tab:higgs:events}}
\end{table}

A SM Higgs boson with mass of $\mH=126\,\GeV$ has a wide range of
decay modes, as listed in \autoref{tab:higgs:decay}, providing the
possibility to test the SM predictions for the couplings of the Higgs
to both gauge bosons and to fermions~\cite{Dittmaier:2012vm}. All the
modes listed in \autoref{tab:higgs:decay} are accessible at CLIC.

\begin{table}[tb]\centering
 \begin{tabular}{lr}
   \toprule 
                  Decay mode                     & Branching ratio \\ \midrule
   $\PH\to\PQb\PAQb$                            &      56.1\,\%        \\ 
   $\PH\to\PW\PW^*$                              &      23.1\,\%     \\ 
   $\PH\to\Pg\Pg$                                    &        8.5\,\%   \\
   $\PH\to\tptm$                                       &        6.2\,\%     \\
   $\PH\to\PQc\PAQc$                             &        2.8\,\%        \\ 
   $\PH\to\PZ\PZ^*$                                &          2.9\,\% \\
   $\PH\to\PGg\PGg$                              &           0.23\,\% \\
   $\PH\to\PZ\PGg$                                &           0.16\,\% \\ 
   $\PH\to\mpmm$                                  &          0.021\,\%  \\
   \midrule
   $\Gamma_{\PH}$                                     & 4.2\,\MeV \\
    \bottomrule
  \end{tabular}
    \caption{The investigated SM Higgs decay modes and their branching ratios as well as the total Higgs width for 
      $\mH=126\,\GeV$~\cite{Dittmaier:2012vm}. \label{tab:higgs:decay}}
\end{table}

\subsection{Motivation for $\roots=350$\,GeV CLIC Operation}

The choice of the CLIC energy stages is motivated by the desire to
pursue a programme of precision Higgs physics and to operate the
machine above $1\,\TeV$ at the earliest possible time; no CLIC
operation is foreseen below the top-pair production threshold. 

From the Higgs physics perspective, 
operation at energies much below $1\,\TeV$ is 
motivated by the direct and model-independent measurement of the
coupling of the Higgs boson to the $\PZ$, which can be obtained from
the recoil mass distribution in $\PZ\PH\to\epem\PH$,
$\PZ\PH\to\mpmm\PH$ and $\PZ\PH\to\qq\PH$ production (see
\autoref{sec:higgsstrahlung_z_ll} and~\autoref{sec:higgsstrahlung_z_qq}).  These measurements play a central
role in the determination of the Higgs couplings at an $\epem$
collider.

However, from a Higgs physics perspective, there is no advantage to running CLIC at around $\roots = 250\,\GeV$ where the $\PZ\PH$ production cross section is larger, compared to running at $\roots = 350\,\GeV$.
Firstly, the
reduction in cross section at $\roots = 350\,\GeV$ is compensated, in 
part, by the increased instantaneous luminosity achievable
at a higher centre-of-mass energy. The instantaneous luminosity scales
approximately linearly with the centre-of-mass energy, ${\cal{L}}
\propto \gamma_{\Pe}$, where $\gamma_{\Pe}$ is the Lorentz factor for
the beam electrons/positrons. For this reason, the precision on the
coupling $g_{\PH\PZ\PZ}$ at $350\,\GeV$ is comparable to that
achievable at $250\,\GeV$ for the same period of operation. Secondly,
the additional boost of the $\PZ$ and $\PH$ at $\roots = 350\,\GeV$
provides greater separation between the final-state jets from $\PZ$
and $\PH$ decays. Consequently, the measurements of
\mbox{$\sigma(\PZ\PH)\times \BR(\PH\to X)$} are more precise at
$\roots = 350\,\GeV$. Thirdly, and most importantly, 
operation of CLIC at $\roots\approx350\,\GeV$ provides access 
to the $\epem\to\PH\PGne\PAGne$ fusion process; this improves 
the precision with which the total decay width $\Gamma_{\PH}$ 
can be determined at CLIC.
For the above
reasons, the preferred option for the first stage of CLIC operation is
$\roots \approx 350\,\GeV$.

Another advantage of $\roots\approx350\,\GeV$ is that 
detailed studies of the top-pair production process can be performed
in the initial stage of CLIC operation.  
Finally, the Higgs boson mass can be measured at $\roots = 350\,\GeV$ with similar precision 
as at $\roots = 250\,\GeV$.

\subsection{Impact of Beam Polarisation}

\label{sec:higgs:polarisation}
The majority of CLIC Higgs physics studies presented in this paper are performed
assuming unpolarised $\Pep$ and $\Pem$ beams. However, in the
baseline CLIC design, the electron beam can be polarised up to
$\pm80\,\%$. There is also the possibility of positron polarisation at a
lower level, although positron polarisation is not part of the baseline CLIC design.
For an electron polarisation of $P_-$ and positron polarisation of
$P_+$, the relative fractions of collisions in the different helicity
states are:
\begin{align*}
  \PemR\PepR\,:\,\ \mbox{$\frac{1}{4}$}(1+P_-)(1+P_+)  & \ \ \text{,} \ \ \PemR\PepL : \  \mbox{$\frac{1}{4}$}(1+P_-)(1-P_+)\,, \\
  \PemL\PepR\,:\,\ \mbox{$\frac{1}{4}$}(1-P_-)(1+P_+)  & \ \ \text{,} \ \ \PemL\PepL : \  \mbox{$\frac{1}{4}$}(1-P_-)(1-P_+)\,.
\end{align*} 
By selecting different beam polarisations it is possible to
enhance/suppress different physical processes. The chiral nature of
the weak coupling to fermions results in significant possible
enhancements in $\PW\PW$-fusion Higgs production, as indicated in
\autoref{tab:higgs:polarisation}. The potential gains for the
$s$-channel \higgsstrahlung process, $\epem\to\PZ\PH$, are less
significant, and the dependence of the $\epem\to\PH\epem$ cross section  on
beam polarisation is even smaller.  In practice, the balance between
operation with different beam polarisations will depend on the CLIC
physics programme taken as a whole, including the searches for and
potential measurements of BSM particle production.

\begin{table}[tb]\centering
  \begin{tabular}{cccc}\toprule
    Polarisation                              & \tabttt{Scaling factor}                        \\ \cmidrule(l){2-4}
    $P(\Pem):P(\Pep)$                  & $\!\!\epem\!\to\PZ\PH\!\!$ & $\!\!\epem\!\to\PH\PGne\PAGne\!\!$& $\!\!\epem\!\to\PH\epem\!\!$ \\ \midrule
    unpolarised                                  & 1.00                & 1.00        &   1.00        \\
    $-80\,\%\,:\phantom{+3}\,0\,\%$   & 1.12                & 1.80        &    1.12             \\
    $-80\,\%\,:\,+30\,\%$                    & 1.40                & 2.34        &    1.17           \\
    $-80\,\%\,:\,-30\,\%$                    & 0.83                & 1.26        &    1.07           \\
    $+80\,\%\,:\phantom{+3}\,0\,\%$  & 0.88                & 0.20        &    0.88             \\
    $+80\,\%\,:\,+30\,\%$                    & 0.69                & 0.26        &    0.92           \\       
    $+80\,\%\,:\,-30\,\%$                    & 1.08                & 0.14        &    0.84           \\\bottomrule    
  \end{tabular}
\caption{The dependence of the event rates for the $s$-channel
  $\epem\to\PZ\PH$ process and the pure $t$-channel
  $\epem\to\PH\PGne\PAGne$ and $\epem\to\PH\epem$ processes for several 
  example beam polarisations. The scale factors assume an effective
  weak mixing angle given by $\sin^2\theta_{\PW}^\text{eff} =
  0.23146$~\cite{Beringer:1900zz}. The numbers are approximate as they do not account for
  interference between $\epem\!\to\PZ\PH\!\to\PGne\PAGne\PH$ and
  $\epem\!\to\PH\PGne\PAGne$.
\label{tab:higgs:polarisation}}
\end{table}

\subsection{\boldmath Overview of Higgs Measurements at {$\boldmath\roots=350$} GeV}

The \higgsstrahlung process, $\epem\to\PZ\PH$, provides an opportunity to study the
couplings of the Higgs boson in an essentially model-independent manner. 
Such a model-independent measurement is
unique to a lepton collider. 
\higgsstrahlung events can be selected based solely
on the measurement of the four-momentum of the $\PZ$ boson through its decay
products, while the invariant mass of the system recoiling against the $\PZ$ boson peaks at $\mH$.
The most distinct event topologies occur for $\PZ\to\epem$
and $\PZ\to\mpmm$ decays, which can be identified by requiring that
the di-lepton invariant mass is consistent with $\mZ$ 
(see \autoref{sec:higgsstrahlung_z_ll}). 
SM background cross sections are relatively low. 
A slightly less clean,
but more precise, measurement is obtained from the recoil mass
analysis for $\PZ\to\qq$ decays (see
\autoref{sec:higgsstrahlung_z_qq}).

Recoil-mass studies
provide an absolute measurement of the total $\PZ\PH$ production cross
section and a model-independent measurement of the coupling of the
Higgs to the $\PZ$ boson, $g_{\PH\PZ\PZ}$. The combination of the
leptonic and hadronic decay channels allows $g_{\PH\PZ\PZ}$ to be
determined with a precision of $0.8\,\%$. In addition, the recoil mass
from $\PZ\to\qq$ decays provides a direct search for possible Higgs
decays to invisible final states, and can be used to constrain the
invisible decay width of the Higgs, $\Gamma_\text{invis}$.

By identifying the individual final states for different Higgs decay
modes, precise measurements of the Higgs boson branching fractions can
be made. Because of the high flavour tagging
efficiencies~\cite{CLIC_PhysDet_CDR} achievable at CLIC, the
$\PH\to\PQb\PAQb$ and $\PH\to\PQc\PAQc$ decays can be cleanly
separated. Neglecting the Higgs decays into light quarks, the
branching ratio of $\PH\to\Pg\Pg$ can also be inferred and
$\PH\to\tptm$ decays can be identified.

Although the cross section is lower, the $t$-channel $\PW\PW$-fusion
process $\epem\to\PH\PGne\PAGne$ is an important part of the CLIC
Higgs physics programme at $\roots \approx 350\,\GeV$. Because the visible 
final state consists of the Higgs boson decay products alone, the
direct reconstruction of the invariant mass of the Higgs boson or
its decay products plays a central role
in the event selection. The combination of Higgs production and decay
data from \higgsstrahlung and $\PW\PW$-fusion processes provides a
model-independent extraction of Higgs couplings.

\subsubsection{Extraction of Higgs Couplings}

At the LHC, only the ratios of the Higgs boson couplings can 
be inferred from the data in a model-independent way. 

In contrast, at an electron-positron collider such as CLIC, 
absolute measurements of the couplings to the Higgs boson can be determined using the
total $\epem\to\PZ\PH$ cross section determined from recoil mass
analyses. This allows the coupling of the Higgs boson to the $\PZ$ to
be determined with a precision of better than $1\,\%$ in an
essentially model-independent manner. Once the coupling to the $\PZ$
is known, the Higgs coupling to the $\PW$ can be determined from, for
example, the ratios of \higgsstrahlung to $\PW\PW$-fusion cross
sections:
\begin{equation*}
    \frac{\sigma(\epem\to\PZ\PH)\times \BR(\PH\to\PQb\PAQb) }{ \sigma{(\epem\to\PGne\PAGne\PH)} \times \BR(\PH\to\PQb\PAQb)  } \propto \left(\frac{g_{\PH\PZ\PZ}}{g_{\PH\PW\PW}} \right)^2 \,.
\end{equation*}

Knowledge of the Higgs total decay width, extracted from the data, 
allows absolute measurements of the other Higgs couplings.

For a Higgs boson mass of around $126\,\GeV$, the total Higgs
decay width in the SM ($\Gamma_{\PH}$) is less than $5\,\MeV$ and
cannot be measured directly at an $\epem$ linear collider. 
However, as the absolute
couplings of the Higgs boson to the $\PZ$ and $\PW$ can be determined, 
the total decay width of the Higgs boson can be
determined from $\PH\to\PW\PW^*$ or $\PH\to\PZ\PZ^*$ decays. For
example, the measurement of the Higgs decay to $\PW\PW^*$ in the
$\PW\PW$-fusion process determines:
\begin{equation*}
                 \sigma(\PH\PGne\PAGne)\times \BR(\PH\to\PW\PW^*)   \propto \frac{g^4_{\PH\PW\PW}}{\Gamma_{\PH}}\,,
\end{equation*}
and thus the total width can be determined utilising the
model-independent measurement of $g_{\PH\PW\PW}$. In practice, a fit
(see \autoref{sec:combined_fits}) is performed to all of the
experimental measurements involving the Higgs boson couplings.

\subsection{\boldmath Overview of Higgs Measurements at $\roots > 1$\,TeV}

For CLIC operation above $1\,\TeV$, the large number of Higgs bosons
produced in the $\PW\PW$-fusion process allow relative couplings of
the Higgs boson to the $\PW$ and $\PZ$ bosons to be determined at the
${\cal{O}}(1\,\%)$ level. These measurements provide a strong test of
the SM prediction for:
\begin{equation*}
g_{\PH\PW\PW} / g_{\PH\PZ\PZ} = \cos^2\theta_\mathrm{W},
\end{equation*} 
where $\theta_\mathrm{W}$ is the weak-mixing
angle. Furthermore, the exclusive Higgs decay modes can be studied
with significantly higher precision than at $\roots=350\,\GeV$. For
example, CLIC operating at $3\,\TeV$ yields a statistical
precision of $2\,\%$ on the ratio $g_{\PH\PQc\PQc}/g_{\PH\PQb\PQb}$,
providing a direct comparison of the SM coupling predictions for
up-type and down-type quarks.  In the
context of the model-independent measurements of the Higgs branching
ratios, the measurement of $\sigma(\PH\PGne\PAGne)\times
\BR(\PH\to\PW\PW^*)$ is particularly important. For CLIC operation at
$\roots\approx1.4\,\TeV$, the large number of events allows this cross
section to be determined with a precision of $1\,\%$ (see
\autoref{sec:ww_fusion_ww}). When combined with the measurements
at $\roots\approx350\,\GeV$, this places a strong constraint on
$\Gamma_{\PH}$.

Although the $\PW\PW$-fusion process has the largest cross section for
Higgs production above $1\,\TeV$, other processes are also important.
For example, measurements of the $\PZ\PZ$-fusion process provide
further constraints on the $g_{\PH\PZ\PZ}$ coupling. Moreover, CLIC
operation at $\roots = 1.4\,\TeV$ enables a determination of the top
Yukawa coupling from the process
$\epem\to\PQt\PAQt\PH\to\PQb\PW^+\PAQb\PW^-\PH$ with a precision of
$4.2\,\%$ (see \autoref{sec:top_yukawa}). Finally, the
self-coupling of the Higgs boson at the $\PH\PH\PH$ vertex is
measurable in $1.4\,\TeV$ and $3\,\TeV$ operation. 

In the SM, the
Higgs boson originates from a doublet of complex scalar fields $\phi$
described by the potential:
\begin{equation*}
       V(\phi) = \mu^2\phi^\dagger\phi + \lambda(\phi^\dagger\phi)^2 \,,
\end{equation*}
where $\mu$ and $\lambda$ are the parameters of the Higgs potential, 
with $\mu^2 < 0$ and $\lambda > 0$.
The measurement of the strength of the Higgs self-coupling
 provides direct access to the coupling $\lambda$ assumed in
the Higgs mechanism. 
For $\mH$ of around $126\,\GeV$, the measurement of the Higgs boson
self-coupling at the LHC will be extremely challenging, even with
$3000\,\fbinv$ of data (see for example~\cite{Dawson:2013bba}). At a
linear collider, the trilinear Higgs self-coupling can be measured
through the $\epem\to\PZ\PH\PH$ and $\epem\to\PH\PH\PGne\PAGne$
processes. The
$\epem\to\PZ\PH\PH$ process at $\roots=500\,\GeV$ has been studied in the context of
the ILC, where the results show that a very large integrated
luminosity is required~\cite{ILCPhysicsDBD}. However for
 $\roots\ge 1\,\TeV$, the sensitivity for the process 
$\epem\to\PH\PH\PGne\PAGne$ increases with increasing centre-of-mass energy and
the measurement of the Higgs boson self-coupling (see
\autoref{sec:higgs_self_coupling}) forms a central part of the CLIC
Higgs physics programme. Ultimately a precision of approximately
$20\,\%$ on $\lambda$ can be achieved.

% Section 2
\section{Event Generation, Detector Simulation and Reconstruction}
\label{sec:software}

The results presented in this paper are based on detailed Monte Carlo
(MC) simulation studies including the generation of a complete set of relevant SM background
processes, \geant~\cite{Agostinelli2003,Allison2006} based
simulations of the CLIC detector concepts, and a full reconstruction
of the simulated events.

\subsection{Event Generation}

Because of the presence of beamstrahlung photons in the colliding
electron and positron beams, it is necessary to generate MC event
samples for $\epem$, $\Pep\PGg$, $\PGg\Pem$, and $\PGg\PGg$
interactions. The main physics backgrounds, with up to six particles
in the final state, are generated using the \whizard
1.95~\cite{Kilian:2007gr} program. In all cases the expected energy
spectra for the CLIC beams, including the effects from beamstrahlung
and the intrinsic machine energy spread, are used for the
initial-state electrons, positrons and beamstrahlung photons. In
addition, low-$Q^2$ processes with quasi-real photons are described
using the Weizs\"{a}cker-Williams approximation as implemented in
\whizard. The process of fragmentation and hadronisation is simulated
using \pythia 6.4 \cite{Sjostrand2006} with a parameter set 
tuned to OPAL $\epem$ data recorded at LEP \cite{Alexander:1995bk} 
(see~\cite{CLIC_PhysDet_CDR} for details). The decays of $\PGt$
leptons are simulated using \tauola~\cite{tauola}. The mass of the
Higgs boson is taken to be $126\,\GeV$\footnote{A Higgs boson of
  $125\,\GeV$ is used in the process $\epem \to \PQt\PAQt\PH$.}
and the decays of the Higgs boson are simulated using \pythia with
the branching fractions listed in \cite{Dittmaier:2012vm}. The events
from the different Higgs production channels are simulated
separately. The background samples do not include Higgs processes. 
MC samples for the measurement of the top Yukawa coupling measurement
(see Section~\ref{sec:top_yukawa}) with eight final-state fermions
are obtained using the \physsim~\cite{gen:physsim} package; again
\pythia is used for fragmentation, hadronisation and the Higgs boson
decays.

\subsection{Simulation and Reconstruction}

\label{sec:simreco}

The \geant detector simulation toolkits \mokka~\cite{Mokka} and
\slic~\cite{Graf:2006ei} are used to simulate the detector response
to the generated events in the \clicild and \clicsid concepts,
respectively. The \texttt{QGSP\_BERT} physics list is used to model
the hadronic interactions of particles in the detectors. The
digitisation, namely the translation of the raw simulated energy
deposits into detector signals, and the event reconstruction are 
performed using the \marlin~\cite{MarlinLCCD} and
\lcsim \cite{Graf:2011zzc} software packages. Particle flow
reconstruction is performed using \pandora~\cite{thomson:pandora, Marshall2013153, Marshall:2015rfa}. 

Vertex reconstruction and heavy flavour tagging
are performed using the \lcfiplus program~\cite{Suehara:2015ura}. 
This consists of a topological vertex finder that reconstructs 
secondary interactions, and a multivariate classifier that 
combines several jet-related variables such as track impact parameter
significance, decay length, number of tracks in vertices, and vertex masses, 
to tag bottom, charm, and light-quark jets.
The detailed
training of the multivariate classifiers for the flavour tagging is performed separately
for each centre-of-mass energy and each final state of interest.

Because of the 0.5\,ns bunch spacing in the CLIC beams, the pile-up of
beam-induced backgrounds can affect the event reconstruction and needs
to be accounted for. Realistic levels of pile-up from the most
important beam-induced background, the \gghadrons process, are 
included in all the simulated event samples to ensure that the impact
on the event reconstruction is correctly modelled. The \gghadrons
events are simulated separately and a randomly chosen subset,
corresponding to 60 bunch crossings, is superimposed on the physics
event before the digitisation step~\cite{LCD:overlay}. 
60 bunch crossings is equivalent to 30\,ns, which is much longer than 
the assumed offline event reconstruction window of 10\,ns around the hard physics event, so this is a 
good approximation \cite{CLIC_PhysDet_CDR}.
For the $\roots = 350\,\GeV$ samples, where the background rates are lower, 
300 bunch crossings are overlaid on the physics event. The impact of the
background is small at $\roots = 350\,\GeV$, and is most significant
at $\roots=3\,\TeV$, where approximately $1.2\,\TeV$ of energy is
deposited in the calorimeters in a time window of 10\,ns. A dedicated
reconstruction algorithm identifies and removes 
approximately $90\,\%$ of these out-of-time background particles using criteria
based on the reconstructed transverse momentum \pT of the particles and
the calorimeter cluster time. A more detailed description can be
found in \cite{CLIC_PhysDet_CDR}.

Jet finding is performed on the objects reconstructed by particle flow,
using the \fastjet~\cite{Fastjet}
package. Because of the presence of pile-up from \gghadrons, it was
found that the Durham~\cite{Catani:1991hj} algorithm employed at LEP is not
optimal for CLIC studies. Instead, the hadron-collider inspired $k_t$
algorithm~\cite{Catani:1993hr, Ellis:1993tq}, with the distance parameter $R$ based on $\Delta\eta$ and
$\Delta\phi$, is found to give better performance since it increases
distances in the forward region, thus reducing the clustering of the
(predominantly low transverse momentum) background particles together
with those from the hard $\epem$ interaction. 
Instead, particles that are found by the $k_t$ algorithm to be 
closer to the beam axis than to any other particles, 
and that are thus likely to have originated from beam-beam backgrounds, 
are removed from the event.
As a result of using the
$R$-based $k_t$ algorithm, the impact of the pile-up from \gghadrons
is largely mitigated, even without the timing and momentum cuts
described above. Further details are given in
\cite{CLIC_PhysDet_CDR}. The choice of $R$ is optimised separately
for different analyses. In many of the following studies, events are
forced into a particular $N$-jet topology. 
The variable $y_{ij}$ is the smallest $k_{t}$ distance when combining $j$ jets to $i = (j-1)$ jets. These resolution 
parameters are widely used in a number of event selections, allowing
events to be categorised into topologically different final states. In
several studies it is found to be advantageous first to apply the
$k_t$ algorithm to reduce the beam-beam 
backgrounds, and then to use only the remaining objects 
as input to the Durham algorithm.
 
\begin{figure}
\begin{centering}
\includegraphics[width=0.9\columnwidth]{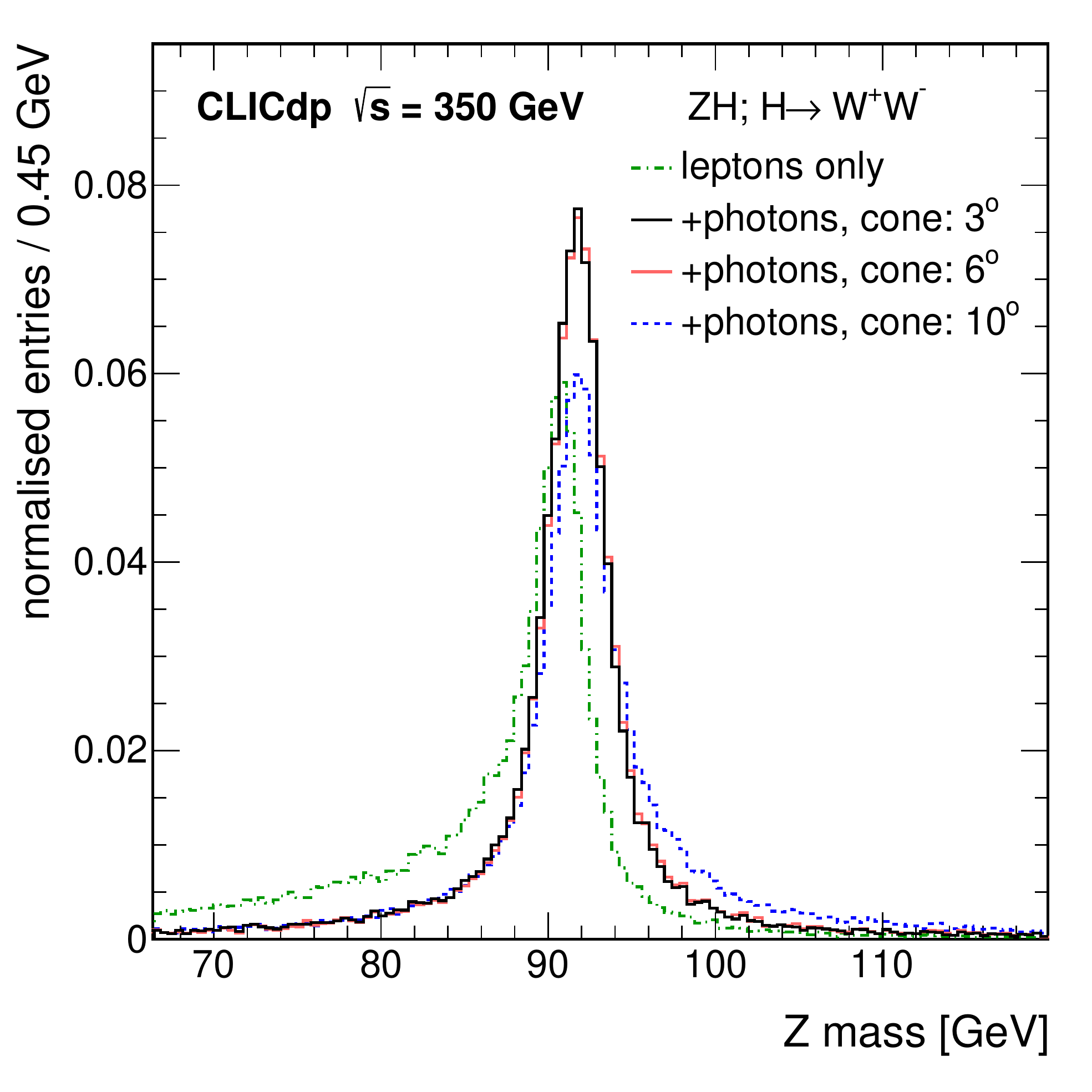}
\caption{\label{fig:bremsstrahlung_recovery} Reconstructed invariant mass of $\PZ\to\epem$ candidates in $\epem\to\PZ\PH\to\PZ\WW^\ast$ events at $\roots=350\,\GeV$. Bremsstrahlung photons in cones of different opening angles around the electron direction are recovered as described in the text. All distributions are normalised to unity.}
\end{centering}
\end{figure}

To
recover the effect of bremsstrahlung photons radiated 
from reconstructed leptons, 
all photons in a cone 
around the flight direction of a lepton candidate are added 
to its four-momentum. The impact of the bremsstrahlung
recovery on the reconstruction of the $\PZ\to\epem$ decays is
illustrated in \autoref{fig:bremsstrahlung_recovery}.
The bremsstrahlung effect leads to a tail at lower values in the 
$\PZ$ candidate invariant mass distribution. This loss can be recovered by 
the procedure described above. It is also visible that a too large 
opening angle of the recovery cone leads to a tail at higher masses;
typically, an opening angle of $3^\circ$ is chosen.

The event simulation and reconstruction of the large data samples used
in this study was performed using the \ilcdirac \cite{Grefe:2014sca, Tsaregorodtsev:2008zz} grid
production tools.

% Section 4
\section{Higgs Production at $\roots=$350\,GeV}
\label{sec:higgsstrahlung}

The study of the \higgsstrahlung process is central to the precision
Higgs physics programme at any future high-energy electron-positron collider~\cite{Thomson:2015jda}.
This section presents studies of
$\epem\to\PZ\PH$ at $\roots=350\,\GeV$ with a focus on
 model-independent measurements of $\PZ\PH$ production from the
kinematic properties of the \PZ decay products. Complementary information 
obtained from Higgs production through WW-fusion at $\roots=350\,\GeV$ is also presented. 
All analyses at $\roots=350\,\GeV$ described in this paper use the \clicild detector model.

% Section 4.1
\subsection{Recoil Mass Measurements of $\epem\to\PZ\PH$}
\label{sec:higgsstrahlung_cross_section}

In the process $\epem\to\PZ\PH$, it is possible to 
identify efficiently $\PZ\to\epem$ and $\PZ\to\mpmm$ decays with a selection
efficiency that is essentially independent of the $\PH$ decay
mode. The four-momentum of the (Higgs boson) system recoiling
against the $\PZ$ can be obtained from $E_{rec} = \roots - E_{\PZ}$
and $\vec{p}_{rec} = -\vec{p}_{\PZ}$, and the recoil mass, $\mrec$,
 peaks sharply around $\mH$. The recoil mass analysis for leptonic
decays of the $\PZ$ is described in
\autoref{sec:higgsstrahlung_z_ll}. While these measurements
provide a clean model-independent probe of $\PZ\PH$ production, they
are limited by the relatively small leptonic branching ratios of the
$\PZ$. Studies of $\PZ\PH$ production with $\PZ\to\qq$ are inherently
less clean, but are statistically more powerful. Despite the
challenges related to the reconstruction of hadronic $\PZ$ decays in
the presence of various Higgs decay modes, a precise and nearly
model-independent probe of $\PZ\PH$ production can be obtained by
analysing the recoil mass in hadronic $\PZ$ decays, as detailed in
\autoref{sec:higgsstrahlung_z_qq}. When all these measurements are
taken together, a model-independent measurement of the $g_{\PH\PZ\PZ}$
coupling constant with a precision of $<1\,\%$ can be inferred~\cite{Thomson:2015jda}.

% Section 4.1.1
\subsubsection{Leptonic Decays: $\PZ\to\epem$ and $\PZ\to\mpmm$} %(John)
\label{sec:higgsstrahlung_z_ll}
{

The signature for $\epem\to\PZ\PH$ production with $\PZ\to\epem$ or
$\PZ\to\mpmm$ is a pair of oppositely charged high-$\pT$ leptons, with
an invariant mass consistent with that of the $\PZ$ boson,
$\mll \approx \mZ$, and a recoil mass, calculated from the four-momenta
of the leptons alone, consistent with the Higgs mass,
$\mrec\approx\mH$~\cite{LCD:recoil_leptonic}. 
Backgrounds from two-fermion final states $\epem\to\Plp\Plm$
($\Pl = \Pe, \PGm, \PGt$) are trivial to remove. The dominant
backgrounds are from four-fermion processes with final states
consisting of a pair of oppositely-charged leptons and any other
possible fermion pair. For both the $\mpmm X$ and $\epem X$ channels,
the total four-fermion background cross section is approximately one thousand
times greater than the signal cross section.

The event selection employs preselection cuts and a
multivariate analysis. 
The preselection requires at least one negatively and one
positively charged lepton of the lepton flavour of interest (muons or
electrons) with an invariant mass loosely consistent with the mass of
the $\PZ$ boson, $40\,\GeV\ < \mll < 126\,\GeV$. For signal events, the
lepton identification efficiencies are $99\,\%$ for muons and $90\,\%$
for electrons. Backgrounds from two-fermion processes are essentially
eliminated by requiring that the di-lepton system has $\pT>60\,\GeV$. 
Four-fermion backgrounds are
suppressed by requiring  $95\,\GeV\ < \mrec < 290\,\GeV$. 
%%VJM - the next two lines could go....
The lower bound suppresses $\epem\to\PZ\PZ$
production. The upper bound is significantly greater than the Higgs
boson mass, to allow for the possibility of $\PZ\PH$ production with
ISR or significant beamstrahlung, which, in the recoil mass analysis, 
results in a tail to the recoil mass distribution, as it is the mass
of the $\PH\PGg$ system that is estimated.

Events passing the preselection cuts are categorised using a
multivariate analysis of seven discriminating variables: the
transverse momentum ($\pT$) and invariant mass ($\mll$) of the
candidate $\PZ$; %as reconstructed from the di-lepton system; 
the cosine
of the polar angle $(|\cos\theta|)$ of the candidate $\PZ$; the
acollinearity and acoplanarity of the leptons; the imbalance between
the transverse momenta of the two selected leptons $({\pT}_1 -
{\pT}_2)$; and the transverse momentum of the highest energy photon in
the event. The event selection employs a Boosted Decision Tree (BDT)
as implemented in \tmva\,\cite{TMVA:2010}. The resulting selection
efficiencies are summarised in \autoref{tab:hz:leptonic}. For both  
final states, the number of selected background events is less than twice the 
number of selected signal events. The impact of the background is  reduced 
using a fit to the recoil mass distribution.

\begin{table}[h]
\begin{center}
   \begin{tabular}{lrrrr}
    \toprule 
         Process                                & \tabt{$\sigma/\text{fb}$} & \tabt{$\varepsilon_\text{presel}$} & \tabt{$\varepsilon_\text{BDT}$} & \tabt{$N_\text{BDT}$} \\ \midrule
            $\PZ\PH; \PZ\to\mpmm$                             & 4.6         &    84\,\%       &  65\,\%                     & 1253  \\
      $\mpmm\text{f}\overline{\text{f}}$             &  4750      &    0.8\,\%       & 10\,\%                   & 1905 \\  
\midrule        
                  $\PZ\PH; \PZ\to\epem$                         & 4.6       &     73\,\%       & 51\,\%                   & 858    \\    
     $\epem\text{f}\overline{\text{f}}$                  &  4847   &      1.2\,\%       &  5.4\,\%                   & 1558     \\ 
%            $\PZ\PH; \PZ\to\mpmm$                             & 4.6         &    83.8\,\%       &  54.1\,\%                     & 1253  \\
%      $\mpmm\text{f}\overline{\text{f}}$             &  4753      &    0.8\,\%       & <0.01\,\%                   & 1905 \\  
%\midrule        
%                  $\PZ\PH; \PZ\to\epem$                         & 4.6       &     73.3\,\%       & 37.1\,\%                   & 858    \\    
%     $\epem\text{f}\overline{\text{f}}$                  &  4847   &      1.2\,\%       &  <0.1\,\%                   & 1558     \\ 
     \bottomrule
    \end{tabular}
    \caption{\label{tab:hz:leptonic}Preselection and selection efficiencies for the $\PZ\PH$ signal and most important background processes in the leptonic recoil mass analysis. The numbers of events correspond to 500\,$\fbinv$ at $\roots=350\,\GeV$.}
\end{center}
\end{table}

\begin{figure*}
\begin{centering}
{\includegraphics[width=0.9\columnwidth]{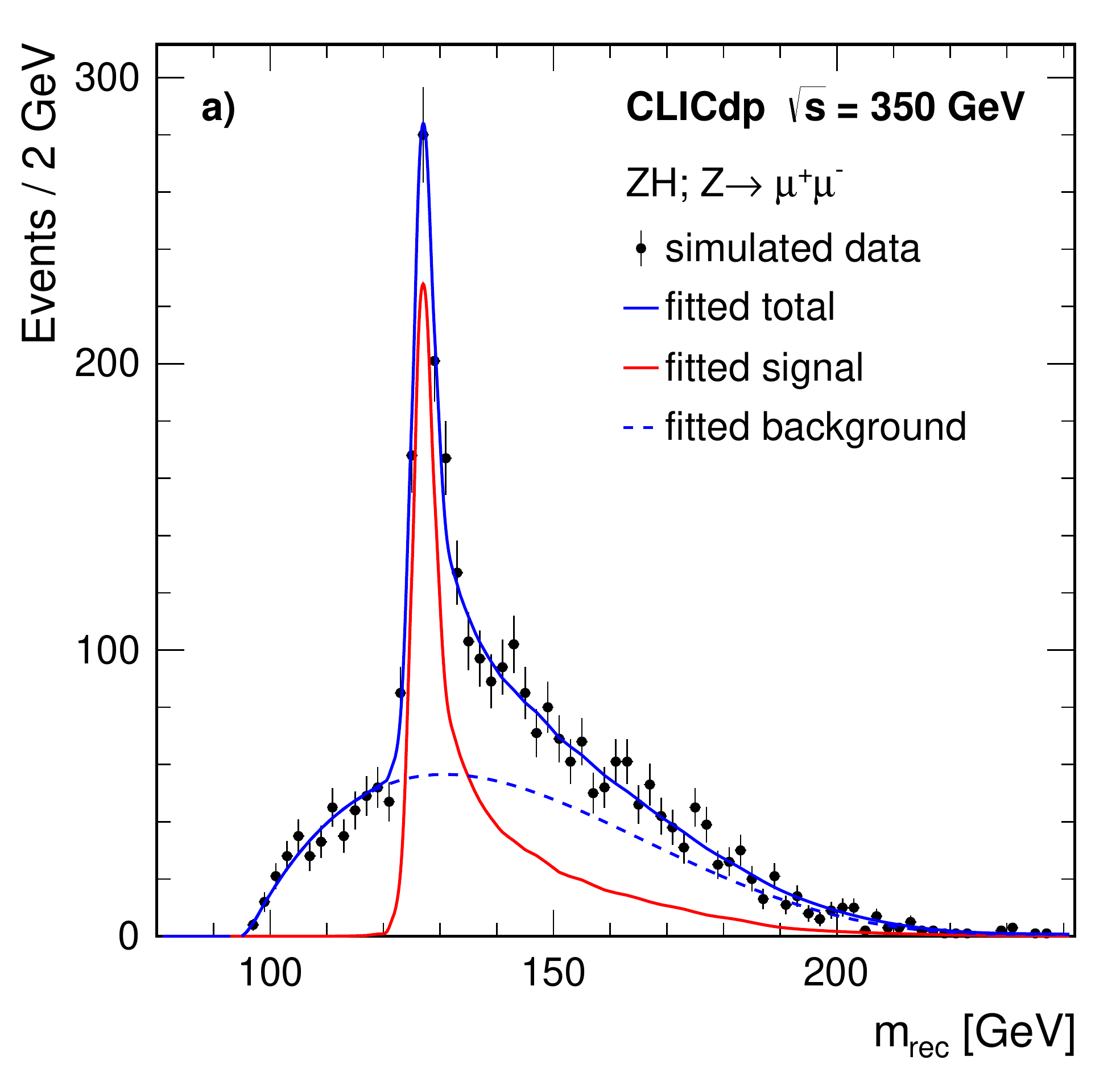}}
\hspace{0.1\columnwidth}
{\includegraphics[width=0.9\columnwidth]{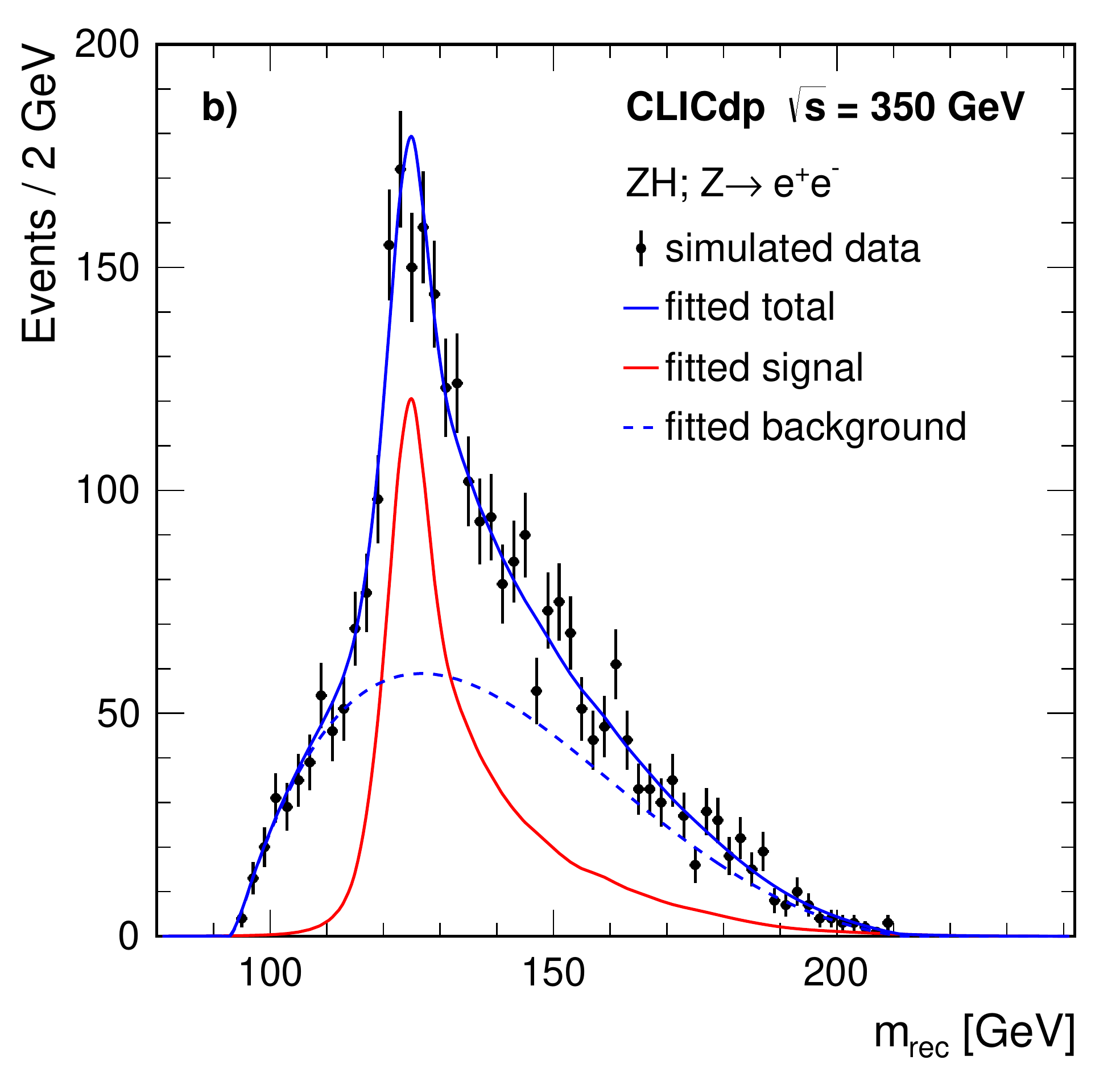}}
\caption{\label{350GeV_Recoil_Fit} Reconstructed recoil mass distributions of $\epem\to\PZ\PH$ events at $\roots=350\,\GeV$, where $\PZ\PH\to\mpmm X$ (a) and $\PZ\PH\to\epem X$ with bremsstrahlung recovery (b). All distributions are normalised to an integrated luminosity of $500\,\fbinv$.}
\end{centering}
\end{figure*}

A fit to the recoil mass distribution of the selected events (in both
the $\PZ\to\epem$ and $\PZ\to\mpmm$ channels) is used to extract
measurements of the $\PZ\PH$ production cross section and the Higgs
boson mass. The shape of the background contribution is parameterised
using a fourth order polynomial and the shape of the signal
distribution is modelled using Simplified Kernel
Estimation~\cite{Cranmer:2000du,CranmerKDE,OPALKDE} that provides a
description of the $\PZ\PH$ recoil mass distribution in which the
Higgs mass can subsequently be varied.  The accuracy
with which the Higgs mass and the number of signal events (and hence
the $\PZ\PH$ production cross section) can be measured, 
is determined using 1000 simulated test
data samples. Each test sample was created by adding the
high statistics selected signal sample (scaled to the correct
normalisation) to the smooth fourth-order polynomial background, then
applying Poisson fluctuations to individual bins %of the resulting smooth distribution 
to create a representative $500\,\fbinv$ data
sample. Each of the 1000 simulated data samples created in this way
is fitted allowing the Higgs mass, the signal normalisation and the
background normalisation to vary. \autoref{350GeV_Recoil_Fit}a
displays the results of fitting a typical test sample for the $\mpmm
X$ channel, while \autoref{350GeV_Recoil_Fit}b displays the
results for the $\epem X$ channel. In the $\epem X$ channel fits are
performed with, and without, applying an algorithm to recover
bremsstrahlung photons. The resulting measurement precisions for the
$\PZ\PH$ cross section and the Higgs boson mass are summarised
in \autoref{Table_350GeV_Recoil_Fit}. In the $\epem X$ channel, the bremsstrahlung recovery 
leads to a moderate improvement on the expected precision for the cross section measurement and a similar degradation in the expected precision for the mass determination, 
because it significantly increases the number of events in the peak of the recoil mass distribution, but also increases the width of this peak.
For an integrated luminosity of $500\,\fbinv$ 
at $\roots=350\,\GeV$, the combined precision on the Higgs boson
mass is:
\begin{align*}
\Delta(m_{\PH}) = 110\,\MeV,
\end{align*}
and the combined precision on the $\PZ\PH$ cross section is:
\begin{align*}
    \frac{\Delta{\sigma(\PZ\PH)}}{\sigma(\PZ\PH)}  &= 3.8\,\% \,.
\end{align*} 
The expected precision with (without) bremsstrahlung recovery in the $\epem X$ channel
was used in the combination for the cross section (mass).

\begin{table}[!h]
\begin{center}
\begin{tabular}{  c c  c }
\toprule
Channel & Quantity & Precision\\
\midrule
\multirow{2}{*}{$\mpmm X$} 
& $\mH$          & 122\,MeV\\
& $\sigma(\PZ\PH)$ & 4.72\,\%    \\
\midrule
\multirow{2}{*}{$\epem X$} 
& $\mH$          & 278\,MeV\\
& $\sigma(\PZ\PH)$ & 7.21\,\%       \\
\
$\epem X$              & $\mH$          & 359\,MeV\\
+ bremsstrahlung recovery        & $\sigma(\PZ\PH)$ & 6.60\,\%   \\
\bottomrule
\end{tabular}
\caption{\label{Table_350GeV_Recoil_Fit} Summary of measurement precisions
from the leptonic recoil mass analyses in the  $\mpmm X$ and $\epem X$ channels 
for an integrated luminosity of $500\,\fbinv$ at $350\,\GeV$.}
\end{center}
\end{table}

}

% Section 4.1.2
\subsubsection{Hadronic Decays: $\PZ \to \qqbar$} %(Mark)
\label{sec:higgsstrahlung_z_qq}
{ 

In the process $\epem\to\PZ\PH$, it is possible to cleanly identify
$\PZ\to\epem$ and $\PZ\to\mpmm$ decays regardless of the decay mode of
the Higgs boson and, consequently, the selection efficiency is almost
independent of the Higgs decay mode. In contrast, for $\PZ\to\qqbar$
decays, the selection efficiency shows a stronger dependence on
the Higgs decay mode~\cite{Thomson:2015jda}. For example,
$\epem\to(\PZ\to\qqbar)(\PH\to\PQb\PAQb)$ events consist of four
jets and the reconstruction of the $\PZ$ boson is complicated by
ambiguities in associations of particles with jets and the three-fold
ambiguity in associating four jets with the hadronic decays of the $\PZ$
and $\PH$. For this reason, it is much more difficult to construct a
selection based only on the reconstructed $\PZ\to\qqbar$ decay that
has a selection efficiency independent of the Higgs decay mode. The
strategy adopted is to first reject events consistent with a number of
clear background topologies using the information from the whole
event; and then to identify $\epem\to(\PZ\to\qqbar)\PH$ events solely
based on the properties from the candidate $\PZ\to\qqbar$ decay.

The $(\PZ\to\qqbar)\PH$ event selection proceeds in three separate
stages. In the first stage, to allow for possible BSM invisible Higgs
decay modes, events are divided into candidate visible Higgs decays
and candidate invisible Higgs decays, in both cases produced along
with a $\PZ\to\qqbar$. Events are categorised as potential visible
Higgs decays if they are not compatible with a clear two-jet topology:
\begin{itemize} 
  \item $\log_{10}(y_{23})> -2.0$ or $\log_{10}(y_{34})> -3.0$\,.
\end{itemize} 
All other events are considered as candidates for an invisible Higgs
decay analysis, based on that described in
\autoref{sec:higgsstrahlung_invisible}, although with looser
requirements to make the overall analysis more inclusive.

Preselection cuts then reduce the backgrounds from large cross section
processes such as \mbox{$\epem\to\qqbar$}
and \mbox{$\epem\to\qqbar\qqbar$}. The preselection variables are
formed by forcing each event into three, four and five jets. In each
case, the best candidate for being a hadronically decaying $\PZ$ boson
is chosen as the jet pair giving the di-jet invariant mass (\mqq)
closest to \mZ,  considering only jets with more than three charged
particles. The invariant mass of the system recoiling against the
$\PZ$ boson candidate, $\mrec$, is calculated
assuming $E_\text{rec} =\roots - E_{\qqbar}$ and $\vec{p}_\text{rec} =
-\vec{p}_{\qqbar}$. In addition, the invariant mass of all the
visible particles not originating from the candidate $\PZ\to\qqbar$
decay, $\mvis$, is calculated. It is important to note that $\mvis$ is
only used to reject specific background topologies in the preselection
and is not used in the main selection as it depends strongly on
the type of Higgs decay. The preselection cuts are:
\begin{itemize}
\item $70\,\GeV < \mqq < 110\,\GeV$ and $80\,\GeV < \mrec < 200\,\GeV$;
\item the background from $\epem\to\qqbar$ is suppressed by removing events with overall $\pT < 20\,\GeV$ and
   either $|\cos\theta_\text{mis}|>0.90$ or $\log_{10}(y_{34})>-2.5$, where $\theta_\text{mis}$ is the polar angle of the missing momentum vector;
\item events with little missing transverse
   momentum ($\pT < 20\,\GeV$) are forced into four jets and are
   rejected if the reconstructed di-jet invariant masses (and particle
   types) are consistent with the expectations for
   $\epem\to\qqbar\Pl\Pl$, $\epem\to\PZ\PZ\to\qqbar\qqbar$,
   $\epem\to\PW\PW\to\qqbar\qqbar$.
\end{itemize}

The final step in the event selection is a multivariate analysis. In
order not to bias the event selection efficiencies for different Higgs
decay modes, only variables related to the candidate $\PZ\to\qqbar$
decay are used in the selection. Forcing the event into four jets is
the right approach for $(\PZ\to\qqbar)\PH$ events where the Higgs
decays to two-body final states, but not necessarily for final states
such as $\PH\to\PW\PW^*\to\qqbar\qqbar$, where there is the chance
that one of the jets from the $\PW\PW^*$ decay will be merged with one
of the jets from the $\PZ\to\qqbar$, potentially biasing the selection
against $\PH\to\PW\PW^*$ decays. To mitigate this effect, the $\PZ$
candidate for the event selection can either be formed from the
four-jet topology as described above, or can be formed from a jet pair
after forcing the event into a five-jet topology. The latter case is
only used when $\log_{10}(y_{45}) > -3.5$ and the five-jet
reconstruction gives better $\PZ$ and $\PH$ candidates than the
four-jet reconstruction.  Attempting to reconstruct events in the
six-jet topology is not found to improve the overall analyses.
Having chosen the best $\PZ$ candidate in the event (from either the
four-jet or five-jet reconstruction), it is used to form variables for
the multivariate selection; information about the remainder of the
event is not used.

A relative likelihood selection is used to classify all events passing
the preselection cuts. Two event categories are considered: the
$\epem\to\PZ\PH\to\qqbar\PH$ signal and all non-Higgs background
processes. The relative likelihood for an event being signal is
defined as:
\begin{equation*}
       {\cal{L}}    = 
       \frac{L_\text{signal} }{L_\text{signal} + L_\text{back}} \,,
\end{equation*}  
where the individual absolute likelihood $L$ for each event type is
estimated from normalised probability distributions, $P_i(x_i)$, of the
discriminating variables $x_i$ for that event type:
\begin{equation*}
    L = \sigma_\text{presel} \times \prod_i^{N} P_i(x_i) \,,
\end{equation*}  
where $\sigma_\text{presel}$ is the cross section after the
preselection cuts. The discriminating variables used, all of which are
based on the candidate $\PZ\to\qqbar$ decay, are: the 2D distribution
of $\mqq$ and $\mrec$; the polar angle of the $\PZ$ candidate,
$|\cos\theta_{\PZ}|$; and the modulus of angle of jets from the $\PZ$
decay relative to its direction after boosting into its rest frame,
$\cosQ$. The clearest separation between signal and background is
obtained from $\mqq$ and the recoil mass $\mrec$, as shown
in \autoref{fig:hzqq:mass} for events passing the preselection. The signal is clearly peaked at $\mqq \approx
\mZ$ and $\mrec\approx \mH$.  The use of 2D mass distributions
accounts for the most significant correlations between the likelihood
variables.

\begin{figure*}
\centering
      \includegraphics[width=0.45\textwidth]{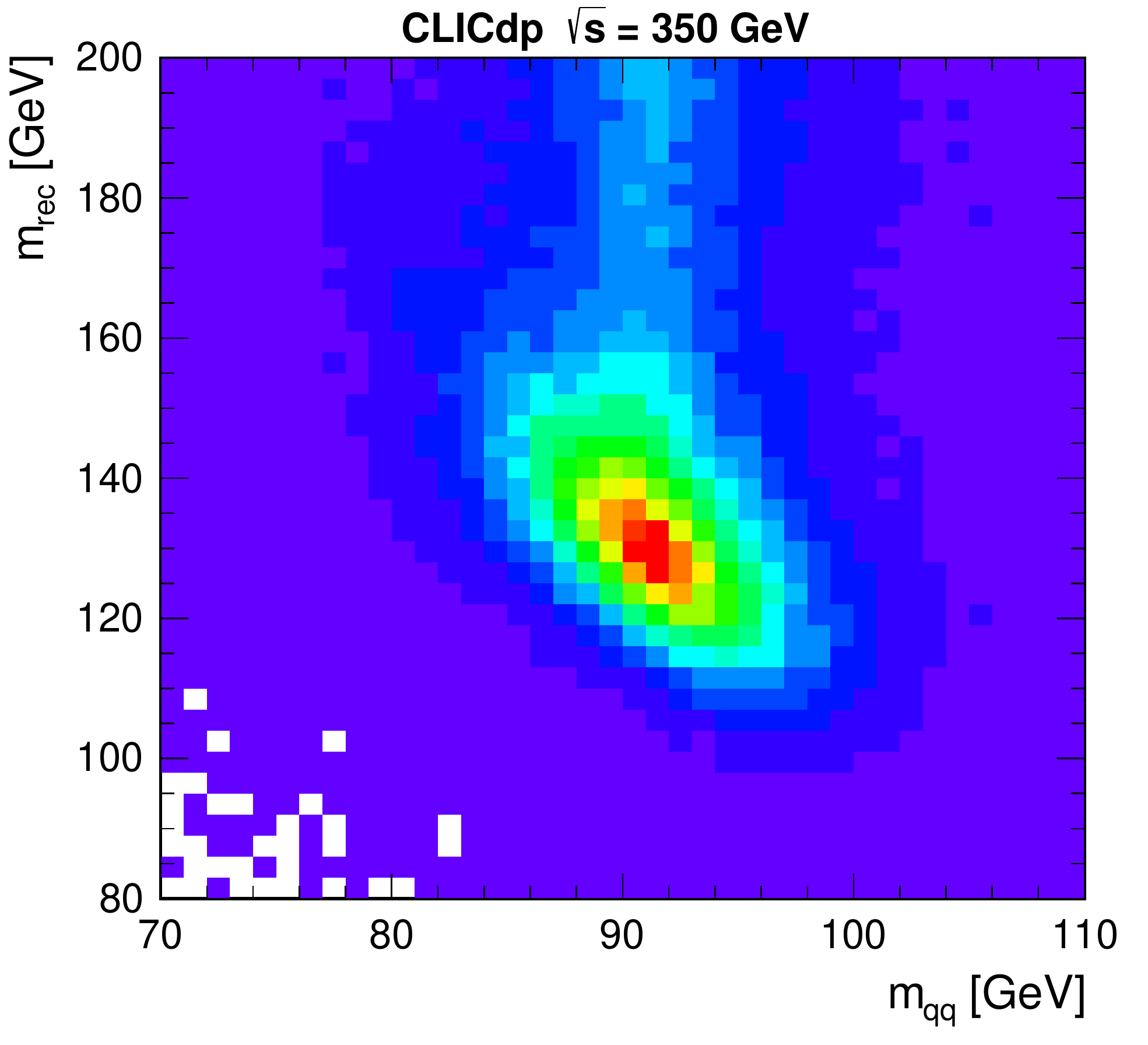}
      \hspace{0.1\columnwidth}
      \includegraphics[width=0.45\textwidth]{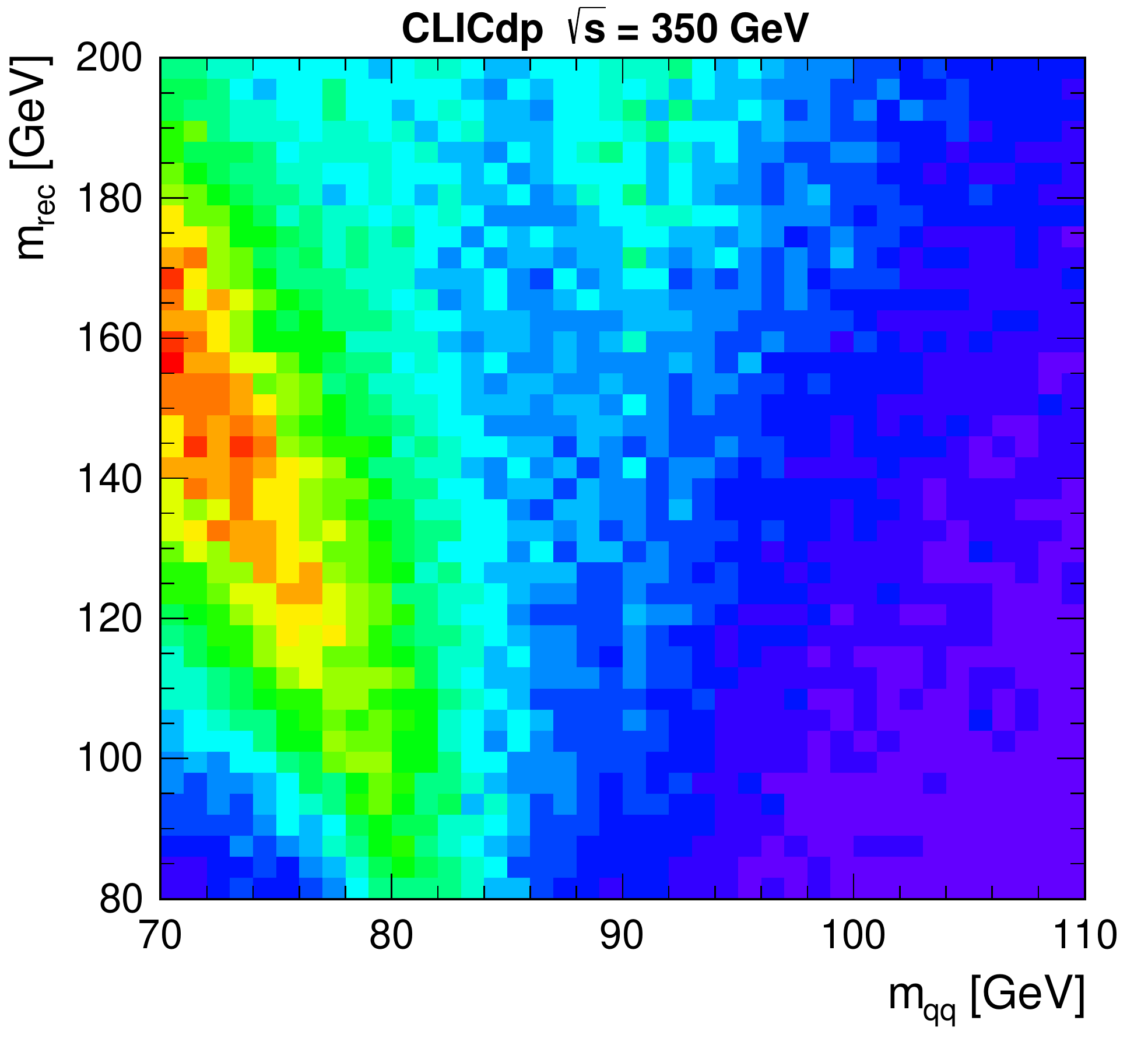}
      \caption{\label{fig:hzqq:mass} Reconstructed di-jet invariant mass versus reconstructed recoil mass distributions for $\PZ\PH\to\qqbar X$ candidate events at $\roots=350\,\GeV$, showing $\PZ\PH$ signal events (a) and all background processes (b). In both cases the plots show all events passing the preselection.}
\end{figure*}

In this high-statistics limit, the fractional error on the number of
signal events (where the Higgs decays to visible final states),
$s_\text{vis}$, given a background $b$ is:
\begin{equation*}
       \frac{\Delta s_\text{vis}}{s_\text{vis}} = \frac{\sqrt{s_\text{vis}+b}}{s_\text{vis}} \,,
\end{equation*}
and this is minimised with the selection requirement ${\cal{L}}>0.65$.
The selection efficiencies and expected numbers of events for the
signal dominated region, ${\cal{L}}>0.65$, are listed in
\autoref{tab:hzqq:events}, corresponding to a fractional error on the number of signal events of $1.9\,\%$. By fitting the shape of the
likelihood distribution to signal and background contributions, this
uncertainty is reduced to:
\begin{equation*}
            \frac{\Delta s_\text{vis} }{s_\text{vis}} = 1.7\,\% \,.
\end{equation*}

This is an example of a measurement for which it will be particularly 
important to tune the background modelling using high-statistics processes.

\begin{table}[tb]
  \centering
   \begin{tabular}{lrrrr}
    \toprule 
         Process                                & \tabt{$\sigma/\text{fb}$} & \tabt{$\varepsilon_\text{presel}$} & \tabt{$\varepsilon_{{\cal{L}}>0.65}$} & \tabt{$N_{{\cal{L}}>0.65}$} \\ \midrule
            $\PQq\PAQq$                                            & 25200   &     0.4\,\%       &  17\,\%                     & 8525  \\
      $\PQq\PAQq\Pl\PGn$                              &  5910      &   11\,\%       & 1.7\,\%                   & 5767 \\   
      $\PQq\PAQq\PQq\PAQq$                        & 5850       &     3.8\,\%       & 13\,\%                   & 14142    \\    
     $\PQq\PAQq\Pl\Pl$                                   &  1700    &      1.5\,\%       &  15\,\%                   & 1961     \\ 
     $\PQq\PAQq\PGn\PAGn$                         &  325       &      0.6\,\%       &  6.2\,\%                    & 60  \\    
     $\PH\PGne\PAGne$                                 &  52       &      2.5\,\%      &   9.2\,\%                      & 60   \\
  $\PZ\PH$; $\PZ\to\PQq\PAQq$                  &   93      &     42.0\,\%      &   54\,\%                   & 10568 \\
%            $\PQq\PAQq$                                            & 25180   &     0.4\,\%       &  0.07\,\%                     & 8525  \\
%      $\PQq\PAQq\Pl\PGn$                              &  5914      &   11.2\,\%       & 0.20\,\%                   & 5767 \\   
%      $\PQq\PAQq\PQq\PAQq$                        & 5847       &     3.8\,\%       & 0.49\,\%                   & 14142    \\    
%     $\PQq\PAQq\Pl\Pl$                                   &  1704    &      1.5\,\%       &  0.22\,\%                   & 1961     \\ 
%     $\PQq\PAQq\PGn\PAGn$                         &  325       &      0.6\,\%       &  0.04\,\%                    & 60  \\    
%     $\PH\PGne\PAGne$                                 &  52       &      2.5\,\%      &   0.23\,\%                      & 60   \\
%  $\PZ\PH$; $\PZ\to\PQq\PAQq$                  &   93      &     42.0\,\%      &   22.6\,\%                   & 10568 \\
    \bottomrule
  \end{tabular}
   \caption{Summary of the $(\PZ\to\qq)(\PH\to\text{vis.})$ event selection at $\roots=350\,\GeV$, giving the raw cross sections, preselection efficiency, overall selection efficiency for a likelihood cut of  ${\cal{L}}>0.65$ and the expected numbers of events passing the 
  event selection for an integrated luminosity of $500\,\fbinv$. 
 \label{tab:hzqq:events}}
\end{table}

}

% Section 4.1.3
\subsubsection{Invisible Higgs Decays} %(Kelvin)
\label{sec:higgsstrahlung_invisible}

The above recoil mass analysis of leptonic decays of the $\PZ$ boson in $\epem\to\PZ\PH$ events provides a measurement of the 
\higgsstrahlung cross section, independent of the Higgs boson decay model. The recoil mass technique can also be used to search for
BSM decay modes of the Higgs boson into long-lived neutral ``invisible'' final states~\cite{Thomson:2015jda}.  At an $\epem$ collider, a search for invisible 
Higgs decays is possible by identification of $\epem\to\PZ\PH$ events with a visible $\PZ\to\qq$ decay and missing energy. Such events would
typically produce a clear two-jet topology with invariant mass consistent with $\mZ$, significant missing energy and a recoil mass 
corresponding to the Higgs mass. \higgsstrahlung events with leptonic $\PZ$ decays, which have a much smaller branching ratio, are not included in the current analysis.

\begin{figure}
\centering
\includegraphics[width=0.9\columnwidth]{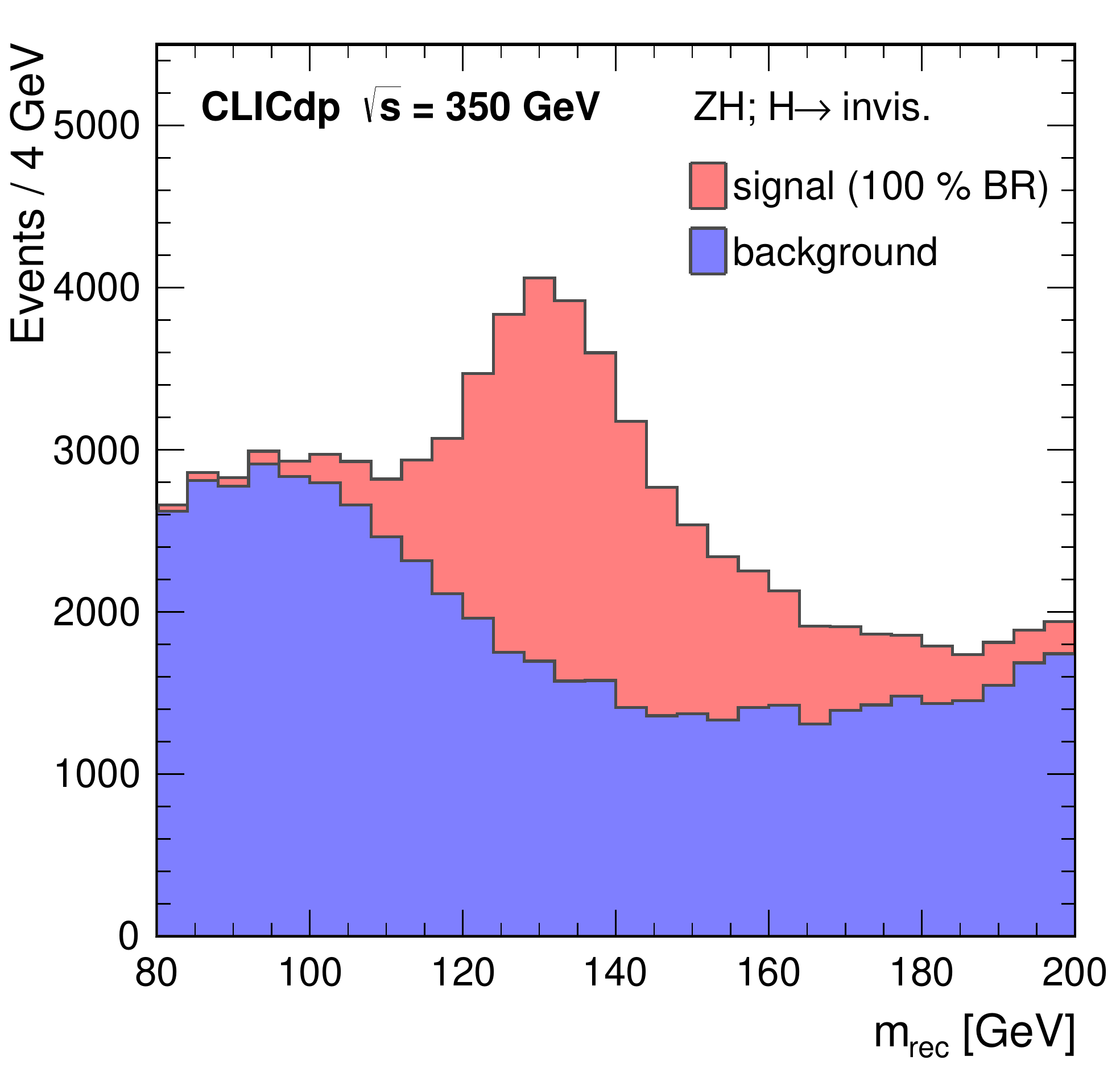}
\caption{Reconstructed recoil mass distributions of $\epem\to\PZ\PH$ events at $\roots=350\,\GeV$, showing the $\PH\to\text{invis.}$ signal, assuming $\BR(\PH\to\text{invis.}) = 100\,\%$, and SM backgrounds as stacked histograms. The distributions are normalised to an integrated luminosity of $500\,\fbinv$.}
\label{invHiggs:fig:recoilMass}
\end{figure}

To identify candidate invisible Higgs decays, a loose preselection is imposed requiring: 
i) a clear two-jet topology, defined by $\log_{10}(y_{23}) < -2.0$ and $\log_{10}(y_{34}) < -3.0$, using the minimal $k_t$ distances discussed 
in \autoref{sec:simreco}; ii) a di-jet invariant mass consistent with $\mZ$,
$84\,\GeV < \mqq < 104\,\GeV$; and iii) the reconstructed momentum of the candidate $\PZ$ boson 
pointing away from the beam direction, $|\cos\theta_{\PZ}| < 0.7$. After the  preselection, a BDT multivariate analysis technique is applied using the \tmva package \cite{TMVA:2010} to further separate the invisible Higgs signal from the SM background. 
In addition to $\mqq$, $|\cos\theta_{\PZ}|$ and $\log_{10}(y_{23})$, four other discriminating variables are employed:
$\mrec$, the recoil mass of the invisible system recoiling against the observed $\PZ$ boson;
$\cosQ$, the decay angle of one of the quarks in the $\PZ$ rest frame, relative to the direction of flight of the $\PZ$ boson;
$\pT$, the magnitude of the transverse momentum of the \PZ boson; and 
$\Evis$, the visible energy in the event. As an example, \autoref{invHiggs:fig:recoilMass} shows the recoil mass distribution for the simulated
invisible Higgs decays and the total SM background. The reconstructed recoil mass for events with invisible Higgs decays peaks near $m_{\PH}$. The cut applied on the BDT output is chosen to minimise the statistical uncertainty on the cross section for invisible Higgs decays. % was found at a BDT value of 0.088.

In the case where the branching ratio to BSM invisible final states 
is zero (or very small), the uncertainty on the invisible branching ratio is determined by the statistical fluctuations on the background after the event selection:
\begin{equation*}
\label{invHiggs:eq:dBR}
\Delta \BR(\PH\to\text{invis.}) = \frac{\sqrt{b} }{s(100\,\%)} \,,
\end{equation*}
where $b$ is the expected number of selected SM background events and $s(100\,\%)$ is the expected number of selected \higgsstrahlung events assuming all Higgs bosons decay invisibly, i.e. $\BR(\PH\to\text{invis.}) = 100\,\%$.
\autoref{invHiggs:table:efftables} summarises the invisible Higgs decay event selection; 
the dominant background processes arise from the final states $\qqbar\Pl\PGn$ and $\qqbar\nunubar$. The
resulting one sigma uncertainty on $\BR(\PH\to\text{invis.})$  is 0.57\,\% (in the case where the invisible Higgs branching ratio is small)
and the corresponding 90\,\% C.L. upper limit  (500\,\fbinv at \roots=350\,GeV) on the invisible Higgs branching ratio in the modified frequentist approach~\cite{Read00} is: 
  \begin{equation*}
       \BR(\PH\to\text{invis.}) < 0.97\,\% \  \ \ \text {at} \ 90\,\% \ \text{C.L.} 
  \end{equation*}
It should be noted that the SM Higgs decay chain $\PH\to\PZ\PZ^*\to\nunubar\nunubar$ has a combined branching ratio of 0.1\,\% and is
not measurable.

\begin{table}[tb]
  \centering
  \begin{tabular}{lrrrr}
    \toprule 
         Process                                & \tabt{$\sigma/\text{fb}$} 
                                                      & \tabt{$\varepsilon_\text{presel}$} 
                                                      & \tabt{$\varepsilon_{\text{BDT}>0.088}$} 
                                                      & \tabt{$N_{\text{BDT}>0.088}$} \\ \midrule
      $\qqbar\Pl\PGn$                              &  5910      &    0.68\,\%   & 4.5\,\%   & 900 \\   
     $\PQq\PAQq\PGn\PAGn$                         &  325       &     17\,\%  &  8.9\,\%   &    2414 \\    
  \midrule
  $\PZ\PH$ (SM decays)                                                    &   93.4      & 0.2\,\%         &   23\,\%  & 21  \\
  \midrule
    $\PH\to\text{invis.}$ 
                                                                     &    & 41\,\%     & 51\,\%       & 9956 \\   
%      $\qqbar\Pl\PGn$                              &  5914      &    <0.7\,\%   & <0.1\,\%   & 900 \\   
%     $\PQq\PAQq\PGn\PAGn$                         &  325       &     16.7\,\%  &  1.5\,\%   &    2414 \\    
%  \midrule
%  $\PZ\PH$ (SM decays)                                                    &   93.4      & 0.2\,\%         &   <0.1\,\%  & 21  \\
%  \midrule
%    $\PH\to\text{invis.}$ 
%                                                                     &    & 41.0\,\%     & 20.7\,\%       & 9956 \\                                                 %                                                                                                                     
    \bottomrule
    \end{tabular}
      \caption{Summary of the invisible Higgs decay event selection at $\roots=350\,\GeV$, giving the raw cross sections, preselection efficiency, selection efficiency for a BDT cut of  $\text{BDT}>0.088$, and the expected numbers of events passing the 
  event selection for an integrated luminosity of $500\,\text{fb}^{-1}$. For the invisible Higgs decay signal the number of selected events corresponds to
  a \BR of 100\,\%. Contributions from all other backgrounds are found to be negligibly small.
 \label{invHiggs:table:efftables}}
\end{table}

% Section 4.1.4
\subsubsection{Model-Independent $\PZ\PH$ Cross Section} %(Mark)
\label{sec:higgsstrahlung_MI}
By combining the two analyses for $\PZ\PH$ production where
$\PZ\to\qqbar$ and the Higgs decays either to invisible final states
(see \autoref{sec:higgsstrahlung_invisible}) or to visible final
states (see \autoref{sec:higgsstrahlung_z_qq}), it is possible to
determine the absolute cross section for $\epem$ $\to\PZ\PH$ in an
essentially model-independent manner:
\begin{equation*}
      \sigma(\PZ\PH) = \frac{\sigma_\text{vis} + \sigma_\text{invis}}{\BR(\PZ\to\qqbar)} \,.
\end{equation*}
Here a slightly modified version of the invisible Higgs analysis is
employed. With the exception of the cuts on $y_{23}$ and $y_{34}$, the
invisible Higgs analysis employs the same preselection as for the
visible Higgs analysis and a likelihood multivariate discriminant is
used.

Since the fractional uncertainties on the total cross
section from the visible and invisible cross sections are $1.7\,\%$ and
$0.6\,\%$ respectively, the fractional uncertainty on the total cross
section will be (at most) the quadrature sum of the two fractional
uncertainties, namely $1.8\,\%$. This measurement is only truly
model-independent if the overall selection efficiencies are
independent of the Higgs decay mode.  For all final state topologies,
the combined (visible + invisible) selection efficiency lies is the
range $19-26\,\%$ regardless of the Higgs decay mode, covering a very
wide range of event topologies. To assess the level of model
independence, the Higgs decay modes in the MC samples are modified and
the total (visible + invisible) cross section is extracted assuming
the SM Higgs branching ratio. \autoref{tab:hzqq:MI:biases} shows the
resulting biases in the extracted total cross section for the case
when a $\BR(\PH\to X) \to \BR(\PH\to X) + 0.05$. Even for these very
large modifications of the Higgs branching ratios over a wide range of
final-state topologies -- including the extreme cases highlighted 
at the bottom of \autoref{tab:hzqq:MI:biases} such as 
$\PH\to\PW\PW^*\to\qqbar\qqbar$, which has six jets in the final state,
and $\PH\to\PW\PW^*\to\PGt\PGn\PGt\PGn$, which has a lot of missing energy
-- the resulting biases in the extracted total
$\PZ\PH$ cross section are less than $1\,\%$ (compared to the $1.8\,\%$
statistical uncertainty). However, such large deviations would have
significant observable effects on exclusive Higgs branching ratio
analyses (at both LHC and CLIC) and it is concluded that the analysis
gives an effectively model-independent measurement
of the $(\PZ\to\qqbar)\PH$ cross section. 
%(unless there are very large
%BSM effects on the Higgs branching ratios which would already be apparent).
\begin{table}[tb]
  \centering
 \begin{tabular}{lrc}
    \toprule 
         \tabt{Decay mode}    &         $\Delta$(\BR) &   $\sigma^{\text{vis}} + \sigma^{\text{invis}}$ Bias \\  
  \midrule       
    $\PH\to\text{invis}$       & $+5\,\%$  &    $-0.01\,\%$     \\
    $\PH\to\qqbar$              &  $+5\,\%$ &    $+0.05\,\%$    \\
    $\PH\to\PW\PW^*$        &  $+5\,\%$ &   $-0.18\,\%$      \\
    $\PH\to\PZ\PZ^*$         &  $+5\,\%$ &     $-0.30\,\%$     \\
    $\PH\to\tptm$                & $+5\,\%$   &   $+0.60\,\%$    \\                                                                         
     $\PH\to\PGg\PGg$      & $+5\,\%$   &    $+0.79\,\%$    \\     
    $\PH\to\PZ\PGg$          & $+5\,\%$  &     $-0.74\,\%$    \\                
    \midrule
    $\PH\to\PW\PW^*\to\qqbar\qqbar$  & $+5\,\%$ & $-0.49\,\%$    \\          
        $\PH\to\PW\PW^*\to\qqbar\Pl\PGn$  & $+5\,\%$ & $+0.10\,\%$    \\    
            $\PH\to\PW\PW^*\to\PGt\PGn\PGt\PGn$  & $+5\,\%$ & $-0.98\,\%$    \\                                                                                                                                                                                                                                                                                                                                                                                    
    \bottomrule
  \end{tabular}
   \caption{Biases in the extracted $\PH(\PZ\to\qqbar)$ cross section if the Higgs branching ratio to a specific final state is increased by 5\,\%, i.e. $\BR(\PH\to X) \to 
  \BR(\PH\to X) + 0.05$.
 \label{tab:hzqq:MI:biases}}
\end{table}

Combining the model-independent measurements of the $\PZ\PH$ cross
section from $\PZ\to\Plp\Plm$ and $\PZ\to\qqbar$ gives an absolute
measurement of the $\PZ\PH$ cross section with a precision of:
\begin{equation*}
          \frac{\Delta \sigma({\PZ\PH})}{\sigma(\PZ\PH)} = 1.65\,\% \,,
\end{equation*}
and, consequently, the absolute coupling of the $\PH$ boson to the $\PZ$ boson is determined to:
\begin{equation*}
          \frac{\Delta g_{\PH\PZ\PZ}}{g_{\PH\PZ\PZ}} = 0.8\,\% \,.
\end{equation*}

The hadronic recoil mass analysis was repeated for collision energies of $\roots=250\,\GeV$ and $\roots=420\,\GeV$~\cite{Thomson:2015jda}. 
Compared with $\roots=350\,\GeV$, the sensitivity is significantly worse in both cases.

% Section 4.2
\subsection{Exclusive Higgs Branching Ratio Measurements at $\roots=350\,\GeV$}
\label{sec:higgsstrahlung_branching_ratios}

The previous section described inclusive measurements of the
$\epem\to\PZ\PH$ production cross section, which provide a
model-independent determination of the coupling at the $\PH\PZ\PZ$
vertex. In contrast, measurements of Higgs production and decay to
exclusive final states provide a determination of the product
$\sigma(\PZ\PH) \times \BR(\PH\to X)$, where $X$ is a particular final
state. This section focuses on the exclusive measurements of the Higgs
decay branching ratios at $\roots=350\,\GeV$. Higgs boson decays to
$\bb$, $\cc$ and $\Pg\Pg$ are studied in
\autoref{sec:higgsstrahlung_bb_cc_gg}. The measurement of
$\PH\to\tptm$ decays is described in
\autoref{sec:higgsstrahlung_tautau}, and the $\PH\to\PW\PW^\ast$
decay mode is described in \autoref{sec:higgsstrahlung_ww}.

% Section 4.2.1
\subsubsection{$\PH \to \PQb \PAQb,~\PQc \PAQc$ and~$\Pg\Pg$}
\label{sec:higgsstrahlung_bb_cc_gg}
As can be seen from \autoref{tab:higgs:events}, at $\roots=350\,\GeV$
the cross section for $\epem\to\PZ\PH$ (\higgsstrahlung) is approximately
four times greater than the $\epem\to\PH\PGne\PAGne$ (mostly $\PW\PW$-fusion)
cross section for unpolarised beams (or approximately a factor 2.5
with $-80\,\%$ electron beam polarisation). For \higgsstrahlung, the
signature of $\PH\to \bb, \cc, \Pg\Pg$ events depends on the $\PZ$
decay mode.
%, resulting in three distinct final state topologies:
%$\text{jjjj}$, $\text{jj}\Pl\Pl$, and $\text{jj}\PGn\PGn$, where $\text{j}$ represents a
%quark or gluon jet from the $\PZ$ or $\PH$ decay. It should be noted that
%the $\text{jj}\PGn\PGn$ final state contains approximately equal
%contributions from \higgsstrahlung and $\PW\PW$-fusion events, although the
%event kinematics are very different.

\begin{table*}
  \centering
  \begin{tabular}{lc|cc|cc}
 	\toprule
 	\multirow{2}{*}{Process} & \multirow{2}{*}{$\sigma$/fb} &  \multicolumn{2}{c|}{$\epsilon_\text{BDT}$, classified as}   &  \multicolumn{2}{c}{$N_\text{BDT}$, classified as} \\
	 & &$\PH\nunubar$  & $\PH\qqbar$ &$\PH\nunubar$  & $\PH\qqbar$ \\
 	\midrule
	$\epem\to\PH\nunubar; \PH\to\PQb\PAQb$ & 28.9 & 55\,\% & 0\,\% & 8000 & 0 \\
	$\epem\to\PH\nunubar; \PH\to\PQc\PAQc$ & 1.46 & 51\,\% & 0\,\% & 372 & 0 \\
	$\epem\to\PH\nunubar; \PH\to\Pg\Pg$ & 4.37 & 58\,\% & 0\,\% & 1270 & 0 \\
	$\epem\to\PH\nunubar; \PH\to\text{other}$ & 16.8 & 6.1\,\% & 0\,\% & 513 & 0 \\
	$\epem\to\PH\qqbar; \PH\to\PQb\PAQb$ & 52.3 & 0\,\% & 42\,\% & 0 & 11100 \\
	$\epem\to\PH\qqbar; \PH\to\PQc\PAQc$ & 2.64 & 0\,\% & 33\,\% & 0 & 434 \\
	$\epem\to\PH\qqbar; \PH\to\Pg\Pg$ & 7.92 & 0\,\% & 37\,\% & 0 & 1480 \\
	$\epem\to\PH\qqbar; \PH\to\text{other}$ & 30.5 & 0.12\,\% & 13\,\% & 20 & 1920 \\
	\midrule
	$\epem\to\qqbar\nunubar$ & 325 & 1.3\,\% & 0\,\% & 2110 & 0 \\ 
	$\epem\to\qqbar\Pl\PGn$ & 5910 & 0.07\,\% & 0.002\,\% & 2090 & 60 \\ 
	$\epem\to\qqbar\Pl\Pl$ & 1700 & 0.012\,\% & 0.01\,\% & 104 & 89 \\ 
	$\epem\to\qqbar\qqbar$ & 5530 & 0.001\,\% & 0.36\,\% & 30 & 9990 \\ 
	$\epem\to\qqbar$ & 24400 & 0.01\,\% & 0.093\,\% & 1230 & 11400 \\ 	
%	$\epem\to\PH\nunubar; \PH\to\PQb\PAQb$ & 28.9 & 55.0\,\% & 0\,\% & 8000 & 0 \\
%	$\epem\to\PH\nunubar; \PH\to\PQc\PAQc$ & 1.46 & 51.0\,\% & 0\,\% & 372 & 0 \\
%	$\epem\to\PH\nunubar; \PH\to\Pg\Pg$ & 4.37 & 58.0\,\% & 0\,\% & 1270 & 0 \\
%	$\epem\to\PH\nunubar; \PH\to\text{other}$ & 17.8 & 6.1\,\% & 0\,\% & 513 & 0 \\
%	$\epem\to\PH\qqbar; \PH\to\PQb\PAQb$ & 52.3 & 0\,\% & 42.3\,\% & 0 & 11100 \\
%	$\epem\to\PH\qqbar; \PH\to\PQc\PAQc$ & 2.64 & 0\,\% & 32.8\,\% & 0 & 434 \\
%	$\epem\to\PH\qqbar; \PH\to\Pg\Pg$ & 7.92 & 0\,\% & 37.4\,\% & 0 & 1480 \\
%	$\epem\to\PH\qqbar; \PH\to\text{other}$ & 30.5 & 12.6\,\% & 0.12\,\% & 20 & 1920 \\
%	\midrule
%	$\epem\to\qqbar\nunubar$ & 325 & 1.3\,\% & 0\,\% & 2110 & 0 \\ 
%	$\epem\to\qqbar\Pl\PGn$ & 5910 & 0.07\,\% & 0.002\,\% & 2090 & 60 \\ 
%	$\epem\to\qqbar\Pl\Pl$ & 1700 & 0.012\,\% & 0.01\,\% & 104 & 89 \\ 
%	$\epem\to\qqbar\qqbar$ & 5530 & 0.01\,\% & 0.36\,\% & 30 & 9990 \\ 
%	$\epem\to\qqbar$ & 24400 & 0.01\,\% & 0.093\,\% & 1230 & 11400 \\ 	
	\bottomrule
  \end{tabular}
       \caption{\label{tab:higgsstrahlung:HbbccggSelection} Summary of
         the expected numbers of events for the different Higgs and
         non-Higgs final states passing the hadronic Higgs decay
         signal selection for $500\,\fbinv$ at $\roots=350\,\GeV$
         (unpolarised beams). No preselection is applied in this analysis.}
\end{table*}

To maximise the statistical power of the $\PH\to \bb, \cc, \Pg\Pg$
branching ratio measurements, two topologies are considered: four
jets, and two jets plus missing momentum (from the unobserved neutrinos). 
The impact of \higgsstrahlung events with leptonic 
$\PZ$ decays is found to be negligible. 
The jets plus missing momentum final state contains approximately equal
contributions from \higgsstrahlung and $\PW\PW$-fusion events, although the
event kinematics are very different.
All events are initially
reconstructed assuming both topologies; at a later stage of the event selection, events are assigned to either $\PH\qqbar$, $\PH\nunubar$, or background.
To minimize the impact of ISR on the jet reconstruction, 
photons with a reconstructed energy higher than $15\,\GeV$ are removed from the events first.

The hadronic final states are reconstructed using the Durham algorithm. 
For the four-jet topology, the most probable \PZ and Higgs boson candidates are 
selected by choosing the jet combination that minimises:
\begin{equation*}
   \chi^2 = (m_{ij}-\mH)^2/\sigma^2_{\PH} + (m_{kl}-\mZ)^2/\sigma^2_{\PZ} \,,
\end{equation*}  
where $m_{ij}$ and $m_{kl}$ are the invariant masses of the jet pairs used to 
reconstruct the Higgs and $\PZ$ boson candidates, respectively, and $\sigma_{\PH, \PZ}$ are the estimated invariant mass resolutions for
Higgs and \PZ boson candidates. In the case of the two jets plus missing energy final state, 
either from $\PZ\PH$ with $\PZ\to\nunubar$ or from $\PH\nunubar$, the
event is clustered into two jets forming the \PH candidate.

\begin{figure*}
  \centering
  \includegraphics[width=0.33\textwidth]{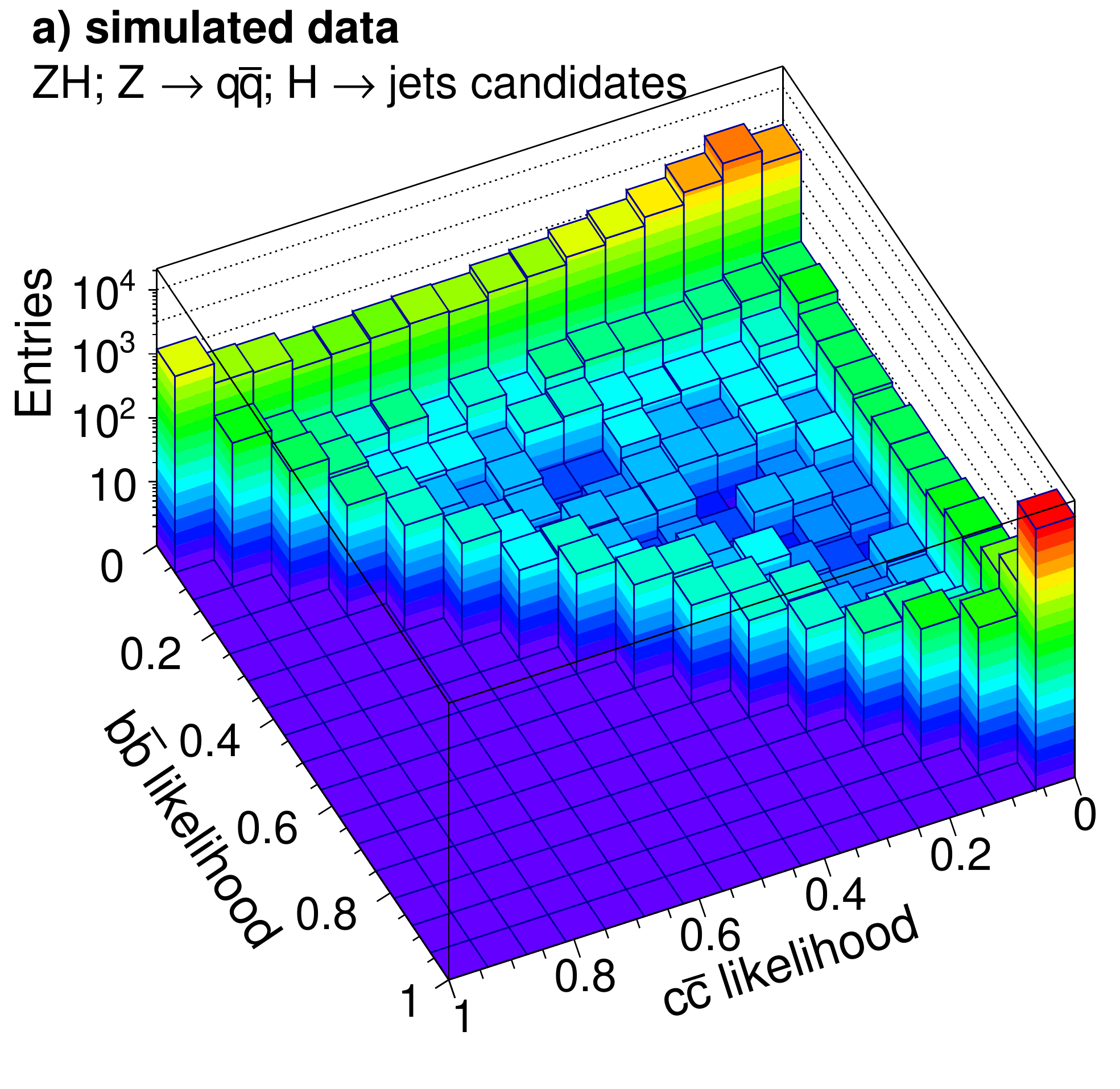}
  \includegraphics[width=0.33\textwidth]{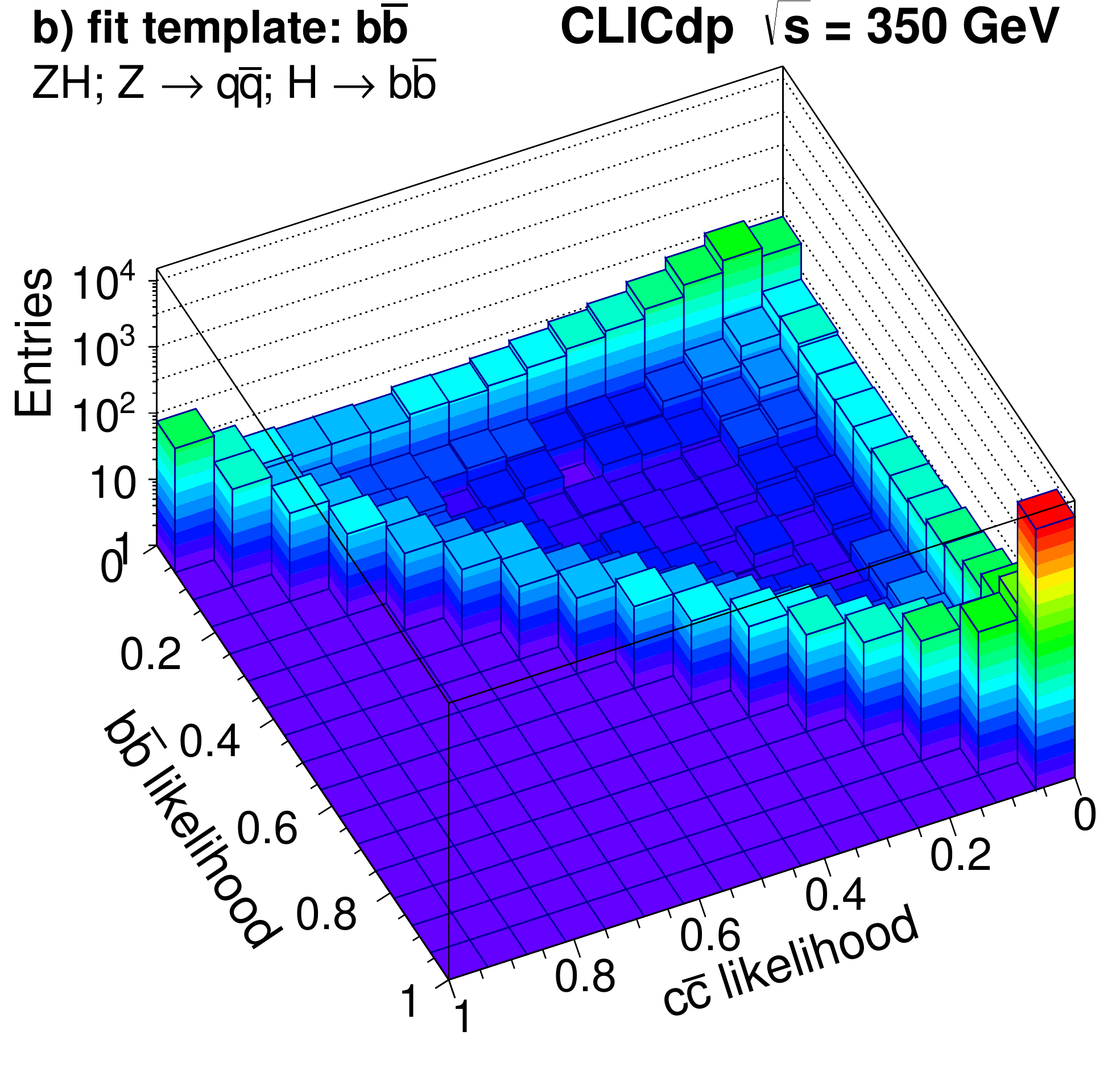}
  \includegraphics[width=0.33\textwidth]{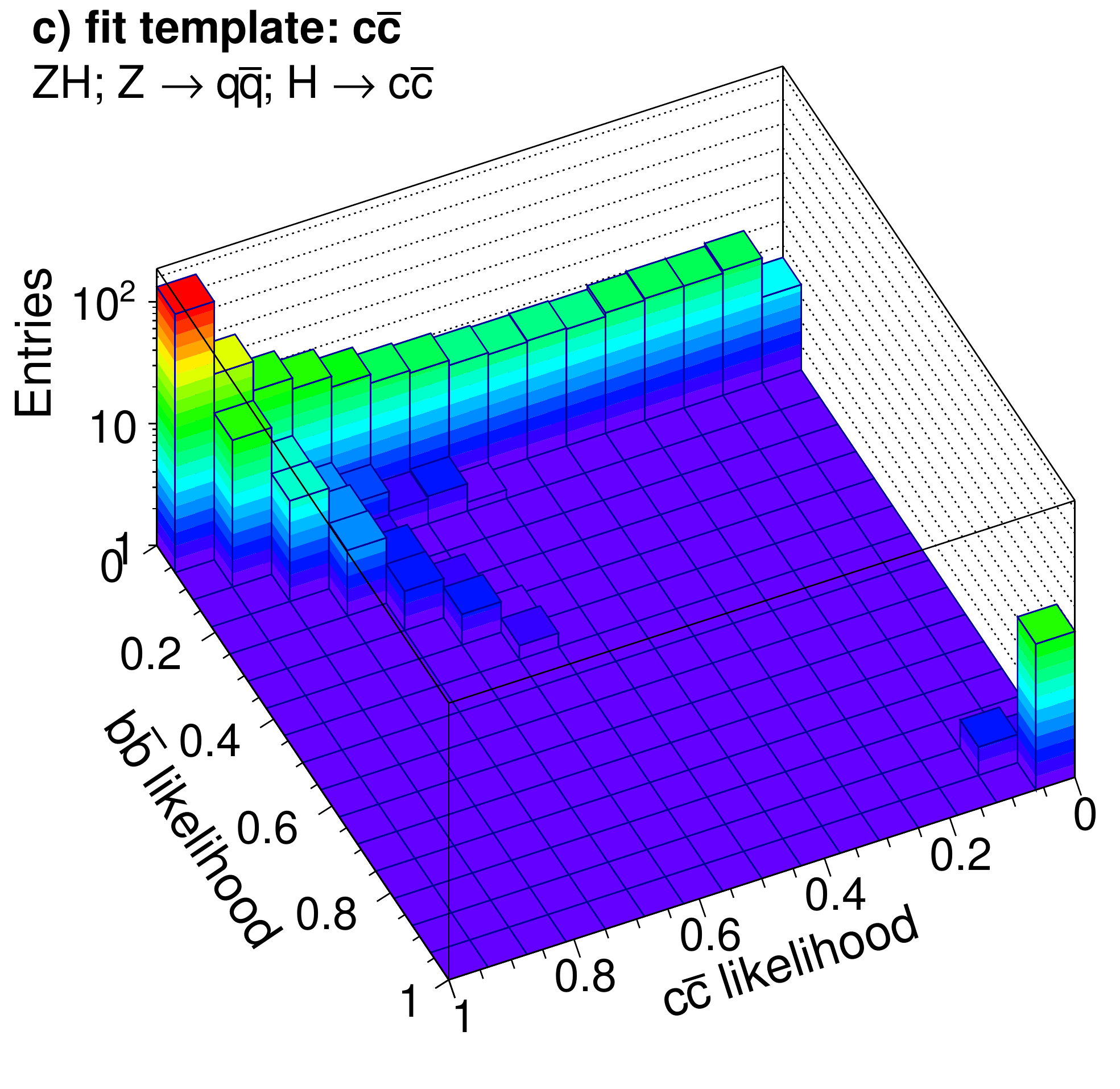}
  \includegraphics[width=0.33\textwidth]{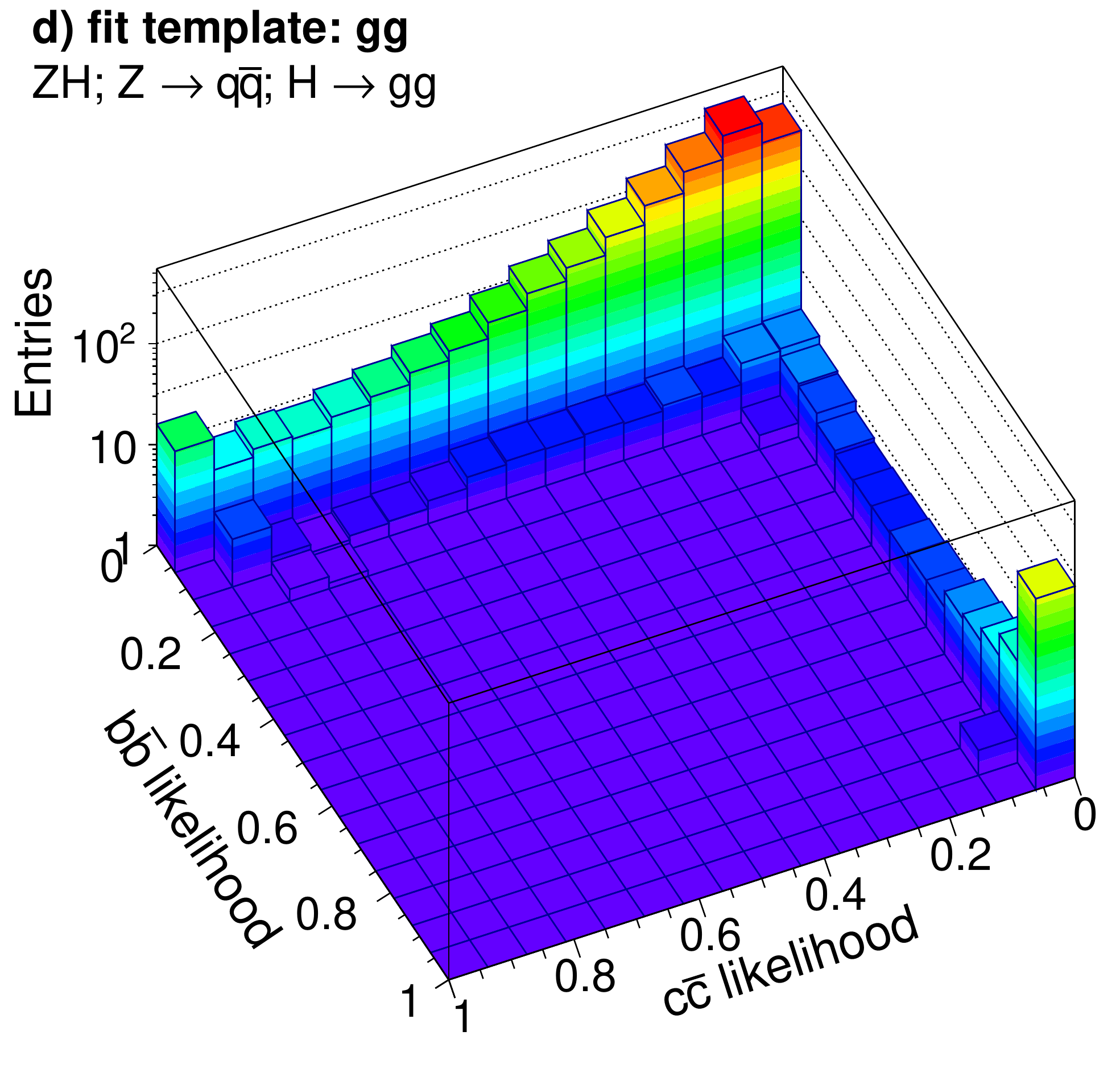}
  \includegraphics[width=0.33\textwidth]{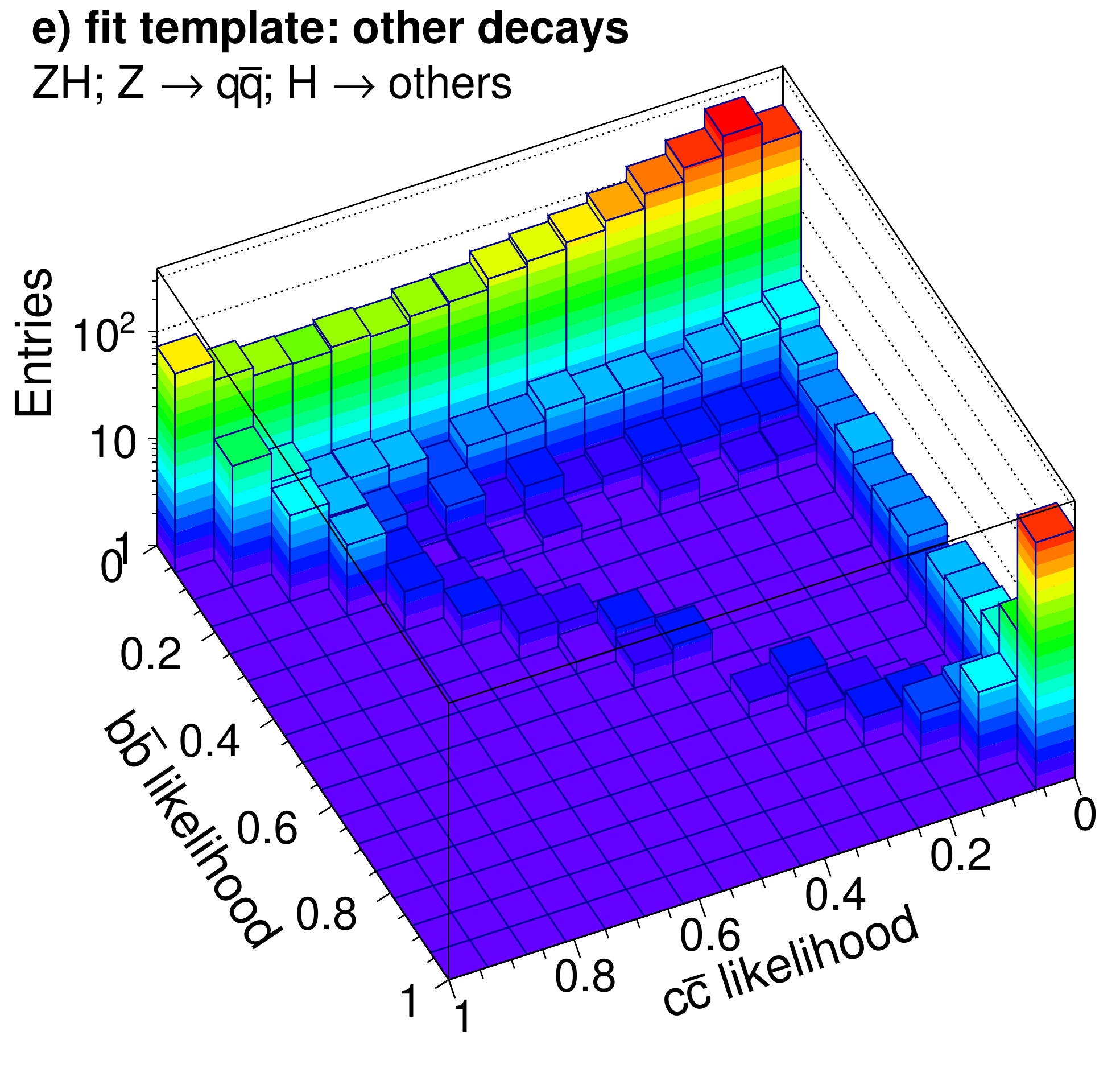}
  \includegraphics[width=0.33\textwidth]{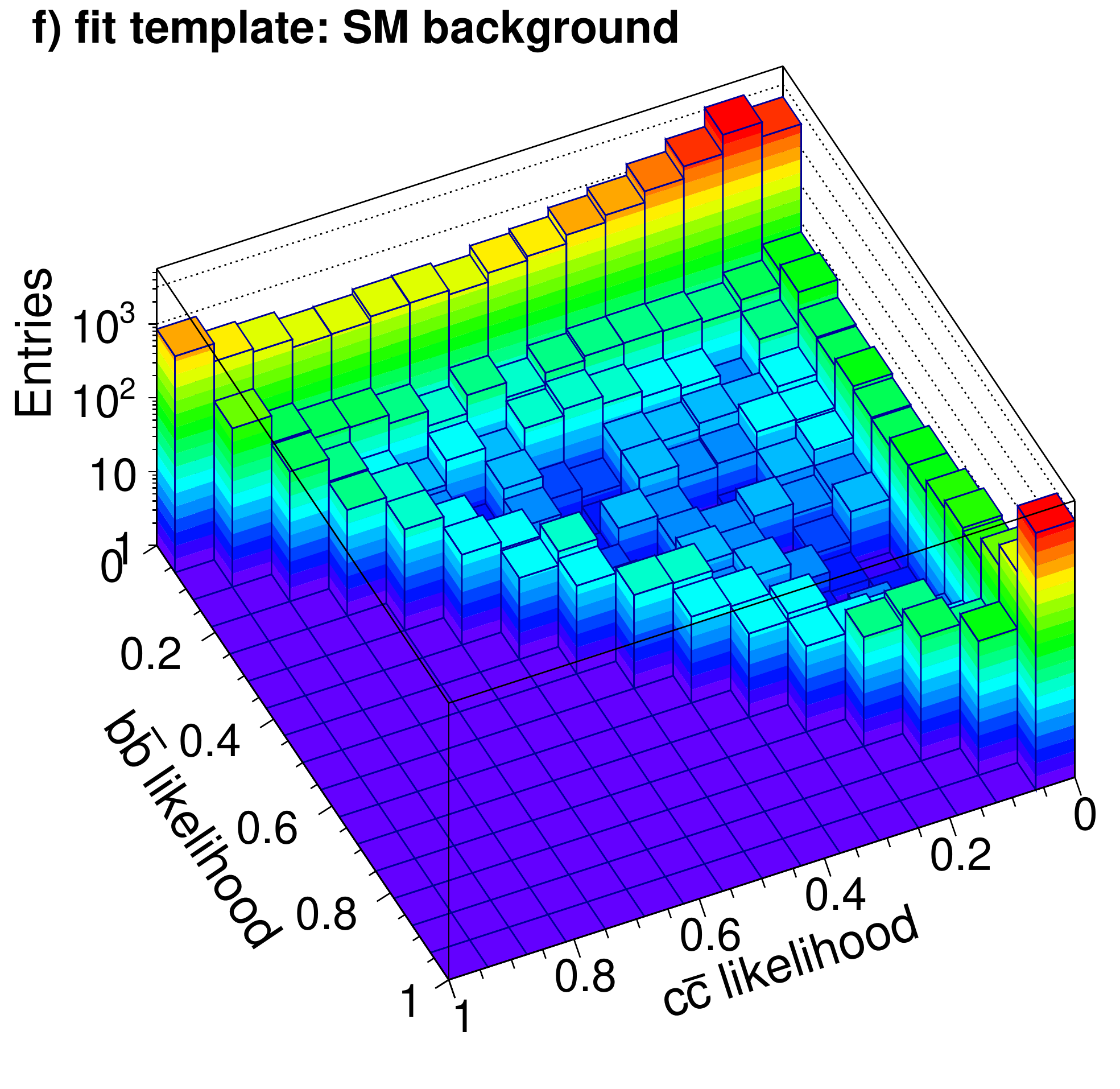}
  \caption{\label{fig:hbbccgg:tag} $\bb$ likelihood versus $\cc$ likelihood distributions for $\epem\to\PZ\PH$ events at $\roots=350\,\GeV$, for (a) all events and for the different event classes: (b) $\PH\to\bb$, (c) $\PH\to\cc$, (d) $\PH\to\Pg\Pg$, background from (e) other Higgs decays and (f) non-Higgs SM background. All distributions are normalised to an integrated luminosity of $500\,\fbinv$.}
\end{figure*}

To help veto backgrounds with leptonic final states, isolated 
electrons or muons with $E > 10\,\GeV$ are identified with the additional requirement that there should 
be less than $20\,\GeV$ of energy from other particles within a cone 
with an opening angle of $20^\circ$ around the lepton direction. All events are 
then classified by gradient boost decision trees employing reconstructed kinematic 
variables from each of the two event topology hypotheses described above. 
The variables used include jet energies, event shape variables (such as 
thrust and sphericity), the masses of \PH and \PZ candidates,
their decay angles and transverse momenta, and the number of isolated leptons in the final state. 
The total number of variables is about 50, which is larger than in other studies 
presented in this paper, because each event is reconstructed assuming two 
different final state configurations and information from 
the $\PH$ candidate decay can be included here, in contrast with the 
recoil mass analyses described in \autoref{sec:higgsstrahlung_cross_section}.

Two separate BDT classifiers are used, one for each signal final state 
($\PH\qqbar$ and $\PH\nunubar$), irrespective of the nature of the hadronic Higgs decay mode. 
Two-fermion ($\qqbar$) and four-fermion ($\qqbar\nunubar$, $\qqbar\Pl\PGn$, $\qqbar\Pl\Pl$ 
and $\qqbar\qqbar$) final states and other Higgs decay modes are taken as background for 
both classifiers. 
In addition, the other signal mode is included in the background for a given classifier. 
The training is performed using a dedicated training sample, simultaneously 
training both classifiers. At this point, no flavour tagging information is used.
 
Each event is evaluated with both classifiers. An event is only accepted if 
exactly one of the signal classifiers is above a positive threshold and 
the other classifier is below a corresponding negative threshold. 
The event is then tagged as a candidate for the corresponding signal process. 
If none of the classifiers passes the selection threshold, the event is 
considered as background and is rejected from the analysis. 
The number of events for which both signal classifiers are above the positive threshold is negligible. 
%Such events with unclear final state classification are rejected from the analysis.
\autoref{tab:higgsstrahlung:HbbccggSelection} summarises the classification 
of all events into the two signal categories, with 
event numbers based on an integrated luminosity of $500\,\fbinv$.

The second stage of the analysis is to measure the contributions of 
the hadronic Higgs decays into the $\PH\to\bb$, 
$\PH\to\cc$ and $\PH\to\Pg\Pg$ exclusive final states, separated into the two production modes  \higgsstrahlung and $\PW\PW$-fusion.  This is achieved by a multi-dimensional template fit using flavour tagging information and, in the case of the $\PH\nunubar$ final state, the transverse momentum of the Higgs candidate.

The jets forming the Higgs candidate 
are classified with the \lcfiplus flavour tagging 
package. Each jet pair is assigned a $\bb$ likelihood and 
a $\cc$ likelihood:
\begin{equation*}
\bb~\text{likelihood} = \frac{b_1 b_2}{b_1 b_2 + (1 - b_1) (1 -
  b_2)},
\label{eq:bb_likelihood}
\end{equation*}
\begin{equation*}
\cc~\text{likelihood} = \frac{c_1 c_2}{c_1 c_2 + (1 - c_1) (1 - c_2)},
\label{eq:cc_likelihood}
\end{equation*}
where $b_1$ and $b_2$ ($c_1$ and $c_2$) are the b-tag (c-tag) values 
obtained for the two jets forming the Higgs candidate. 

The resulting two-dimensional distributions of the 
$\bb$ and $\cc$ likelihoods in $\PH\qqbar$ events are shown in \autoref{fig:hbbccgg:tag}, 
where separation between the different event categories can be 
seen. These distributions form the templates used to determine the contribution of the different signal categories for the $\PH\qqbar$ final states.

\begin{figure}
  \centering
  \includegraphics[width=0.9\columnwidth]{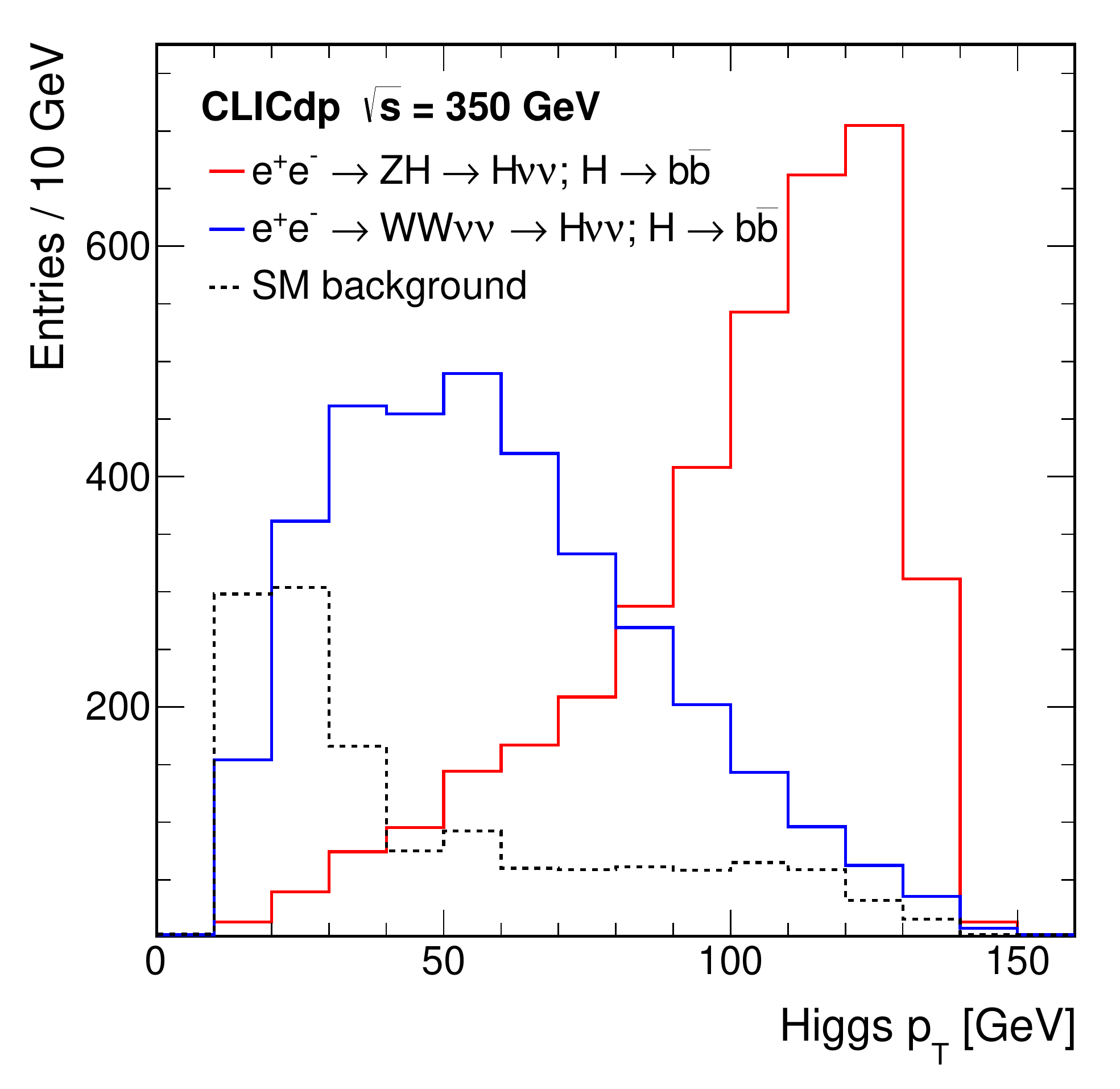}
  \caption{\label{fig:hbbccgg:pT} Reconstructed Higgs candidate transverse momentum distributions for selected $\PH\nunubar$ events at $\roots=350\,\GeV$, showing the contributions from Higgsstrahlung, $\PW\PW$-fusion and non-Higgs background. The distributions are normalised to an integrated luminosity of $500\,\fbinv$.}
\end{figure}

Signal and background templates are also obtained for the $\PH\nunubar$ final state. 
As $\PH\nunubar$ has roughly equal contributions from the \higgsstrahlung and the 
$\PW\PW$-fusion process, separation into the two production processes is required,
in addition to separation into the different signal and background final states.
This is achieved by adding the transverse momentum of the Higgs candidate 
to the templates as a third dimension.
This exploits the fact that the transverse momentum of the Higgs candidate 
is substantially different for \higgsstrahlung and $\PW\PW$-fusion events, 
as illustrated in \autoref{fig:hbbccgg:pT} for events with a 
high $\bb$ likelihood, which provides a high signal purity. 
%Here, the \higgsstrahlung and WW-fusion contributions are shown separately.

Contributions from events with $\PH\to\bb$, $\PH\to\cc$ and $\PH\to\Pg\Pg$ decays, 
separated by production mode, are extracted in a template fit maximizing 
the combined likelihood of the $\PH\qqbar$ and $\PH\nunubar$ templates.
It is assumed that the contributions from other Higgs decay modes are 
determined from independent measurements and therefore these contributions are fixed in the fit. 

The results of the above analysis are summarised
in \autoref{tab:higgsstrahlung:HbbccggResults}, giving the statistical
uncertainties of the various $\sigma \times \BR$ measurements. Since the parameters in this
analysis are determined in a combined extraction from overlapping
distributions, the results are correlated. In particular the \higgsstrahlung and $\PW\PW$-fusion results for the same final states show sizeable anti-correlations, as large as $-38\%$ for the cases of $\PH\to\cc$ and $\PH\to\Pg\Pg$. These correlations are taken into account 
in the global fits described in \autoref{sec:combined_fits}.

\begin{table}[tb]
  \centering
 \begin{tabular}{lcc}
 	\toprule
 	\multirow{2}{*}{Decay} & \multicolumn{2}{c}{Statistical uncertainty}\\
	& \higgsstrahlung & $\PW\PW$-fusion\\
 	\midrule
	$\PH\to\bb$ &0.86\,\%& 1.9\,\% \\ 
	$\PH\to\cc$ &14\,\% & 26\,\%\\ 
	$\PH\to\Pg\Pg$ & 6.1\,\% & 10\,\%\\ 
	\bottomrule
  \end{tabular}
    \caption{\label{tab:higgsstrahlung:HbbccggResults} Summary of statistical uncertainties for events with a $\PH\to\bb$, $\PH\to\cc$ or $\PH\to\Pg\Pg$ decay, where the Higgs boson is produced by Higgsstrahlung or WW-fusion, at $\roots=350\,\GeV$ derived from the template fit as described in the text. All numbers correspond to an integrated luminosity of $500\,\fbinv$.} 
\end{table}

% Section 4.2.2
\subsubsection{$\PH \to \tptm$}
\label{sec:higgsstrahlung_tautau}
Because of the neutrino(s) produced in $\PGt$ decays, the signature
for $\PH\to\tptm$ is less distinct than that for other decay
modes. The invariant mass of the visible decay products of the $\tptm$
system will be less than $\mH$, and it is difficult to identify
$\PH\to\tptm$ decays from the $\PW\PW$-fusion process or
from \higgsstrahlung events where $\PZ\to\PGn\PAGn$.  For this reason,
the product of $\sigma(\PZ\PH)\times \BR(\PH\to\tptm)$ is only
determined for the case of hadronic $\PZ$ decays at $\roots=350\,\GeV$. 
In this analysis only hadronic \PGt decays are considered, so the 
experimental signature is two hadronic jets from $\PZ\to\qq$ and two
isolated low-multiplicity narrow jets from the two tau
decays~\cite{LCD:tautau_350}. Candidate \PGt leptons are identified using the \taufinder
algorithm \cite{LCDnote_TauFinder}, which is a seeded-cone based
jet-clustering algorithm. The algorithm was optimised to distinguish
the tau lepton decay products from hadronic gluon or quark jets. Tau
cones are seeded from single tracks ($\pT>5\,\GeV$). The seeds are
used to define narrow cones of $0.05$\,rad. The cones
are required to contain either one or three charged particles (from
one- and three-prong tau decays) and further rejection of background
from hadronic jets is implemented using cuts on isolation-related
variables. Tau cones which contain identified electrons or muons are
rejected and only the hadronic one- and three-prong \PGt decays are
retained.  The \PGt identification efficiency for hadronic tau decays
is found to be $73\,\%$ and the fake rate to mistake a quark for
a \PGt is $5\,\%$. The fake rate is relatively high, but is acceptable
as the background from final states with quarks can be suppressed
using global event properties.

\begin{figure}
  \centering
  \includegraphics[width=0.9\columnwidth]{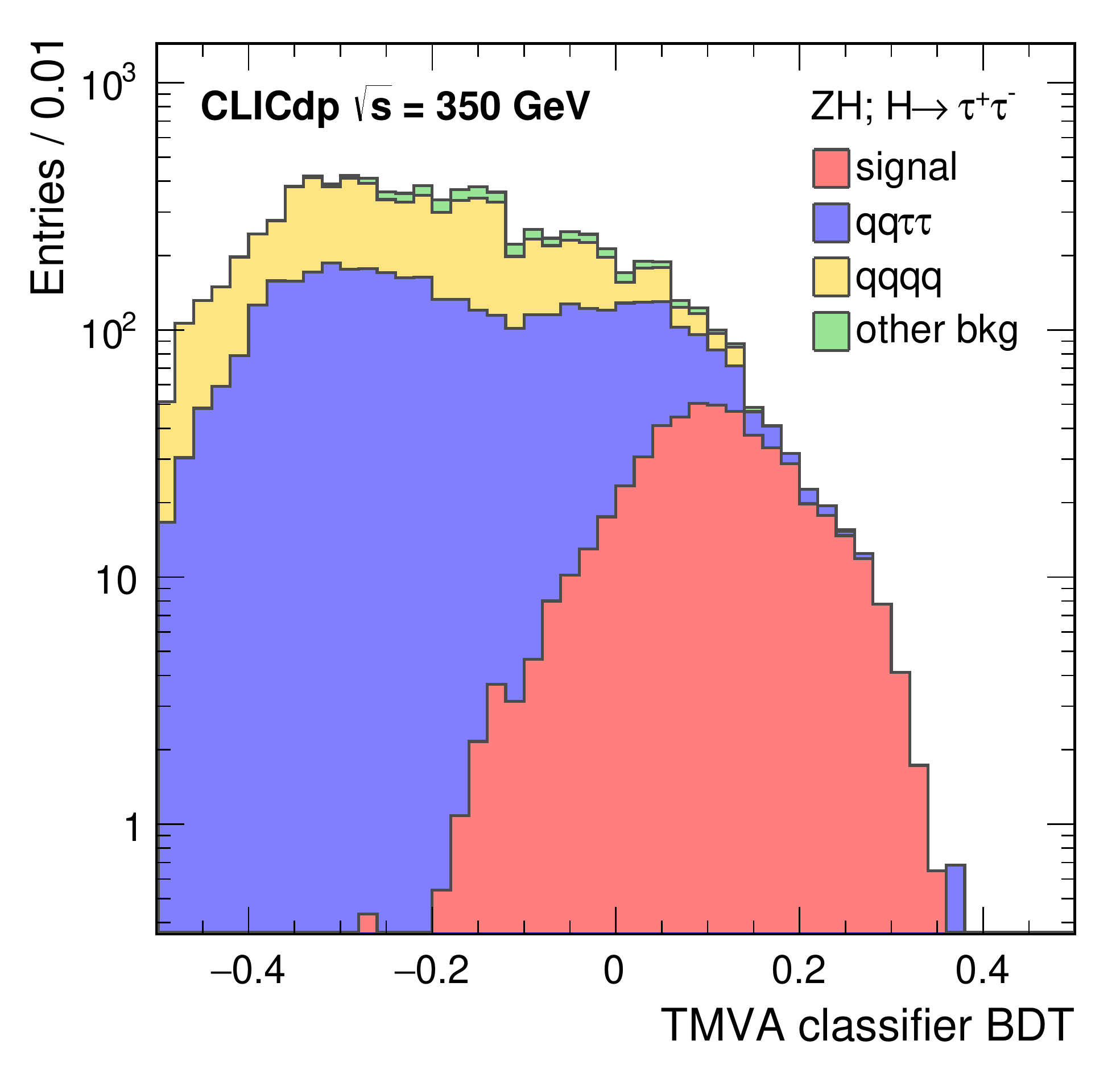}
  \caption{\label{htautau-plots} BDT classifier distributions for $\PH\to\tptm$ events at $\roots=350\,\GeV$, showing the signal and main backgrounds as stacked histograms. The distributions are normalised to an integrated luminosity of $500\,\fbinv$.}
\end{figure}

Events with two identified hadronic tau candidates (with opposite net
charge) are considered as $\PH\to\tptm$ decays. Further separation of
the signal and background events is achieved using a BDT classifier
based on the properties of the tau candidates and global event
properties. 
Seventeen discriminating variables are used as BDT inputs, including 
the thrust and oblateness of the quark and tau systems, and masses, 
transverse momenta, and angles in the events.  A full list is given 
in~\cite{LCD:tautau_350}.
%The seventeen input discriminating variables are: the
%total \pT of the full event; the event thrust; the thrust and
%oblateness of the \tptm system; the thrust and oblateness of the quark
%system; the sum of the transverse momenta of both \PGt candidates and
%of the quark jets; the cosines of the polar angle of both \PGt
%candidates; the invariant mass of the \tptm system; the invariant mass
%of the quark system; the angle between the two \PGt candidates; the
%angle between the two quark jets; the polar angle of the missing
%momentum vector; the azimuthal angle between the two \PGt candidates;
%the azimuthal angle between the two quark jets; and the visible energy
%in the event. 
The resulting BDT distributions for the signal and the 
backgrounds are shown in \autoref{htautau-plots}. Events passing a 
cut on the BDT output maximising the significance of the measurement are selected.
The cross sections and numbers of selected events for the signal and the dominant background
processes are listed in \autoref{tab:higgsstrahlung_tautau_numbers}. 
The contribution from background processes with photons in the initial state is negligible after the event selection. 
A template fit to the BDT output distributions leads to:
\begin{align*}
\frac{\Delta[\sigma(\PZ\PH)\times\BR(\PH\to\tptm)]}{\sigma(\PZ\PH)\times\BR(\PH\to\tptm)} = 6.2\,\%\,.
\end{align*}

\begin{table}[t]
\begin{center}
\begin{tabular}{lrrrr}
\toprule
Process & $\sigma/\fb$ & $\varepsilon_\text{presel}$ & $\varepsilon_\text{BDT}$ & $N_\text{BDT}$ \\
\midrule
$\epem \to \PZ\PH;$ & 5.8 & 18\,\% & 59\,\% & 312 \\
\ $\PZ \to \qq, \PH \to \tptm$ & & & \\
\midrule
$\epem \to \PZ\PH;$ & 4.6 & 15\,\% & 2.6\,\% & 9 \\
\ $\PZ \to \tptm, \PH \to X$ & & & \\ 
$\epem \to \PQq\PQq\PGt\PGt (\text{non-Higgs})$ & 70 & 10\,\% & 3.3\,\% & 117 \\
$\epem \to \PQq\PQq\PGt\PGt\PGn\PGn$ & 1.6 & 9.7\,\% & 5.1\,\% & 4 \\
%$\epem \to \PQq\PQq\PQq\PQq$ & 5900 & 0.13\,\% & 0.54\,\% & 21 \\
$\epem \to \PQq\PAQq\PQq\PAQq$ & 5850 & 0.13\,\% & 0.54\,\% & 21 \\

\bottomrule
\end{tabular}
\end{center}
\caption{
Cross sections and numbers of preselected and selected events with BDT > 0.08 (see \autoref{htautau-plots}) for $\epem \to \PZ\PH (\PZ \to \qq,\PH \to \tptm)$ signal events and the dominant backgrounds at $\roots=350\,\GeV$ assuming an integrated luminosity of $500\,\fbinv$. \label{tab:higgsstrahlung_tautau_numbers}}
\end{table}

% Section 4.2.3
\subsubsection{$\PH \to \PW\PW^\ast$}
\label{sec:higgsstrahlung_ww}
In case the Higgs boson decays to a pair of \PW bosons, only
the fully hadronic channel, $\PH \to \WW^\ast \to \qqqq$, allows the
reconstruction of the Higgs invariant mass. Two final states in $\epem\to\PZ\PH$ events 
have been studied depending on the \PZ boson decay mode: $\PZ\to\Plp\Plm$, where 
$\Pl$ is an electron or muon, and $\PZ\to\qqbar$.

First, isolated electrons and muons from \PZ decays are identified. 
Photons in a cone with an opening angle of $3^\circ$ around the lepton 
candidates are added to their four-momentum as described in \autoref{sec:simreco}.
%To
%recover the effect of bremsstrahlung photons radiated off the leptons, 
%all photons in a cone with an opening angle of $3^\circ$
%around the flight direction of a lepton candidate are added 
%to its four-momentum. The impact of the bremsstrahlung
%recovery on the recontruction of the $\PZ\to\epem$ decays is
%illustrated in \autoref{fig:bremsstrahlung_recovery}.
%The bremsstrahlung effect leads to a tail at lower values in the 
%$\PZ$ candidate invariant mass distribution. This loss can be recovered by 
%the procedure described above. It is also visible that a too large 
%opening angle of the recovery cone leads to a tail at higher masses.

If a leptonic \PZ candidate is found, four jets are reconstructed from 
all particles not originating from the \PZ decay. 
The jets are paired, with the pair that gives the mass closest to the \PW boson
mass being taken as one \PW boson candidate, and the other pair taken as the $\PW^\ast$.
The events are considered further if the invariant mass of 
the \PZ boson candidate is in the range between $70$ and $110\,\GeV$ and at least 20 particles 
are reconstructed.

In events without a leptonic \PZ candidate, six jets are reconstructed. The jets are grouped 
into \PW, \PZ and Higgs boson candidates by minimising:
\begin{equation*}
\chi^2=\frac{(m_{ij}-m_{\PW})^{2}}{\sigma_{\PW}^2}+\frac{(m_{kl}-m_{\PZ})^{2}}{\sigma_{\PZ}^2}+\frac{(m_{ijmn}-m_{\PH})^{2}}{\sigma_{\PH}^2},
\end{equation*}
where $m_{ij}$ is the invariant mass of the jet pair used to reconstruct the $\PW$ candidate, $m_{kl}$ is the invariant mass 
of the jet pair used to reconstruct the $\PZ$ candidate, $m_{ijmn}$ is the invariant mass of the four jets used to 
reconstruct the Higgs candidate and $\sigma_{\PW, \PZ, \PH}$ are the estimated invariant mass resolutions for \PW, \PZ and 
Higgs boson candidates. The preselection cuts for this final state are:
\begin{itemize}
\item invariant mass of the \PZ candidate greater than $40\,\GeV$;
\item at least 50 reconstructed particles;
\item event thrust of less than 0.95;
\item no jet with a b-tag probability of more than 0.95;
\item topology of the hadronic system consistent with six jets: $\log_{10}(y_{12})>-2.0$, $\log_{10}(y_{23})>-2.6$, $\log_{10}(y_{34})>-3.0$, $\log_{10}(y_{45})>-3.5$ and $\log_{10}(y_{56})>-4.0$.
\end{itemize}

%\begin{figure}
%\begin{centering}
%\includegraphics[width=0.9\columnwidth]{higgsstrahlung/InvMassZsignal_BeamRecoil.pdf}
%\caption{\label{fig:bremsstrahlung_recovery} Reconstructed invariant mass of $\PZ\to\epem$ candidates in $\epem\to\PZ\PH\to\PZ\WW^\ast$ events at $\roots=350\,\GeV$. Bremsstrahlung photons in cones of different opening angles around the electron direction are recovered as described in the text. The normalisation is arbitrary.}
%\end{centering}
%\end{figure}

\begin{table}[tb]
\begin{center}
  \begin{tabular}{lrrrr}
    \toprule 
      Process & $\sigma/\fb$ & $\varepsilon_\text{presel}$ & $\varepsilon_\text{BDT}$ & $N_\text{BDT}$ \\ \midrule
      $\epem\to\PZ\PH;\PZ\to\epem;$ & 0.45 & 80\,\% & 53\,\% & 95 \\
      $\ \PH\to\WW^\ast\to\qqqq$ & & & & \\
    \midrule
      $\epem\to\PZ\PH;\PZ\to\epem;$ & 4.1 & 69\,\% & 3.4\,\% & 48 \\
      $\ \PH\to\textnormal{other}$ & & & & \\ 
      $\epem\to\qqbar\Pl\Pl$ & 1700 & 3.6\,\% & 0.24\,\% & 75 \\
      $\epem\to\WW\PZ$ & 10 & 3.1\,\% & 5.9\,\% & 9 \\
      & & & & \\
    \bottomrule
      $\epem\to\PZ\PH;\PZ\to\mpmm;$ & 0.45 & 87\,\% & 65\,\% & 125 \\
      $\ \PH\to\WW^\ast\to\qqqq$ & & & & \\
    \midrule
      $\epem\to\PZ\PH;\PZ\to\mpmm;$ & 4.1 & 69\,\% & 5.2\,\% & 74 \\
      $\ \PH\to\textnormal{other}$ & & & & \\
      $\epem\to\qqbar\Pl\Pl$ & 1700 & 1.7\,\% & 0.35\,\% & 51 \\
      $\epem\to\WW\PZ$ & 10 & 2.6\,\% & 7.1\,\% & 9 \\
      & & & & \\
    \bottomrule
      $\epem\to\PZ\PH;\PZ\to\qq;$ & 9.2 & 71\,\% & 41\,\% & 1328 \\
      $\ \PH\to\WW^\ast\to\qqqq$ & & & & \\
    \midrule
      $\epem\to\PZ\PH;\PZ\to\qq;$ & 84 & 17\,\% & 10\,\% & 730 \\
      $\ \PH\to\textnormal{other}$ & & & & \\
      $\epem\to\qqqq$ & 5850 & 18\,\% & 0.54\,\% & 2849 \\
      $\epem\to\ttbar$ & 450 & 19\,\% & 2.5\,\% & 1071 \\
      $\epem\to\WW\PZ$ & 10 & 20\,\% & 18\,\% & 179 \\
    \bottomrule
    %\toprule 
    %  Process & $\sigma/\fb$ & $\varepsilon_\text{presel}$ & $\varepsilon_\text{BDT}$ & $N_\text{BDT}$ \\ \midrule
    %  $\epem\to\PZ\PH;\PZ\to\epem;$ & 0.45 & 80.2\,\% & 45.5\,\% & 95 \\
    %  $\ \PH\to\WW^\ast\to\qqqq$ & & & & \\
    %\midrule
    %  $\epem\to\PZ\PH;\PZ\to\epem;$ & 4.1 & 68.8\,\% & 2.3\,\% & 48 \\
    %  $\ \PH\to\textnormal{other}$ & & & & \\ 
    %  $\epem\to\qqbar\Pl\Pl$ & 1700 & 3.60\,\% & 0.24\,\% & 75 \\
    %  $\epem\to\WW\PZ$ & 10 & 3.1\,\% & 5.9\,\% & 9 \\
    %\bottomrule
    %  $\epem\to\PZ\PH;\PZ\to\mpmm;$ & 0.45 & 86.8\,\% & 64.8\,\% & 125 \\
    %  $\ \PH\to\WW^\ast\to\qqqq$ & & & & \\
    %\midrule
    %  $\epem\to\PZ\PH;\PZ\to\mpmm;$ & 4.1 & 68.8\,\% & 5.6\,\% & 74 \\
    %  $\ \PH\to\textnormal{other}$ & & & & \\
    %  $\epem\to\qqbar\Pl\Pl$ & 1700 & 1.69\,\% & 0.35\,\% & 51 \\
    %  $\epem\to\WW\PZ$ & 10 & 2.6\,\% & 7.1\,\% & 9 \\
    %\bottomrule
    %  $\epem\to\PZ\PH;\PZ\to\qq;$ & 9.2 & 70.9\,\% & 40.9\,\% & 1328 \\
    %  $\ \PH\to\WW^\ast\to\qqqq$ & & & & \\
    %\midrule
    %  $\epem\to\PZ\PH;\PZ\to\qq;$ & 84 & 16.6\,\% & 10.4\,\% & 730 \\
    %  $\ \PH\to\textnormal{other}$ & & & & \\
    %  $\epem\to\qqqq$ & 5850 & 18.1\,\% & 0.098\,\% & 2849 \\
    %  $\epem\to\ttbar$ & 450 & 19.0\,\% & 2.5\,\% & 1071 \\
    %  $\epem\to\WW\PZ$ & 10 & 19.7\,\% & 18.2\,\% & 179 \\
    %\bottomrule
  \end{tabular}
  \caption{\label{tab:HZ_hWW_channels} Preselection and selection efficiencies for the $\PZ\PH$ signal and most important background processes of the $\PH \to \WW^\ast$ analysis in all three considered \PZ decay channels. The numbers assume an integrated luminosity of $500\,\fbinv$ at $\roots=350\,\GeV$.}
\end{center}
\end{table}

For both final states, BDT classifiers are used to suppress the backgrounds 
further. The event selection for the signal processes and the most relevant 
background samples is summarised in \autoref{tab:HZ_hWW_channels}. The expected 
precisions for the measurement of the investigated processes are summarised in \autoref{tab:higgsstrahlung_hww_results_350}. 
The best precision is achieved using the $\PZ\to\qq$ decay due to its large branching ratio compared to leptonic decays.
The selection of $\PZ\to\epem$ events is more difficult compared to $\PZ\to\mpmm$ events because 
the $\epem\to\qq\Pl\Pl$ background sample contains more events with electron pairs than events with 
muon pairs. Hence the precision achieved using $\PZ\to\mpmm$ decays is somewhat better compared to that obtained 
using $\PZ\to\epem$ decays.
The combined precision for an integrated luminosity of 500\,$\fbinv$ is:
\begin{equation*}
\frac{\Delta[\sigma(\PZ\PH)\times \BR(\PH\to\PW\PW^*)]}{\sigma(\PZ\PH)\times \BR(\PH\to\PW\PW^*)} = 5.1\,\% \,,
\end{equation*}
which is dominated by the final state with hadronic \PZ boson decays.

\begin{table}[tb]
\begin{center}  
\begin{tabular}{lr}
\toprule
Process & Stat. uncertainty \\ \midrule
%$\epem\to\PZ\PH;\PZ\to\epem; \PH\to\WW^{\ast}\to\qqqq$ & 16.1\,\% \\
$\epem\to\PZ\PH;\PZ\to\epem; \PH\to\WW^{\ast}\to\qqqq$ & 16\,\% \\
%$\epem\to\PZ\PH;\PZ\to\mpmm; \PH\to\WW^{\ast}\to\qqqq$ & 13.1\,\% \\
$\epem\to\PZ\PH;\PZ\to\mpmm; \PH\to\WW^{\ast}\to\qqqq$ & 13\,\% \\
$\epem\to\PZ\PH;\PZ\to\qq; \PH\to\WW^{\ast}\to\qqqq$ & 5.9\,\% \\
\bottomrule
\end{tabular}
\end{center}
\caption{\label{tab:higgsstrahlung_hww_results_350} Statistical precisions for the listed processes at $\roots=350\,\GeV$ for an integrated luminosity of $500\,\fbinv$.}
\end{table}

% Section 5
\section{WW-fusion at $\roots>1\,\TeV$}
\label{sec:ww_fusion}

This section presents measurements of Higgs decays from the $\PW\PW$-fusion process at CLIC with centre-of-mass energies of 
 1.4\,\TeV and 3\,\TeV. The Higgs self-coupling measurement, which is also accessed in $\PW\PW$-fusion production, is discussed in \autoref{sec:higgs_self_coupling}. 
The cross section of the Higgs production via the vector boson fusion process $\epem \to \PH\nuenuebar$ scales with $\log(s)$ and becomes the dominating Higgs production process in \epem collisions with $\roots>500\,\GeV$. 
The respective cross sections for $\epem \to \PH\nuenuebar$ at $\roots = 1.4\,\TeV$ 
and 3\,\TeV are approximately 244\,\fb and 415\,\fb, respectively, including the effects of the CLIC beamstrahlung spectrum and ISR.  The relatively large cross sections at the higher energies allow the 
Higgs decay modes to be probed with high statistical precision and provide access to rarer Higgs decays, such as $\PH\to\mpmm$.

Since $\PW\PW$-fusion $\epem \to \PH\nuenuebar$ proceeds through the $t$-channel, the Higgs boson is typically boosted along the beam direction and the presence of neutrinos  in the final state can result in significant missing $\pT$.  Because of the missing transverse and longitudinal momentum, the experimental 
signatures for $\PH\nuenuebar$ production are relatively well separated from most SM backgrounds.
 At $\roots = 350\,$GeV, the main SM background processes are two- and four-fermion production, $\epem\to 2f$ and $\epem\to 4f$. At higher energies, backgrounds from $\PGg\PGg$ and $\PGg\Pepm$ hard interactions become increasingly relevant for measurements of Higgs boson production in $\PW\PW$-fusion. Additionally, pile-up of relatively soft $\gghadrons$ events with the primary interaction occurs. However, this background of relatively low-$\pT$ particles is largely mitigated through the timing cuts and jet finding strategy outlined in \autoref{sec:software}.

% Section 5.1.1
\subsection{$\PH \to \bb, \cc, \Pg\Pg$}
\label{sec:ww_fusion_bbccgg}
The physics potential for the measurement of hadronic Higgs decays at the 
centre-of-mass energies of 1.4\,\TeV and 3\,\TeV was studied using the 
\clicsid detector model. The signatures for $\PH\to\PQb\PAQb$, $\PH\to\PQc\PAQc$ 
and $\PH\to\Pg\Pg$ decays in $\epem\to\PH\nuenuebar$ events are two jets and missing energy. 
Flavour tagging information from \lcfiplus is used to separate the investigated Higgs boson 
decay modes in the selected event sample. The invariant mass of the reconstructed 
di-jet system provides rejection against background processes, e.g. hadronic $\PZ$ boson decays.

At both centre-of-mass energies, an invariant mass of the di-jet system in the range from 
$60$ to $160\,\GeV$ and a distance between both jets in the $\eta-\phi$ plane of less than 
4 are required. The energy sum of the two jets must exceed 75\,\GeV and 
a missing momentum of at least 20\,\GeV is required. The efficiencies of these preselection cuts on the 
signal and dominant background samples are listed in \autoref{tab:ww_fusion_hbb_numbers_1400} and 
\autoref{tab:ww_fusion_hbb_numbers_3000} for the centre-of-mass energies of 1.4 and 3\,\TeV, respectively.

\begin{table}[t]
\begin{center}
\begin{tabular}{lrrrr}
\toprule
Process & $\sigma/\fb$ & $\varepsilon_\text{presel}$ & $\varepsilon_\text{BDT}$ & $N_\text{BDT}$ \\
\midrule
$\epem \to \PH\nuenuebar; \PH\to\PQb\PAQb$ & 137 & 85\,\% & 38\,\% & 65400 \\
$\epem \to \PH\nuenuebar; \PH\to\PQc\PAQc$ & 6.9 & 87\,\% & 42\,\% & 3790 \\
$\epem \to \PH\nuenuebar; \PH\to\Pg\Pg$ & 20.7 & 82\,\% & 40\,\% & 10100 \\
\midrule
$\epem \to \PQq\PAQq\PGn\PAGn$ & 788 & 76\,\% & 2.1\,\% & 18500 \\
$\epem \to \PQq\PAQq\Pl\PGn$ & 4310 & 40\,\% & 0.91\,\% & 23600 \\
$\Pepm\PGg \to \PQq\PAQq\Pe$ & 16600 & 14\,\% & 0.54\,\% & 18500 \\
$\Pepm\PGg \to \PQq\PAQq\PGn$ & 29300 & 60\,\% & 0.64\,\% & 170000 \\
$\PGg\PGg \to \PQq\PAQq$ & 76600 & 4.2\,\% & 0.47\,\% & 22200 \\
%$\epem \to \PH\nuenuebar; \PH\to\PQb\PAQb$ & 136.9 & 84.5\,\% & 37.7\,\% & 65400 \\
%$\epem \to \PH\nuenuebar; \PH\to\PQc\PAQc$ & 6.91 & 86.8\,\% & 42.1\,\% & 3740 \\
%$\epem \to \PH\nuenuebar; \PH\to\Pg\Pg$ & 20.7 & 82.2\,\% & 39.6\,\% & 10100 \\
%\midrule
%$\epem \to \PQq\PAQq\PGn\PAGn$ & 788 & 75.6\,\% & 2.1\,\% & 18500 \\
%$\epem \to \PQq\PAQq\Pl\PGn$ & 4313 & 40.0\,\% & 0.91\,\% & 23600 \\
%$\Pepm\PGg \to \PQq\PAQq\Pe$ & 16600 & 13.7\,\% & 0.54\,\% & 18500 \\
%$\Pepm\PGg \to \PQq\PAQq\PGn$ & 29300 & 60.2\,\% & 0.64\,\% & 170000 \\
%$\PGg\PGg \to \PQq\PAQq$ & 76600 & 4.2\,\% & 0.47\,\% & 22200 \\
\bottomrule
\end{tabular}
\end{center}
\caption{Preselection and selection efficiencies for the signal and most important background processes in the $\PH\to\PQb\PAQb$, $\PH\to\PQc\PAQc$ and $\PH\to\Pg\Pg$ analysis. The numbers of events correspond to 1.5\,$\abinv$ at $\roots=1.4\,\TeV$.
\label{tab:ww_fusion_hbb_numbers_1400}}
\end{table}

\begin{table}[t]
\begin{center}
\begin{tabular}{lrrrr}
\toprule
Process & $\sigma/\fb$ & $\varepsilon_\text{presel}$ & $\varepsilon_\text{BDT}$ & $N_\text{BDT}$ \\
\midrule
$\epem \to \PH\nuenuebar; \PH\to\PQb\PAQb$ & 233 & 74\,\% & 35\,\% & 120000 \\
$\epem \to \PH\nuenuebar; \PH\to\PQc\PAQc$ & 11.7 & 75\,\% & 36\,\% & 6380 \\
$\epem \to \PH\nuenuebar; \PH\to\Pg\Pg$ & 35.2 & 69\,\% & 35\,\% & 16800 \\
\midrule
$\epem \to \PQq\PAQq\PGn\PAGn$ & 1300 & 67\,\% & 2.7\,\% & 47400 \\
$\epem \to \PQq\PAQq\Pe\PGn$ & 5260 & 45\,\% & 1.1\,\% & 52200 \\
$\Pepm\PGg \to \PQq\PAQq\Pe$ & 20500 & 13\,\% & 2.3\,\% & 118000 \\
$\Pepm\PGg \to \PQq\PAQq\PGn$ & 46400 & 46\,\% & 0.92\,\% & 394000 \\
$\PGg\PGg \to \PQq\PAQq$ & 92200 & 7.0\,\% & 1.6\,\% & 207000 \\
%$\epem \to \PH\nuenuebar; \PH\to\PQb\PAQb$ & 232.8 & 74.4\,\% & 34.7\,\% & 120000 \\
%$\epem \to \PH\nuenuebar; \PH\to\PQc\PAQc$ & 11.7 & 75.3\,\% & 36.2\,\% & 6340 \\
%$\epem \to \PH\nuenuebar; \PH\to\Pg\Pg$ & 35.2 & 68.8\,\% & 34.6\,\% & 16800 \\
%\midrule
%$\epem \to \PQq\PAQq\PGn\PAGn$ & 1305 & 67.2\,\% & 2.7\,\% & 47400 \\
%$\epem \to \PQq\PAQq\Pe\PGn$ & 5255 & 45.0\,\% & 1.1\,\% & 52200 \\
%$\Pepm\PGg \to \PQq\PAQq\Pe$ & 20500 & 12.7\,\% & 2.3\,\% & 118000 \\
%$\Pepm\PGg \to \PQq\PAQq\PGn$ & 46400 & 45.9\,\% & 0.92\,\% & 394000 \\
%$\PGg\PGg \to \PQq\PAQq$ & 92200 & 7.0\,\% & 1.6\,\% & 207000 \\
\bottomrule
\end{tabular}
\end{center}
\caption{Preselection and selection efficiencies for the signal and most important background processes in the $\PH\to\PQb\PAQb$, $\PH\to\PQc\PAQc$ and $\PH\to\Pg\Pg$ analysis. The numbers of events correspond to 2\,$\abinv$ at $\roots=3\,\TeV$.
\label{tab:ww_fusion_hbb_numbers_3000}}
\end{table}

The backgrounds are suppressed further using a single BDT at each energy. The samples of signal events 
used to train these classifiers consist of equal amounts of $\PH\to\PQb\PAQb$, $\PH\to\PQc\PAQc$, 
and $\PH\to\Pg\Pg$ events, while the different processes in the background sample were normalised according 
to their respective cross sections. No flavour tagging information is used in the event selection. 
This leads to classifiers with similar selection efficiencies for events with the different signal Higgs decays.

The fractions of signal events with $\PH\to\PQb\PAQb$, $\PH\to\PQc\PAQc$ and $\PH\to\Pg\Pg$ decays in the selected event samples are extracted from the two-dimensional distributions of the $\bb$ versus $\cc$ likelihood variables for the two reconstructed jets as defined in \autoref{sec:higgsstrahlung_branching_ratios}. The normalisations of the backgrounds from other Higgs decays and non-Higgs events are fixed and expected to be provided by other measurements. The results of these fits are shown in \autoref{tab:ww_fusion_hbb_results_1400} and \autoref{tab:ww_fusion_hbb_results_3000} at 1.4 and 3\,\TeV, respectively.

The expected precisions obtained at 1.4 and 3\,\TeV are similar although the number of signal events is about twice as large at 3\,\TeV compared to 1.4\,\TeV. The main reasons for this are that the jet reconstruction and flavour tagging are more challenging at 3\,\TeV, since the jets from the Higgs decay tend more towards the beam axis, and the impact of the beam-induced backgrounds is larger compared to 1.4\,\TeV. In addition, the cross sections for the most important background processes rise with \roots (see \autoref{tab:ww_fusion_hbb_numbers_1400} and \autoref{tab:ww_fusion_hbb_numbers_3000}).

\begin{table}[t]
\begin{center}
\begin{tabular}{lr}
\toprule
Process & Statistical uncertainty \\
\midrule
$\epem \to \PH\nuenuebar; \PH\to\PQb\PAQb$ & 0.4\,\% \\
$\epem \to \PH\nuenuebar; \PH\to\PQc\PAQc$ & 6.1\,\% \\
$\epem \to \PH\nuenuebar; \PH\to\Pg\Pg$ & 5.0\,\% \\
\bottomrule
\end{tabular}
\end{center}
\caption{Statistical precisions for the listed processes from the fit described in the text at $\roots=1.4\,\TeV$ for an integrated luminosity of 1.5\,$\abinv$.
\label{tab:ww_fusion_hbb_results_1400}}
\end{table}

\begin{table}[t]
\begin{center}
\begin{tabular}{lrrrr}
\toprule
Process & Statistical uncertainty \\
\midrule
$\epem \to \PH\nuenuebar; \PH\to\PQb\PAQb$ & 0.3\,\% \\
$\epem \to \PH\nuenuebar; \PH\to\PQc\PAQc$ & 6.9\,\% \\
$\epem \to \PH\nuenuebar; \PH\to\Pg\Pg$ & 4.3\,\% \\
\bottomrule
\end{tabular}
\end{center}
\caption{Statistical precisions for the listed processes from the fit described in the text at $\roots=3\,\TeV$ for an integrated luminosity of 2\,$\abinv$.
\label{tab:ww_fusion_hbb_results_3000}}
\end{table}

% Section 5.1.2
\subsection{$\PH \to \tptm$}
\label{sec:ww_fusion_tautau}
\newcommand{\htautau}{\ensuremath{\PH\to\tptm}\xspace}

The sensitivity for the measurement of
$\sigma(\epem\to\PH\nuenuebar)\times\BR(\htautau)$ at CLIC has been studied
using the \clicild detector model at centre-of-mass energies of
1.4\,\TeV and 3\,\TeV~\cite{LCD:tautau_1400}. For a SM Higgs with a mass of
126\,\GeV, $\BR(\PH\to\tptm)=6.2\,\%$, resulting in an effective
signal cross section of 15.0\,\fb at $\roots = 1.4\,\TeV$ and
25.5\,\fb at $\roots = 3\,\TeV$.

\begin{figure}
\begin{center}
\includegraphics[width=0.45\textwidth]{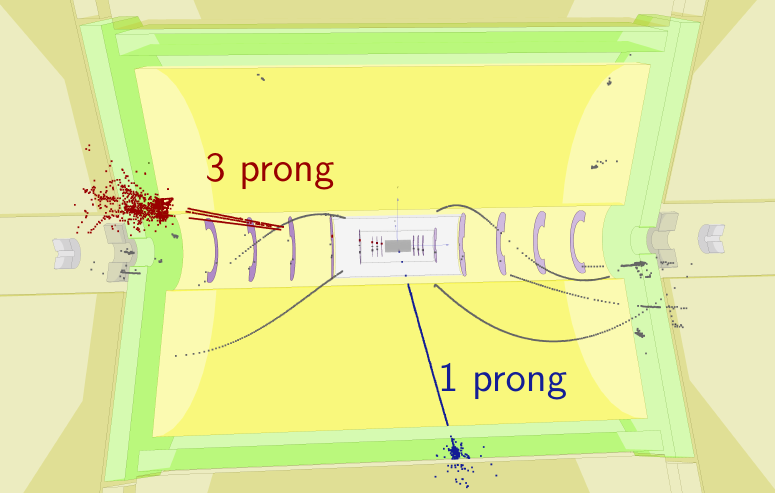}
\caption{Event display of a \htautau event at $\roots=1.4\,\TeV$ in the \clicild detector. A 1-prong tau decay is visible in the central part of the detector (blue). The other tau lepton decays to three charged particles and is reconstructed in the forward direction (red). A few soft particles from beam-induced backgrounds are also visible (grey).\label{fig:event_htautau}}
\end{center}
\end{figure}

The experimental signature is two relatively high-momentum narrow jets from the two tau decays and 
significant missing transverse and longitudinal momenta. 
A typical event display is shown in \autoref{fig:event_htautau}.
The analysis is restricted to hadronic \PGt decays, 
which are identified using the \taufinder algorithm, as described in \autoref{sec:higgsstrahlung_tautau}.
The \taufinder algorithm parameters were tuned using the $\PH\to\tptm$ signal events and 
$\epem \to \qqbar\nunubar$ background events. 
The working point has a $\PGt$ selection efficiency of 70\,\% (60\,\%) 
with a quark jet fake rate of 7\,\% (9\,\%) at $\roots = 1.4\,\TeV$ ($\roots = 3\,\TeV$). 
All relevant SM backgrounds are taken into account, including $\PGg\PGg$ and $\PGg\Pepm$ collisions.
The most significant 
backgrounds are $\epem \to \tptm \nunubar$, $\Pepm\PGg \to \tptm \Pepm$ and $\gamgam \to \tptm \nunubar$. 
The latter two processes become increasingly important at higher $\roots$, 
due to the increasing number of beamstrahlung photons.
Backgrounds from Higgs decays other than $\PH\to\tptm$ are expected to
be negligible~\cite{Kawada:2015wea}.

\begin{table}[t]
\begin{center}
\begin{tabular}{lrrrr}
\toprule
Process & $\sigma/\fb$ & $\varepsilon_\text{presel}$ & $\varepsilon_\text{BDT}$ & $N_\text{BDT}$ \\
\midrule
$\epem \to \PH\nuenuebar; \PH \to \tptm$ & 15.0 & 9.3\,\% & 39\,\% & 814 \\
\midrule
$\epem \to \tptm\nunubar$ & 38.5 & 5.0\,\% & 18\,\% & 528 \\
$\Pepm\PGg \to \tptm\Pepm$ & 2140 & 1.9\,\% & 0.075\,\% & 45 \\
$\PGg\PGg \to \tptm(\nunubar~\textnormal{or}~\Plm\Plp)$ & 86.7 & 2.7\,\% & 2.3\,\% & 79 \\
%$\epem \to \PH\nuenuebar; \PH \to \tptm$ & 15.0 & 9.3\,\% & 38.9\,\% & 814 \\
%\midrule
%$\epem \to \tptm\nunubar$ & 38.5 & 5.0\,\% & 18.0\,\% & 528 \\
%$\Pepm\PGg \to \tptm\Pepm$ & 2580 & 1.9\,\% & 0.075\,\% & 45 \\
%$\PGg\PGg \to \tptm(\nunubar~\textnormal{or}~\Plm\Plp)$ & 128.1 & 2.7\,\% & 2.25\,\% & 79 \\
\bottomrule
\end{tabular}
\end{center}
\caption{Preselection and selection efficiencies for the signal and most important background processes in the $\htautau$ analysis. The numbers of events correspond to 1.5\,$\abinv$ at $\roots=1.4\,\TeV$. The cross sections for the backgrounds include cuts on the kinematic properties of the tau lepton pair applied at generator level. The preselection efficiencies include the reconstruction of two hadronic tau lepton decays per event.
\label{tab:ww_fusion_htautau_numbers_1400}}
\end{table}

\begin{table}[t]
\begin{center}
\begin{tabular}{lrrrr}
\toprule
Process & $\sigma/\fb$ & $\varepsilon_\text{presel}$ & $\varepsilon_\text{BDT}$ & $N_\text{BDT}$ \\
\midrule
$\epem \to \PH\nuenuebar; \PH \to \tptm$ & 25.5 & 6.7\,\% & 23\,\% & 787 \\
\midrule
$\epem \to \tptm\nunubar$ & 39.2 & 5.7\,\% & 11\,\% & 498 \\
$\Pepm\PGg \to \tptm\Pepm$ & 2393 & 2.0\,\% & 0.26\,\% & 246 \\
$\PGg\PGg \to \tptm(\nunubar~\textnormal{or}~\Plm\Plp)$ & 158 & 2.0\,\% & 0.14\,\% & 9 \\
\bottomrule
\end{tabular}
\end{center}
\caption{Preselection and selection efficiencies for the signal and most important background processes in the $\htautau$ analysis. The numbers of events correspond to 2\,$\abinv$ at $\roots=3\,\TeV$. The cross sections for the backgrounds include cuts on the kinematic properties of the tau lepton pair applied at generator level. The preselection efficiencies include the reconstruction of two hadronic tau lepton decays per event.
\label{tab:ww_fusion_htautau_numbers_3000}}
\end{table}

The event preselection requires two identified $\PGt$ leptons, both of which must be within the polar angle range
$15^\circ < \theta(\PGt) < 165^\circ$ and have $\pT(\PGt) > 25\,\GeV$. To reject back-to-back or nearby tau leptons, the 
angle between the two tau candidates must satisfy $29^\circ < \Delta\theta(\PGt\PGt) < 177^\circ$. The visible invariant mass $m(\PGt\PGt)$ and the visible transverse mass 
$m_\text{T}(\PGt\PGt)$ of the two tau candidates must satisfy
$45\,\GeV < m(\PGt\PGt) < 130\,\GeV$ and 
$m_\text{T}(\PGt\PGt) < 20\,\GeV$. Finally the event thrust must be less than 0.99. 

Events passing the preselection are classified as either signal or SM background using 
a BDT classifier. 
The kinematic variables used in the classifier are $m(\PGt\PGt)$, $m_\text{T}(\PGt\PGt)$, 
event shape variables (such as thrust and oblateness),  the missing $\pT$, the polar angle of the missing momentum vector $\costhetamis$
and the total reconstructed energy excluding the Higgs candidate. 
The event selection for the signal and the most relevant background processes is summarised in \autoref{tab:ww_fusion_htautau_numbers_1400} for $\roots = 1.4\,\TeV$ and in \autoref{tab:ww_fusion_htautau_numbers_3000} for $\roots = 3\,\TeV$.
Rather than applying a simple cut, the full BDT shape information is used in a template fit. The resulting statistical uncertainties for 
$1.5\,\abinv$  at $\roots=1.4\,\TeV$ and $2.0\,\abinv$ at $\roots=3\,\TeV$ are:
\begin{align*}
\frac{\Delta[\sigma(\PH\nuenuebar)\times\BR(\PH\to\tptm)]}{\sigma(\PH\nuenuebar)\times\BR(\PH\to\tptm)} & = 4.2\,\% \ \text{at} \ 1.4\,\TeV\,, \\
\frac{\Delta[\sigma(\PH\nuenuebar)\times\BR(\PH\to\tptm)]}{\sigma(\PH\nuenuebar)\times\BR(\PH\to\tptm)} & = 4.4\,\% \ \text{at} \ 3\,\TeV\,. \\
\end{align*}

Similar to the observations described in \autoref{sec:ww_fusion_bbccgg}, the expected precisions at 1.4\,\TeV and 3\,\TeV are similar. The identification of tau leptons is more challenging at 3 TeV where the impact of the beam-induced backgrounds is larger and the tau leptons from Higgs decays in signal events tend more towards the beam axis.

% Section 5.2.1
\subsection{$\PH \to \PW\PW^\ast$}
\label{sec:ww_fusion_ww}
The signature for $\PH\to\PW\PW^*$ decays in $\epem\to\PH\nuenuebar$
depends on the $\PW\PW^*$ decay modes. As $\mH < 2\mW$, at least one of 
the $\PW$-bosons is off mass-shell. Studies for two different final 
states are described in the following.
The presence of a charged lepton in the $\PW\PW^\ast\to\PQq\PAQq\Pl\PGn$ final state 
suppresses backgrounds from other Higgs decays. However, the invariant mass of
the Higgs boson in $\PH\to\PW\PW^*$ decays can be reconstructed for
fully-hadronic decays alone, $\PW\PW^\ast\to\PQq\PAQq\PQq\PAQq$.

\subsubsection{$\PW\PW^\ast\to\PQq\PAQq\PQq\PAQq$}

The experimental signature for
$\PH\PGne\PAGne$ production with $\PH\to\PW\PW^*\to\PQq\PAQq\PQq\PAQq$
is a four-jet final state with missing $\pT$ and a total invariant
mass consistent with the Higgs mass, where one pair of jets has a mass
consistent with $\mW$.
%There are two main sources of potential
%backgrounds. The first is other Higgs decays, in particular
%$\PH\to\PQb\PAQb$, $\PH\to\PQc\PAQc$ and $\PH\to\Pg\Pg$, which produce
%hadronic final states with an invariant mass consistent with the Higgs
%mass; here QCD radiation in the parton shower can lead to a four-jet
%topology. The second main source of potential background comes from
%non-Higgs processes such as $\epem\to\PQq\PAQq\PGn\PAGn$ and $\PGg\Pepm\to\PQq\PAQq\PQq\PAQq\PGn$.
 
The $\PH\to\PW\PW^*$ event selection has been studied at
$\roots=1.4\,\TeV$ using the \clicild detector model. It proceeds in two separate stages: a set of
preselection cuts designed to reduce the backgrounds from large cross
section processes such as \mbox{$\epem\to\PQq\PAQq$}
and \mbox{$\epem\to\PQq\PAQq\PQq\PAQq$}; followed by a
like\-lihood-based multivariate event selection. The preselection
variables are formed by forcing each event into four jets using the
Durham jet finder. Of the three possible jet associations with
candidate $\PW$ bosons, (12)(34), (13)(24) or (14)(23), the one giving
a di-jet invariant mass closest to $\mW$ is selected. 
The preselection requires that there is no high-energy electron or muon with $E_{\ell}>30\,\GeV$.
Further preselection cuts are made on the properties of the jets, the invariant masses of the off-shell and on-shell $\PW$ boson candidates, the Higgs boson candidate, the total visible energy and the missing transverse momentum. 
%The
%preselection cuts require: $\log_{10}(y_{23})>-2.75$ and
%$\log_{10}(y_{34})>-3.5$; visible energy,
%$125\,\GeV<E_\text{vis}<600\,\GeV$; missing transverse momentum, $\pT
%> 65\,\GeV$; $\cos\theta_\text{miss} < 0.99$; one candidate on-shell
%$\PW$ boson, $ 50\,\GeV < m_{\PW 1} < 95\,\GeV$; one off-shell $\PW$
%boson, $m_{\PW 2} < 65\,\GeV$; total invariant mass consistent with a
%Higgs decay, $90\,\GeV<\mH<150\,\GeV$; and the absence of a
%high-energy electron or muon, $E_\text{lept}<30\,\GeV$.  
In addition,
in order to reject $\PH\to\PQb\PAQb$ decays, the event is forced into
a two-jet topology and flavour tagging is applied to the two
jets. Events where at least one jet has a $\PQb$-tag probability 
above 0.95 are rejected as part of the
preselection. The cross sections and preselection efficiencies for the
signal and main background processes are listed in
\autoref{tab:ww_fusion:hww:events}. After the preselection, the main backgrounds are $\epem\to\PQq\PAQq\PGn\PAGn$, $\PGg\Pepm\to\PQq\PAQq\PQq\PAQq\PGn$ and other Higgs decay modes, predominantly
$\PH\to\PQb\PAQb$ and $\PH\to\Pg\Pg$, 
where QCD radiation in the parton shower can lead to a four-jet topology.

\begin{table}[tb]
  \centering
   \begin{tabular}{lrrrr}
    \toprule 
         Process                                & \tabt{$\sigma/\text{fb}$} & \tabt{$\varepsilon_\text{presel}$} & \tabt{$\varepsilon_{{\cal{L}}>0.35}$} & \tabt{$N_{{\cal{L}}>0.35}$} \\ \midrule
  All $\PH\PGne\PAGne$                                    &   244    & 14.6\,\%     & 21\,\% & 11101  \\
  $\PH\to\PW\PW^*\to\PQq\PAQq\PQq\PAQq$  
                                                                     &               & 32\,\%     & 56\,\% & 7518  \\
  $\PH\to\PW\PW^*\to\PQq\PAQq\Pl\PGn$  
                                                                     &               & 4.4\,\%     & 14\,\% &     253  \\
   $\PH\to\PQb\PAQb$  
                                                                     &               & 1.9\,\%      & 21\,\% & 774  \\       
   $\PH\to\PQc\PAQc$  
                                                                     &               & 8.1\,\%     & 26\,\% &    209  \\                                                                    
    $\PH\to\Pg\Pg$  
                                                                     &               & 19\,\%     & 37\,\% &  1736  \\          
    $\PH\to\PZ\PZ^*$  
                                                                     &               & 12\,\%     & 42\,\% &    556  \\                     
    $\PH\to\text{other}$  
                                                                     &               & 0.7\,\%     & 29\,\% &   55  \\
\midrule
    $\epem\to\PQq\PAQq\PGn\PAGn$                          &  788     & 4.6\,\%      & 4.1\,\%    & 2225  \\    
      $\epem\to\PQq\PAQq\PQq\PAQq\Pl\PGn$           &  115     & 0.1\,\%     & 25\,\%    &    43  \\ 
  $\epem\to\PQq\PAQq\PQq\PAQq\PGn\PAGn$         &   24.7     & 0.8\,\%     & 44\,\%    &     130 \\
  $\PGg\Pepm\to\PQq\PAQq\PQq\PAQq\PGn$  &    254   &  1.8\,\%  & 20\,\%  & 1389 \\   

    \bottomrule
  \end{tabular}
   \caption{Summary of the $\PH\to\PW\PW^\ast\to\PQq\PAQq\PQq\PAQq$ event selection at $\roots=1.4\,\TeV$, giving the raw cross sections, preselection efficiency, selection efficiency for a likelihood cut of  ${\cal{L}}>0.35$, and the expected numbers of events passing the 
  event selection for an integrated luminosity of $1.5\,\abinv$. 
 \label{tab:ww_fusion:hww:events}}
\end{table}

\begin{figure}
	  \centering
      \includegraphics[width=0.9\columnwidth]{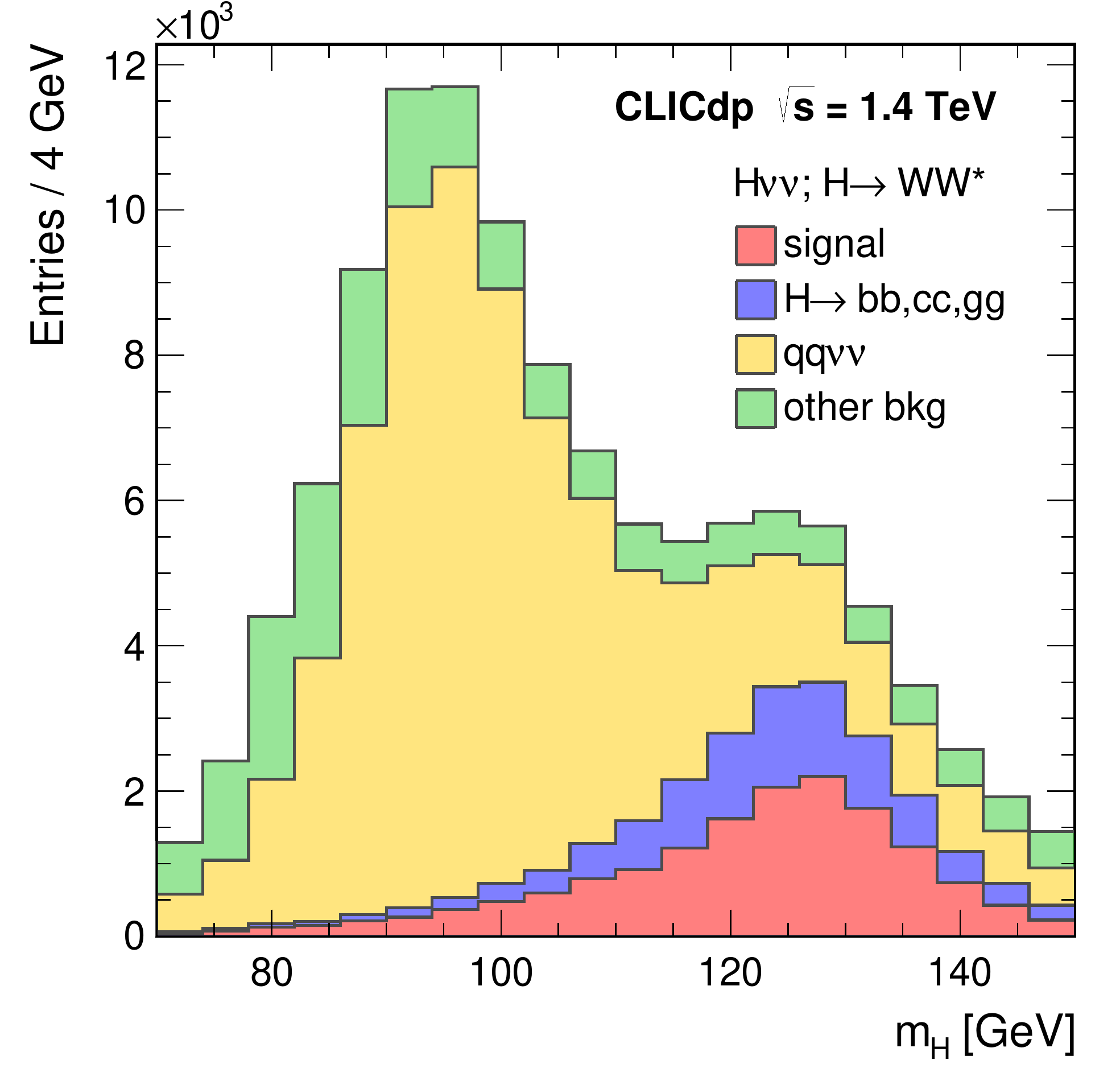}
      \caption{\label{fig:ww_fusion:hww:mH_HWW} Reconstructed Higgs invariant mass distributions for preselected $\PH \to \PW\PW^\ast\to\PQq\PAQq\PQq\PAQq$ events at $\roots = 1.4\,\TeV$, showing the signal and main backgrounds as stacked histograms. The distributions are normalised to an integrated luminosity of $1.5\,\abinv$.}
\end{figure}

A relative likelihood selection is used to classify all events passing
the preselection cuts. Five event categories including the signal are considered. 
The relative likelihood of an event being signal is estimated as:
\begin{equation*}
{\cal{L}} = \frac{L(\PH\to\PW\PW^*\to\PQq\PAQq\PQq\PAQq)}{L(\PH\to\PW\PW^*\to\PQq\PAQq\PQq\PAQq)+ L_1 + L_2 + L_3 + L_4}\,,
\end{equation*}  
where $L_i$ represents the likelihood for four background categories: $\PH\to\PQb\PAQb$, $\PH\to\Pg\Pg$, 
$\epem\to\PQq\PAQq\PGn\PAGn$ and $\PGg\Pepm\to\PQq\PAQq\PQq\PAQq\PGn$.
The absolute likelihood $L$ for each event type is
formed from normalised probability distributions $P_i(x_i)$ of the $N$
likelihood discriminating variables $x_i$ for that event type. For
example, the distribution of the reconstructed Higgs mass for all
events passing the preselection is shown
in \autoref{fig:ww_fusion:hww:mH_HWW}; it can be seen that good
separation between signal and background is achievable. The
discriminating variables are: the 2D distribution of reconstructed
invariant masses $\mH$ and $\mW$, the 2D distribution of minimal 
$k_t$ distances $y_{23}$, $y_{34}$, and 2D distribution of $\PQb$-tag
probabilities when the event is forced into two jets. The use of 2D distributions accounts
for the most significant correlations between the likelihood
variables.
The selection efficiencies and expected numbers of events for the
signal dominated region, ${\cal{L}}>0.35$, are listed in
\autoref{tab:ww_fusion:hww:events}.

The expected precision on $\BR(\PH\to\PW\PW^*)$ is extracted from a
fit to the likelihood distribution. Given the non-negligible
backgrounds from other Higgs decays, it is necessary to simultaneously
fit the different components. A $\chi^2$ fit to the expected
${\cal{L}}$ distribution is performed by scaling independently five
components: the $\PH\to\PW\PW^*$ signal, the $\PH\to\PQb\PAQb$,
$\PH\to\PQc\PAQc$ and $\PH\to\Pg\Pg$ backgrounds, and all other
backgrounds (dominated by $\PQq\PAQq\PGn\PAGn$ and
$\PQq\PAQq\PQq\PAQq\PGn$). The constraints on the $\PH\to\PQb\PAQb$,
$\PH\to\PQc\PAQc$ and $\PH\to\Pg\Pg$ branching ratios, as described in
\autoref{sec:ww_fusion_bbccgg}, are implemented by modifying the
$\chi^2$ function to include penalty terms:
\begin{align*}
   \chi^2  \to  \chi^2 + & \frac{(s_{\PQb\PAQb}-1)^2}{\sigma^2_{\PQb\PAQb}} +  \frac{(s_{\PQc\PAQc}-1)^2}{\sigma^2_{\PQc\PAQc}} 
  + \frac{(s_{\Pg\Pg}-1)^2}{\sigma^2_{\Pg\Pg}}   + \\
     & \frac{(s_{\PZ\PZ^*}-1)^2}{\sigma^2_{\PZ\PZ^*}} +     \frac{(b-1)^2}{\sigma^2_b}  \,.
\end{align*}
Here, for example, $s_{\Pg\Pg}$ is the amount by which the
$\PH\to\Pg\Pg$ complement is scaled in the fit and $\sigma_{\Pg\Pg}$
is the expected statistical error on $\BR(\PH\to\Pg\Pg)$ from the
analysis of \autoref{sec:ww_fusion_bbccgg}. The expected uncertainties on 
the contributions from $\PH\to\PQb\PAQb$ and $\PH\to\PQc\PAQc$ are taken from 
the same analysis. The background from $\PH\to\PZ\PZ^*$ is 
assumed here to be known to 1\,\% from other measurements of $\gHZZ^2$ and $\gHZZ^2/\gHWW^2$.
The systematic uncertainty in the non-$\PH$ background, denoted by $b$, is taken to
be 1\,\%. This has a small effect on the resulting uncertainty on the
$\PH\to\PW\PW^*$ branching ratio, which is:
\begin{equation*}
\frac{\Delta[\sigma(\PH\nuenuebar)\times\BR(\PH\to\PW\PW^*)]}{\sigma(\PH\nuenuebar)\times\BR(\PH\to\PW\PW^*)} = 1.5\,\% \,.
\end{equation*}

\subsubsection{$\PW\PW^\ast\to\PQq\PAQq\Pl\PGn$}
 
As a second channel, the $\PH\to\PW\PW^\ast\to\qqbar\Pl\PGn$ decay is investigated~\cite{lcd:ww_qqlv_1400}. The study 
is performed at $\roots=1.4\,\TeV$ using the \clicild detector model.

As a first step,  isolated electrons or muons from  \PW boson decay are 
identified. An efficiency of 93\,\% is achieved for the identification of electrons and 
muons in signal events including the geometrical acceptance of the detector. Two jets are 
reconstructed from the remaining particles, excluding the isolated electron or muon. Flavour tagging information 
is obtained from the \lcfiplus package.

The following preselection cuts are imposed:
\begin{itemize}
\item energy of the \PW candidate less than 590\,\GeV;
\item mass of the \PW candidate less than 230\,\GeV;
\item energy of the \PH candidate less than 310\,\GeV;
\item total missing energy of the event in the range between 670\,\GeV and 1.4\,\TeV.
\end{itemize}
Nearly all signal events pass this preselection, while more than 30\,\% of the  critical 
$\epem\to\qqbar\Pl\PGn$ background events are rejected. The background processes are suppressed further 
using a BDT classifier with 19 input variables including the number of isolated leptons. The event selection is 
summarised in \autoref{tab:ww_fusion_hww_qqlv_numbers_1400}. The resulting statistical precision for $1.5\,\abinv$ is:
\begin{equation*}
\frac{\Delta[\sigma(\PH\nuenuebar)\times\BR(\PH\to\PW\PW^*)]}{\sigma(\PH\nuenuebar)\times\BR(\PH\to\PW\PW^*)} = 1.3\,\% \,.
\end{equation*}
The combined precision for  $\PH\to\PW\PW^\ast\to\PQq\PAQq\PQq\PAQq$ and $\PH\to\PW\PW^\ast\to\qqbar\Pl\PGn$ decays at $\roots=1.4\,\TeV$ for an integrated luminosity of $1.5\,\abinv$ is 1.0\,\%.

\begin{table}[t]
\begin{center}
\begin{tabular}{lrrrr}
\toprule
Process & $\sigma/\fb$ & $\varepsilon_\text{presel}$ & $\varepsilon_\text{BDT}$ & $N_\text{BDT}$ \\
\midrule
$\epem \to \PH\nuenuebar;$ & 18.9 & 100\,\% & 42\,\% & 11900 \\
$\PH\to\PW\PW^\ast\to\PQq\PAQq\Pl\PGn$ & & & \\
\midrule
$\epem \to \PH\nuenuebar;$ & 25.6 & 100\,\% & 1.9\,\% & 721 \\
$\PH\to\PW\PW^\ast\to\PQq\PAQq\PQq\PAQq$ & & & \\
$\epem \to \PH\nuenuebar;$ & 200 & 99.6\,\% & 1.2\,\% & 3660 \\
$\PH\to\textnormal{other}$ & & & \\
\midrule
$\epem\to\PQq\PAQq\PGn\PAGn$ & 788 & 97\,\% & 0.07\,\% & 841 \\
$\epem\to\PQq\PAQq\Pl\Pl$ & 2730 & 90\,\% & 0.005\,\% & 178 \\
$\epem\to\PQq\PAQq\Pl\PGn$ & 4310 & 67\,\% & 0.11\,\% & 4730 \\
$\PGg\Pepm\to\PQq\PAQq\Pepm$ & 88400 & 86\,\% & 0.0013\,\% & 1430 \\
%$\epem \to \PH\nuenuebar;$ & 18.9 & 100.0\,\% & 42.7\,\% & 12100 \\
%$\PH\to\PW\PW^\ast\to\PQq\PAQq\Pl\PGn$ & & & \\
%\midrule
%$\epem \to \PH\nuenuebar;$ & 25.6 & 100.0\,\% & 1.79\,\% & 687 \\
%$\PH\to\PW\PW^\ast\to\PQq\PAQq\PQq\PAQq$ & & & \\
%$\epem \to \PH\nuenuebar;$ & 199.6 & 99.6\,\% & 1.26\,\% & 3767 \\
%$\PH\to\textnormal{other}$ & & & \\
%\midrule
%$\epem\to\PQq\PAQq\PGn\PAGn$ & 788 & 96.6\,\% & 0.07\,\% & 777 \\
%$\epem\to\PQq\PAQq\Pl\Pl$ & 2726 & 89.8\,\% & <0.01\,\% & 257 \\
%$\epem\to\PQq\PAQq\Pl\PGn$ & 4310 & 66.4\,\% & 0.07\,\% & 4810 \\
%$\PGg\Pep(\PGg\Pem)\to\PQq\PAQq\Pep(\PQq\PAQq\Pem)$ & 88400 & 85.3\,\% & <0.01\,\% & 1580 \\
\bottomrule
\end{tabular}
\end{center}
\caption{Preselection and selection efficiencies for the signal and most important background processes in the $\PH\to\PW\PW^\ast\to\PQq\PAQq\Pl\PGn$ analysis. Numbers of events correspond to 1.5\,$\abinv$ at $\roots=1.4\,\TeV$. \label{tab:ww_fusion_hww_qqlv_numbers_1400}}
\end{table}

% Section 5.2.2
\subsection{$\PH \to \PZ\PZ^\ast$}
\label{sec:ww_fusion_zz}
The decay $\PH\to\PZ\PZst$ in $\epem\to\PH\nuenuebar$ events is 
studied using $\PZ^{(*)}\to\qqbar$ and $\PZ^{(*)}\to\Plp\Plm$ decays at $\roots = 1.4\,\TeV$ using the \clicild detector model. 
The experimental signature is two jets, a pair of oppositely charged leptons and 
missing $\pT$. The total invariant mass of all visible final-state particles is equal to the Higgs mass, 
while either the quarks or the charged lepton pair have a mass consistent with $\mZ$. 
Due to the large background from $\PH\to\PW\PW^*$, the $\PZ\PZst\to\PQq\PAQq\PQq\PAQq$ final 
state is not considered here. The $\PZ\PZst\to\Plp\Plm\Plp\Plm$ signature is not expected to be competitive 
at CLIC due to the small number of expected events and is not further considered.

\begin{table}[t]
\begin{center}
\begin{tabular}{lrrrr}
\toprule
Process & $\sigma/\fb$ & $\varepsilon_\text{presel}$ & $\varepsilon_\text{BDT}$ & $N_\text{BDT}$ \\
\midrule
$\epem \to \PH\nuenuebar;$ & 0.995 & 62\,\% & 46\,\% & 425 \\
$\PH\to\PZ\PZst\to\qqbar\Plp\Plm$ & & & \\
\midrule
$\epem \to \PH\nuenuebar;$ & 25.6 & 32\,\% & 0.2\,\% & 24 \\
$ \PH\to\PW\PW^{\ast}\to\qqbar\qqbar$ & & & \\
$\epem \to \PH\nuenuebar;\PH\to\PQb\PAQb$ & 137 & 20\,\% & 0.06\,\% & 23 \\
$\epem \to \PH\nuenuebar;\PH\to\Pg\Pg$ & 21 & 25\,\% & 0.05\,\% & 4 \\
$\epem \to \PH\nuenuebar;\PH\to\PQc\PAQc$ & 6.9 & 23\,\% & 0.0\,\% & 0 \\
$\epem \to \PH\nuenuebar;\PH\to\textnormal{other}$ & 51 & 50\,\% & 0.3\,\% & 98 \\
%$\epem \to \PH\nuenuebar;$ & 0.995 & 62\,\% & 46\,\% & 425 \\
%$\PH\to\PZ\PZst\to\qqbar\Plp\Plm$ & & & \\
%\midrule
%$\epem \to \PH\nuenuebar;$ & 25.6 & 32\,\% & 0.002\,\% & 24 \\
%$ \PH\to\PW\PW^{\ast}\to\qqbar\qqbar$ & & & \\
%$\epem \to \PH\nuenuebar;\PH\to\PQb\PAQb$ & 137 & 20.3\,\% & 0.0006\,\% & 23 \\
%$\epem \to \PH\nuenuebar;\PH\to\Pg\Pg$ & 21 & 25\,\% & 0.0005\,\% & 4 \\
%$\epem \to \PH\nuenuebar;\PH\to\PQc\PAQc$ & 6.9 & 23\,\% & 0.0\,\% & 0 \\
%$\epem \to \PH\nuenuebar;\PH\to\textnormal{other}$ & 51 & 50\,\% & 0.003\,\% & 98 \\
\bottomrule
\end{tabular}
\end{center}
\caption{\label{tab:ww_fusion_hzz_numbers} Preselection and selection efficiencies for the signal and the relevant background processes in the $\PH \to \PZ\PZst$ analysis. The numbers of events correspond to 1.5\,$\abinv$ at $\roots=1.4\,\TeV$. All background processes other than Higgs production are completely rejected by the event selection.}
\end{table}

The analysis is performed in several steps. First, 
isolated electrons and muons with an impact parameter of less than 0.02~mm 
are searched for. Hadronic \PGt lepton decays are identified using the \taufinder algorithm described in
\autoref{sec:higgsstrahlung_tautau}, with the requirement $\pT >
10\,\GeV$ for the seed track and $\pT > 4\,\GeV$ for all other tracks within
a search cone of 0.15\,radian. In signal events, 87\,\% of the electron or muon 
pairs and 37\,\% of the tau lepton pairs are found, including the effect of the 
geometrical acceptance of the detector in the forward direction.

In events with exactly two identified leptons of the same flavour and opposite charge, two jets are 
reconstructed from the remaining particles. No other preselection cuts are applied. 
Flavour tagging information is obtained from the \lcfiplus package.

A BDT classifier is used to suppress the background processes using 17 input variables, including:
\begin{itemize}
\item the invariant masses of the \PH, \PZ and \PZst candidates;
\item the topology of the hadronic system: $-\log_{10}(y_{34})$, \\ 
$-\log_{10}(y_{23})$ and $-\log_{10}(y_{12})$;
\item the b-tag and c-tag probabilities for both jets;
\item the visible energy and the missing transverse momentum of the event;
\item the number of particles in the event.
\end{itemize}

\begin{figure*}
  \centering
  \includegraphics[width=0.9\columnwidth]{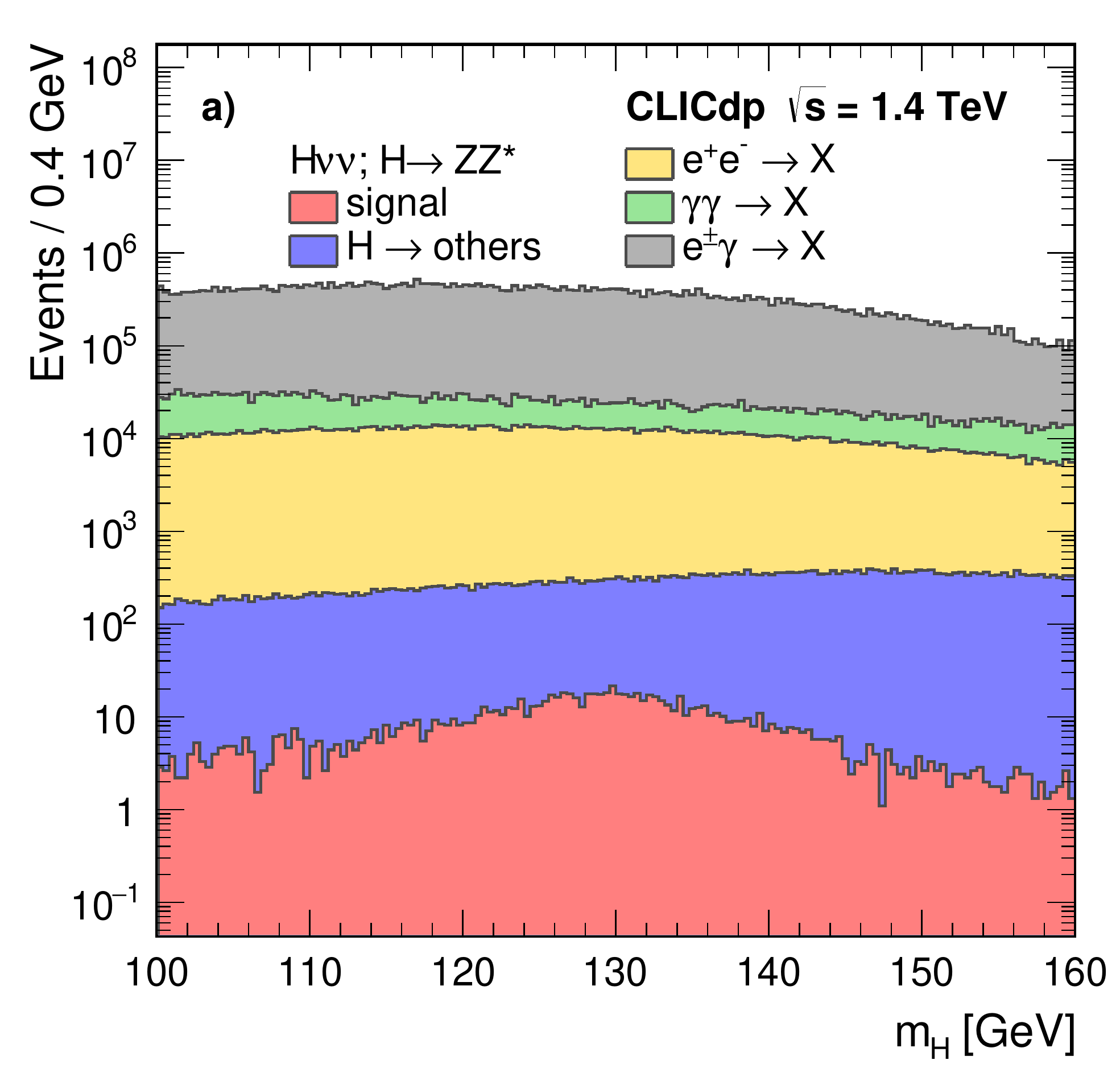}
  \hspace{0.1\columnwidth}
    \includegraphics[width=0.9\columnwidth]{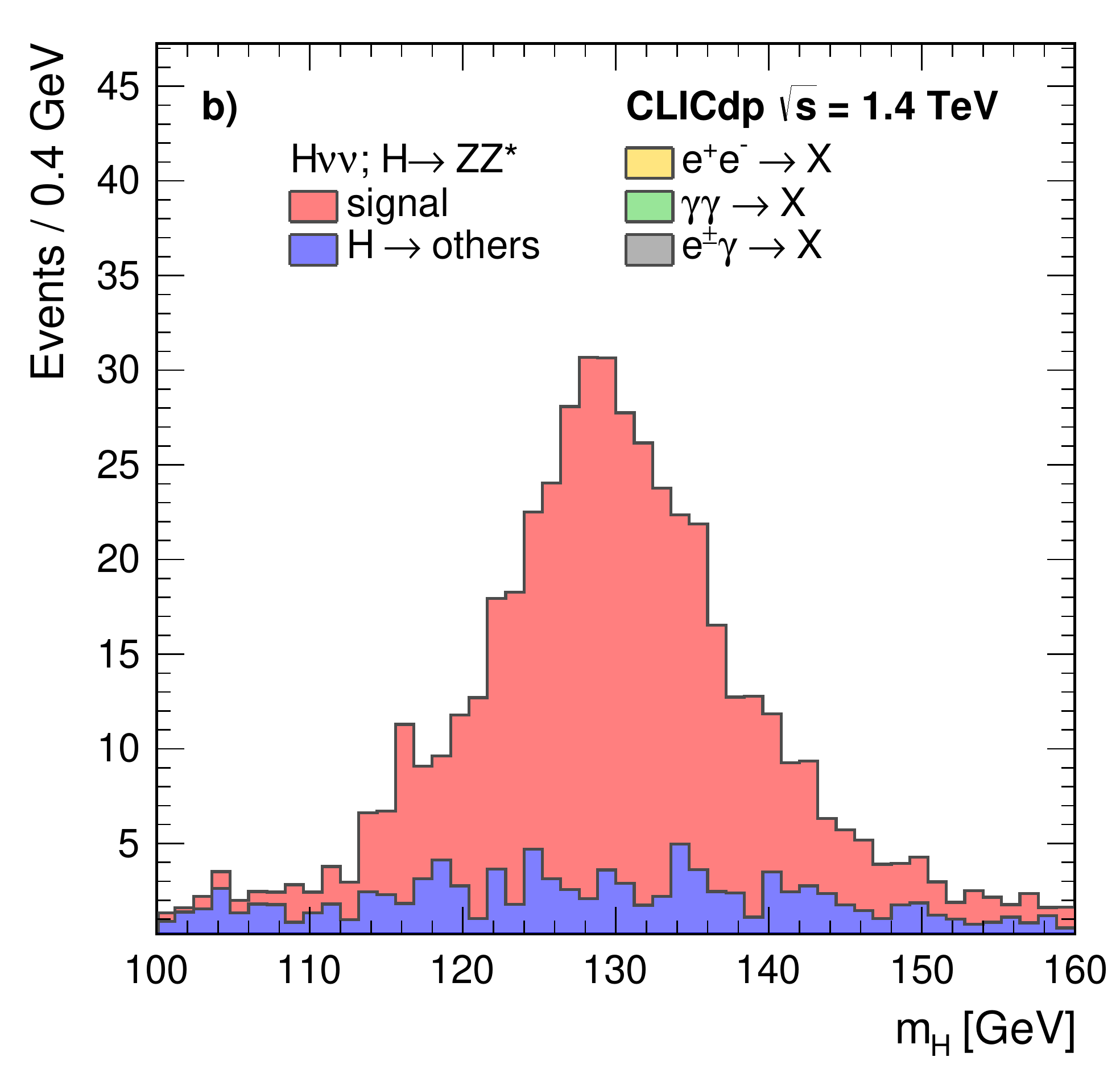}
  \caption{\label{fig:hzz} Reconstructed Higgs invariant mass distributions of $\PH\to\PZ\PZst\to\qq\lplm$ events at $\roots = 1.4\,\TeV$, showing the signal and main backgrounds as stacked histograms a) after preselection, and b) after the full event selection including a cut on the BDT classifier. The distributions are normalised to an integrated luminosity of $1.5\,\abinv$.}
\end{figure*}

The event selection is summarised in \autoref{tab:ww_fusion_hzz_numbers}. Only backgrounds from other Higgs decays pass the event selection, while all 
other background processes are fully rejected. The invariant mass distribution of the Higgs candidates in events with two isolated leptons after 
the full selection chain, including the BDT classifier, is shown in \autoref{fig:hzz}. The resulting statistical uncertainty is:
\begin{equation*}
\frac{\Delta[\sigma(\PH\nuenuebar)\times\BR(\PH\to\PZ\PZ^\ast)]}{\sigma(\PH\nuenuebar)\times\BR(\PH\to\PZ\PZ^\ast)} = 5.6\,\% \,.
\end{equation*}

% Section 5.3.1
\subsection{$\PH \to \PGg\PGg$}
\label{sec:ww_fusion_gammagamma}
\newcommand{\hgamgam}{\ensuremath{\PH\to\gamgam}\xspace}

The measurement of the $\PH\to\gamgam$ decay played a central role in the discovery of the Higgs boson at the LHC~\cite{Aad:2012tfa, Chatrchyan:2012xdj}. In the SM, this decay is induced via loops of heavy charged particles, with dominant contributions from \PW bosons and \PQt quarks. For BSM
scenarios, other heavy charged particles can appear in the loops, modifying the expected effective $\PH\to\PGg\PGg$ branching ratio. 
The sensitivity for the measurement of $\BR(\hgamgam)$ at CLIC has been studied using the \clicsid detector model for
$\roots = 1.4\,\TeV$ and an integrated luminosity of 1.5\,\abinv. The SM branching ratio for $\mH=126\,\GeV$ is 0.23\,\% which results in approximately 840 signal events. The experimental signature for $\epem\to\PH\nuenuebar;$ $\hgamgam$ is two high $\pT$ photons with invariant mass $m(\PGg\PGg)$ consistent with $\mH$, and missing momentum from the $\nuenuebar$ system.  
All relevant SM background processes with one or two photons in the final state have been considered. In addition to the photons from the hard interaction, the MC samples 
include additional ISR and FSR photons.

The following preselection cuts are applied to restrict the analysis
to relevant events. At least two reconstructed photons each with
energy $E_{\PGg} > 15\,\GeV$ and $\pT > 10\,\GeV$ are required. The
two highest energy photons passing these requirements are used to form
the $\PH$ candidate and the preselection requires an invariant mass
consistent with $\mH$, $115\,\GeV < m(\gamgam) < 140\,\GeV$. The
highest energy photon in the event is required to have $\pT>40\,\GeV$.
In addition, to remove contributions from FSR, both photons are
required to be isolated with no reconstructed particle with $\pT >
5\,\GeV$ within a cone of radius 500\,\mrad centred on the
photon. Furthermore, the remaining reconstructed energy after
excluding the Higgs candidate has to be less than 250\,\GeV. The 
cross sections and efficiencies of the preselection cuts for the signal and the main backgrounds are listed in
\autoref{tab:ww_fusion_hgamgam_channels}. At this stage in the event
selection the background dominates.

\begin{table}[t]
\begin{center}
\begin{tabular}{lrrrr}
\toprule
Process & $\sigma/\fb$ & $\varepsilon_\text{presel}$ & $\varepsilon_\text{BDT}$ & $N_\text{BDT}$ \\
\midrule
$\epem\to\PH\nuenuebar;\,\hgamgam$ & 0.56 & $85\,\%$ & $47\,\%$ & 337 \\
\midrule
$\epem \to \nunubar\PGg$    &  29.5 & $34\,\%$ & $7.3 \,\%$ & 1110 \\
$\epem \to \nunubar\gamgam$ &  17.3 & $31\,\%$ & $8.6 \,\%$ &  688 \\
$\epem \to \gamgam$         &  27.2 & $20\,\%$ & $0.68\,\%$ &   55 \\
$\epem \to \epem\PGg$       &   289 & $ 9.2\,\%$ & $0.66\,\%$ &  265 \\
$\epem \to \epem\gamgam$    &  12.6 & $ 5.2\,\%$ & $0.2\,\%$ &    2 \\
$\epem \to \qqbar\PGg$      &  67.0 & $ 0.8\,\%$ & $0.0 \,\%$ &    0 \\
$\epem \to \qqbar\gamgam$   &  16.6 & $ 1.4\,\%$ & $0.57\,\%$ &    2 \\
%$\epem\to\PH\nuenuebar;\,\hgamgam$ & 0.56 & $84.9\,\%$ & $40.4\,\%$ & 337 \\
%\midrule
%$\epem \to \nunubar\PGg$    &  29.5 & $34.2\,\%$ & $2.5 \,\%$ & 1110 \\
%$\epem \to \nunubar\gamgam$ &  17.3 & $31.0\,\%$ & $2.6 \,\%$ &  688 \\
%$\epem \to \gamgam$         &  27.2 & $19.8\,\%$ & $0.14\,\%$ &   55 \\
%$\epem \to \epem\PGg$       & 289.0 & $ 9.2\,\%$ & $0.06\,\%$ &  265 \\
%$\epem \to \epem\gamgam$    &  12.6 & $ 5.2\,\%$ & $0.01\,\%$ &    2 \\
%$\epem \to \qqbar\PGg$      &  67.0 & $ 0.8\,\%$ & $0.0 \,\%$ &    0 \\
%$\epem \to \qqbar\gamgam$   &  16.6 & $ 1.4\,\%$ & $0.01\,\%$ &    2 \\
\bottomrule
\end{tabular}
\end{center}
\caption{Signal and relevant background processes used in the $\hgamgam$ analysis. Additional photons from ISR and FSR are present in each sample. The cross sections for the backgrounds include cuts applied at generator level that are slightly looser than the preselection described in the text.  The numbers of events correspond to 1.5\,$\abinv$ at $\roots=1.4\,\TeV$.
\label{tab:ww_fusion_hgamgam_channels}}
\end{table}
%generator level cuts at https://edms.cern.ch/ui/file/1403842/1/LCWS_eva_sicking_higgs.pdf

\begin{figure}
\includegraphics[width=0.9\columnwidth]{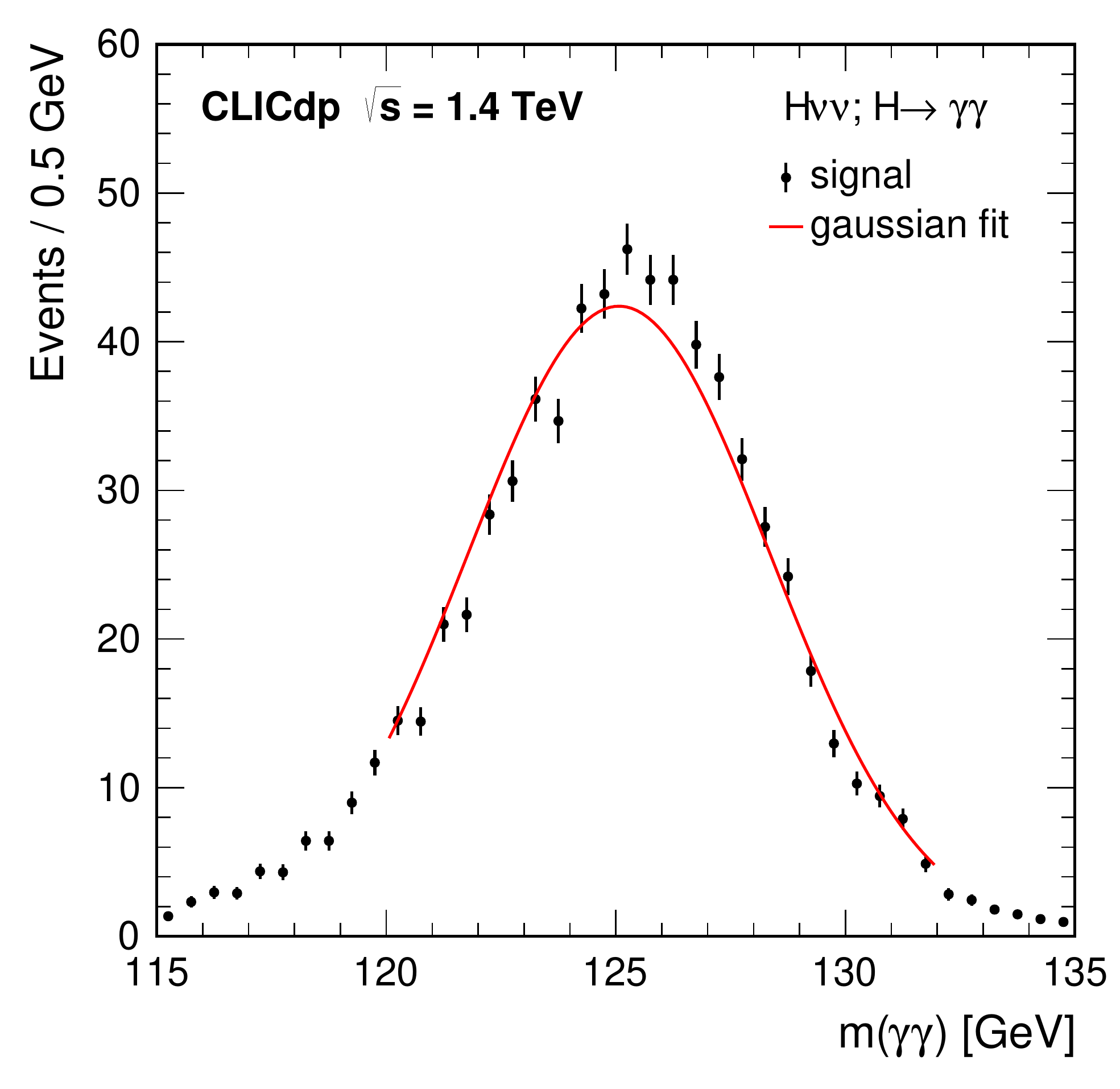}
\caption{Reconstructed di-photon invariant mass distribution of preselected signal \hgamgam events at $\roots=1.4\,\TeV$. The distribution is normalised to an integrated luminosity of $1.5\,\abinv$. The statistical uncertainties correspond to the size of the simulated event sample. The line shows the fit described in the text.}
\label{fig:ww_fusion_hgamgam_mass}
\end{figure}

To illustrate the photon reconstruction capabilities of the \clicsid detector concept, the invariant mass of Higgs candidates in signal events after the preselection is shown in \autoref{fig:ww_fusion_hgamgam_mass}. A fit to the distribution using a Gaussian function indicates a mass resolution in the signal sample of $\sigma = 3.3\,\GeV$.

The signal and background events are classified using a BDT. The 13 variables used to distinguish the signal from the backgrounds include:
\begin{itemize}
\item the invariant mass of the Higgs candidate;
\item kinematic properties of the Higgs candidate;
\item kinematic properties of the two photons;
\item the angle between the two photons and the helicity angle of the Higgs candidate;
\item the remaining reconstructed energy excluding the Higgs candidate. %The response of the BDT for all channels is shown in \autoref{fig:ww_fusion_hgamgam_bdt}.
\end{itemize}
For the optimal BDT cut,  the total signal selection efficiency is 40\%, corresponding to 337 selected signal events in 1.5\,\abinv. 
The event selection for the signal and the main backgrounds is summarised in \autoref{tab:ww_fusion_hgamgam_channels}, leading to a statistical uncertainty of: 
\begin{equation*}
\frac{\Delta[\sigma(\PH\nuenuebar)\times\BR(\PH\to\PGg\PGg)]}{\sigma(\PH\nuenuebar)\times\BR(\PH\to\PGg\PGg)} = 15\,\%\,.
\end{equation*}

% Section 5.3.2
\subsection{$\PH \to \PZ\PGg$}
\label{sec:ww_fusion_zgamma}
\newcommand{\HZG}{\ensuremath{\PH \to \PZ \PGg}}
\newcommand{\Zqq}{\ensuremath{\PZ \to \qqbar}}
\newcommand{\Zll}{\ensuremath{\PZ \to \lplm}}
\newcommand{\Zee}{\ensuremath{\PZ \to \epem}}
\newcommand{\Zmumu}{\ensuremath{\PZ \to \mpmm}}
\newcommand{\Ztt}{\ensuremath{\PZ \to \tptm}}
\newcommand{\CxTBr}{\ensuremath{\sigma(\epem \to \PH \nuenuebar) \times \BR(\HZG)}}

\begin{figure}
\centering
\includegraphics[width=0.45\textwidth]{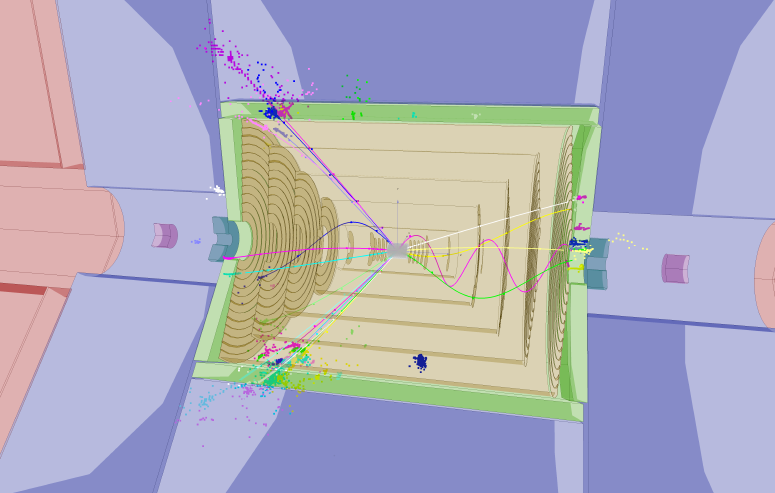}
\caption{Event display of a $\HZG \to \PQq\PAQq\PGg$ event at $\roots=1.4\,\TeV$ in the \clicsid detector. Both jets are visible. The photon creates a cluster in the central part of the electromagnetic calorimeter (blue).\label{fig:event_zgamma}}
\end{figure}

As is the case for $\PH\to\PGg\PGg$, at lowest order, the SM decay
$\HZG$ is induced by loops of heavy charged particles. Contributions
from BSM particles would lead to deviations from the SM expectation
for $\BR(\HZG)$. For $\mH=126\,\GeV$, the decay $\HZG$ is expected to
have a branching ratio of $\BR(\HZG)= 0.16\,\%$.  The potential to
measure $\CxTBr$ at CLIC has been studied at $\roots=1.4\,\TeV$ with
the \clicsid detector model, where 585 $\HZG$ events are expected
in 1.5\,\abinv of data~\cite{LCD:zgamma_1400}. For the purpose of the event selection, only
$\PZ\to\PQq\PAQq$ and $\PZ\to\Plp\Plm$ (with $\Pl = \Pe, \PGm$) are
useful, giving small event samples of 409 $\qqbar\PGg$, 21 $\epem\PGg$
and 21 $\mpmm\PGg$ events from $\HZG$ in 1.5\,\abinv at
$\roots=1.4\,\TeV$. A typical event display is shown in
\autoref{fig:event_zgamma}.

The visible final states of the signal channels $\qqbar\PGg$ or $\lplm\PGg$ are also produced in several background processes, some 
of which have much larger cross sections than the signal.  In addition to background with photons from the hard process,  
$\epem\to\qqbar$ or $\epem\to\lplm$ events with a FSR or ISR photon can mimic the signal.  

The $\HZG$ event selection requires at least one identified high-$\pT$ photon and either two electrons, muons or quarks
consistent with a $\PZ$ decay. The  photon with the highest energy in the event is identified. Events are 
considered as either $\epem\PGg$, $\mpmm\PGg$ or $\qq\PGg$ candidates. 
In the case where an $\epem$ or $\mpmm$ pair is found, photons nearly collinear with the lepton trajectories (within $0.3^\circ$) are combined with the leptons under the assumption that these photons originate from bremsstrahlung.
If neither an $\epem$ nor a $\mpmm$ pair is found, all reconstructed particles except for the photon of highest energy are clustered into two jets % assuming that the $\PZ$ decayed into two quarks, 
using a jet radius of $R=1.2$.
In all cases, the selected $\PZ$ decay candidate and the highest energy photon are combined to form the $\PH$ candidate.

In order to reduce the number of background process events, two selection steps are performed.
First, preselection cuts are applied: the Higgs candidate daughter photon and jets, electrons, or muons are only accepted if they have an energy of $E > 20\,\GeV$ and $\pT >15\,\GeV$. In the $\qq\PGg$ channel, only jets with at least 5 particles are considered in order to suppress hadronic $\PGt$ decays. In addition, the reconstructed $\PZ$ and $\PH$ masses in the event are required to be consistent with a $\HZG$ decay.
The second step in the event selection is three BDT selections (one for each signal final state).
The input variables are the properties of the reconstructed $\PH$, $\PZ$, and $\PGg$ such as mass, energy, momentum, and polar angle, event shapes such as sphericity and aplanarity, as well as missing energy distributions and particle multiplicity distributions.

\begin{table}[h]
\begin{center}
\begin{tabular}{lrrrr}
\toprule
Process                        &  $\sigma/\fb$  & $\epsilon_{\text{presel}}$  & $\epsilon_{\text{BDT}}$  & $N_{\text{BDT}}$ \\
\midrule
 $\epem \to \PH \nuenuebar;$ & 0.27 & 45\,\% & 41\,\% & 75 \\
  \ $\PH\to\PZ\PGg;\,\Zqq$ & & & & \\
\midrule
 $\epem\to\nunubar\qq\PGg$      & 37.3           & 12\,\%   & 7.3\,\%        & 504  \\
 $\epem\to\nunubar\qq$              & 122         &  8.4\,\%    & 3.0\,\%        & 463  \\
 $\Pepm\PGg\,\to\Pepm\qq$         & $978$  & 2.4\,\%     & 0.2\,\% &   70    \\
\bottomrule
 $\epem \to \PH \nuenuebar;$ & 0.014 & 38\,\% & 50\,\% & 4 \\
  \ $\PH\to\PZ\PGg;\,\Zee$ & & & & \\
\midrule
 $\epem\to\nunubar\lplm\PGg$ & 9.6 & 1.6\,\% & 6.5\,\% & 15 \\ 
 $\epem\to\nunubar\lplm$ & 23.3 & 1.0\,\% & 34\,\% & 12 \\ 
 $\Pepm\PGg\, \to\Pepm\lplm$ & $1940$ & 0.22\,\% & 0.1\,\% & 7 \\
\bottomrule
 $\epem \to \PH \nuenuebar;$ & 0.014 & 54\,\% & 44\,\% & 5 \\
  \ $\PH\to\PZ\PGg;\,\Zmumu$ & & & & \\
\midrule
 $\epem\to\nunubar\lplm\PGg$ & 9.6 & 1.2\,\% & 8.1\,\% & 14 \\ 
 $\epem\to\nunubar\lplm$ & 23.3 & 0.45\,\% & 8.3\,\% & 13 \\ 
 $\Pepm\PGg\, \to\Pepm\lplm$ & $1940$ & 0.27\,\% & 1.1\,\% & 9 \\
% $\epem \to \PH \nuenuebar;$ & 0.27 & 45.1\,\% & 18.2\,\% & 75 \\
%  \ $\PH\to\PZ\PGg;\,\Zqq$ & & & & \\
%\midrule
% $\epem\to\nunubar\qq\PGg$      & 37.3           & 12.3\,\%   & 0.9\,\%        & 504  \\
% $\epem\to\nunubar\qq$              & 121.8         &  8.4\,\%    & 0.2\,\%        & 463  \\
% $\Pepm\PGg\,\to\Pepm\qq$         & $977.8$  & 2.4\,\%     & <0.01\,\% &   70    \\
%\bottomrule
% $\epem \to \PH \nuenuebar;$ & 0.014 & 38.3\,\% & 17.3\,\% & 4 \\
%  \ $\PH\to\PZ\PGg;\,\Zee$ & & & & \\
%\midrule
% $\epem\to\nunubar\lplm\PGg$ & 9.6 & 1.6\,\% & 0.10\,\% & 15 \\ 
% $\epem\to\nunubar\lplm$ & 23.3 & 1.0\,\% & 0.03\,\% & 12 \\ 
% $\Pepm\PGg\, \to\Pepm\lplm$ & $1942.1$ & 0.22\,\% & <0.01\,\% & 7 \\
%\bottomrule
% $\epem \to \PH \nuenuebar;$ & 0.014 & 53.7\,\% & 24.5\,\% & 5 \\
%  \ $\PH\to\PZ\PGg;\,\Zmumu$ & & & & \\
%\midrule
% $\epem\to\nunubar\lplm\PGg$ & 9.6 & 1.2\,\% & 0.10\,\% & 14 \\ 
% $\epem\to\nunubar\lplm$ & 23.3 & 0.45\,\% & 0.04\,\% & 13 \\ 
% $\Pepm\PGg\, \to\Pepm\lplm$ & $1942.1$ & 0.27\,\% & <0.01\,\% & 9 \\
\bottomrule
\end{tabular}
\caption{
\label{tab:zg:background}
Preselection and selection efficiencies for $\HZG$ events in all three considered \PZ decay channels. The cross sections for the backgrounds include kinematic cuts applied at generator level.
All numbers assume an integrated luminosity of $1.5\,\abinv$ at $1.4\TeV$.}
\end{center}
\end{table}

For the optimal BDT cuts, expected statistical significances of 2.2, 0.54 and 0.78 (in units of standard deviations) are found for the $\qq\PGg$, $\epem\PGg$ and $\mpmm\PGg$ channels respectively. 
The signal selection efficiencies and contributions from the most important backgrounds are summarised in \autoref{tab:zg:background}.
When the results from all three channels are combined, the expected statistical precision at $\roots=1.4\,\TeV$ for an integrated luminosity of 1.5\,\abinv is:
\begin{equation*}
\frac{\Delta[\sigma(\PH\nuenuebar)\times\BR(\HZG)]}{\sigma(\PH\nuenuebar)\times\BR(\HZG)} = 42\,\%\,.
\end{equation*}
With electron polarisation the statistical precision can be increased, for
example with $-80\,\%$ electron polarisation, $\Delta [ \CxTBr ] \approx
31\,\%$.  Further gains are expected at higher
centre-of-mass energies, as the Higgs production cross section at
$\roots=3\,\TeV$ is 70\,\% higher than at 1.4\,\TeV.

% Section 5.3.3
\subsection{$\PH \to \mpmm$}
\label{sec:ww_fusion_mumu}
The measurement of the rare $\PH \to \mpmm$ decay is challenging due to the very low SM branching ratio of $2\times10^{-4}$. 
In $\epem\to\PH\nuenuebar$ production, the signature for  $\PH \to \mpmm$ decay is a $\mpmm$ pair with invariant mass consistent with $\mH$ 
and missing momentum. The efficient rejection of background relies on the excellent detector momentum resolution, which directly influences the width of the reconstructed di-muon invariant mass peak. 
Signal and background events have been simulated at $\roots=1.4\,\TeV$ and $3\,\TeV$ using the $\clicild$ and $\clicsid$ detector models respectively~\cite{Grefe:2012gj, Milutinovic-Dumbelovic:2015fba}. In contrast with other studies presented in this paper, an electron beam polarisation of $-80\,\%$ is assumed owing to the very small branching ratio for the $\PH\to\mpmm$ decay. The two analyses were performed independently. They follow the same strategy but differ in some of the observables that are used in the event selection.

The most important background processes include $\mpmm\nunubar$ in the final state, as shown in \autoref{tab:Hvv:mumu1.4} for $1.4\,\TeV$ and in \autoref{tab:Hvv:mumu3} for $3\,\TeV$. A significant fraction of these events are also produced from interactions involving beamstrahlung photons. Another important background is $\epem\to\epem\mpmm$, where both electrons are usually emitted at very low polar angles and thus might not be detected. Tagging of these low angle electrons in the very forward calorimeters---LumiCal and BeamCal---is essential to keep this background under control.

The event selection requires two reconstructed, oppositely charged muons with a di-muon invariant mass within the relevant mass region of $105-145\,\GeV$. Events with one or more detected high-energy electrons ($E>200\,\GeV$ at $1.4\,\TeV$, $E>250\,\GeV$ at $3\,\TeV$) in the very forward calorimeters are vetoed. This introduces the possibility of vetoing signal events if they coincide with Bhabha scattering events. The $\epem\to\epem$ cross section is sufficiently high that the probability of such a coincidence within 20 bunch crossings ($10\,\text{ns}$) is about $7\,\%$ in both analyses. The cuts on the minimum energy and the minimum polar angle for vetoing forward electrons need to be chosen carefully. $\epem\to\epem\mpmm$ and $\Pepm\PGg\to\Pepm\mpmm$ events need to be rejected efficiently while a low probability for coincidence with Bhabha scattering events needs to be maintained.

The $3\,\TeV$ analysis includes some additional preselection cuts to remove phase space regions that do not include any signal events. These cuts reject events that contain a reconstructed non-muon object with an energy greater than $100\,\GeV$; in addition, events containing electrons in the central region of the detector with an energy above $20\,\GeV$ are also rejected.
% for any reconstructed non-muon object and a maximum energy of $20\,\GeV$ for reconstructed electrons in the central parts of the detector. 
 The sum of the transverse momenta of the two muons, $\pT(\PGmm) + \pT(\PGmp)$, is required to be above $50\,\GeV$ and the transverse momentum of the di-muon system should be above $25\,\GeV$.

The final event selection uses a BDT classifier using various kinematic variables, excluding the invariant mass of the di-muon system. The $1.4\,\TeV$ analysis uses the visible energy of the event after removal of the di-muon system $E_{\text{vis}}$, the transverse momentum of the di-muon system $\pT(\PGm\PGm)$, the sum of the transverse momenta of the two muons $p_T(\PGmm) + p_T(\PGmp)$, the polar angle of the di-muon system $\theta_{\PGm\PGm}$, the 
boost of the di-muon system $\beta_{\PGm\PGm}$, and the cosine of the helicity angle $\cos{\theta^{\ast}}$. The $3\,\TeV$ analysis uses the energy of the hardest non-muon object instead of the total visible energy and also includes the energy, transverse momentum, polar angle and azimuthal angle of both individual muons. This event selection reduces background from four-fermion processes by several orders of magnitude, while maintaining an overall signal selection efficiency of $\epsilon = 30.5\,\%$ and $\epsilon = 26.3\,\%$ at $1.4\,\TeV$ and $3\,\TeV$ respectively.

\begin{table}[h]
\centering
\begin{tabular}{ l  c  c  c  c }
\hline
Process & $\sigma/\fb$ & $\epsilon_\text{presel}$ & $\epsilon_\text{BDT}$ & $N_\text{BDT}$ \\
\hline
$\epem \to \PH\nuenuebar;\,\PH\to\mpmm$ & 0.094   & 83\,\% & 37\,\%   &  43 \\ \hline
$\epem \to \nuenuebar \mpmm$                   & 232      &  1.1\,\% &  27\,\%   & 1030 \\ %
$\Pepm\PGg\to\Pepm\PGnGm\PAGnGm\mpmm$                & 35       &  8.5\,\% & 1.3\,\%   & 57 \\ %
$\PGg\PGg\to\PGnGm\PAGnGm\mpmm$                  & 162      & 10.6\,\% & 2.2\,\%   & 560 \\ %
%$\epem \to \PH\nuenuebar;\,\PH\to\mpmm$ & 0.094   & 82.5\,\% & 30.5\,\%   &  43 \\ \hline
%$\epem \to \nuenuebar \mpmm$                   & 232      &  1.1\,\% &  0.30\,\%   & 1030 \\ %
%$\Pepm\PGg\to\Pepm\PGnGm\PAGnGm\mpmm$                & 35       &  8.5\,\% & 0.11\,\%   & 57 \\ %
%$\PGg\PGg\to\PGnGm\PAGnGm\mpmm$                  & 162      & 10.6\,\% & 0.23\,\%   & 560 \\ %
\hline
\end{tabular}
\caption{\label{tab:Hvv:mumu1.4} The signal and main backgrounds in the $\PH\to\mpmm$ analysis at $\roots=1.4\,\TeV$ with the corresponding cross sections. The numbers of selected events assume an integrated luminosity of $1.5\,\abinv$ and $-80\,\%$ polarisation of the electron beam. Other processes, including $\epem\to\mpmm$ and $\Pepm\PGg\to\Pepm\mpmm$, contribute a total of less than 10 events to the final selection.}
\end{table}
\begin{table}[h]
\centering
\begin{tabular}{ l  c  c  c  c }
\hline
Process & $\sigma/\fb$ & $\epsilon_\text{presel}$ & $\epsilon_\text{BDT}$ & $N_\text{BDT}$ \\
\hline
$\epem \to \PH\nuenuebar$;\,$\PH\to\mpmm$ & 0.16   &  64\,\%  &  41\,\%   &    84 \\ \hline
$\epem \to \nuenuebar \mpmm$             & 6.6    &  33\,\%  &  41\,\%   &  1797 \\ %
$\Pepm\PGg\to\Pepm\mpmm$                 & 1210   &  6.9\,\% & 0.16\,\% &   262 \\ %
$\PGg\PGg\to\PGnGm\PAGnGm\mpmm$          & 413    &  4.3\,\% & 0.50\,\% &   176 \\ %
%$\epem \to \PH\nuenuebar$;\,$\PH\to\mpmm$ & 0.16   &  64\,\%  &  26\,\%   &    84 \\ \hline
%$\epem \to \nuenuebar \mpmm$             & 6.6    &  33\,\%  &  14\,\%   &  1797 \\ %
%$\Pepm\PGg\to\Pepm\mpmm$                 & 1210   &  6.9\,\% & 0.011\,\% &   262 \\ %
%$\PGg\PGg\to\PGnGm\PAGnGm\mpmm$          & 413    &  4.3\,\% & 0.021\,\% &   176 \\ %
\hline
\end{tabular}
\caption{\label{tab:Hvv:mumu3} The signal and most important background processes in the $\PH\to\mpmm$ analysis at $\roots=3\,\TeV$ with the corresponding cross sections. The numbers of selected events assume an integrated luminosity of $2\,\abinv$ and $-80\,\%$ polarisation of the electron beam. All other processes contribute of the order of 10 events to the final event selection. The cross sections are calculated for events with invariant mass of the di-muon system between $100\,\GeV$ and $140\,\GeV$.}
\end{table}

\begin{figure}
	  \centering
      \includegraphics[width=0.9\columnwidth]{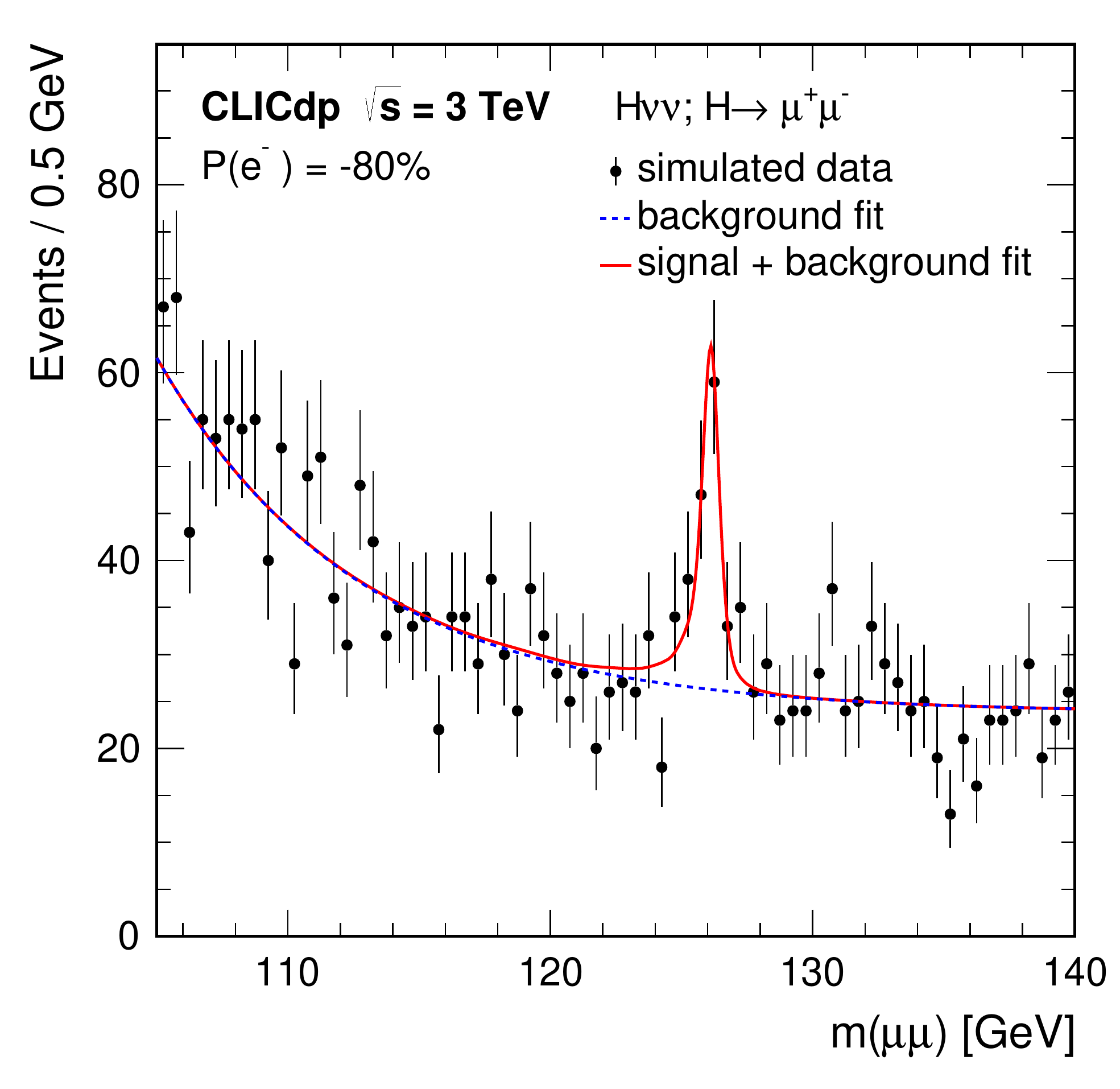}
      \caption{Reconstructed di-muon invariant mass distribution of selected $\PH\to\mpmm$ events at $\roots=3\,\TeV$. The simulated data are shown as dots while the solid line represents the fit function described in the text. The dotted line shows the background contribution of the fit function. The distribution is normalised to an integrated luminosity of $2\,\abinv$, assuming $-80\,\%$ electron polarisation.}
      \label{fig:Hvv:mumu}
  \end{figure}

The number of signal events is extracted from the reconstructed invariant mass distribution after the event selection, as shown in \autoref{fig:Hvv:mumu}. Using a large MC sample, the signal and background shapes are extracted. The signal is described by a Gaussian distribution with asymmetric exponential tails. The combined background is parameterised as the sum of an exponential and a constant function. To assess the expected statistical precision, a large number of trial samples are generated from the expected reconstructed mass distributions of signal and background and are then fitted to the signal and background components. For $P(\Pem) = -80\,\%$, the expected relative uncertainty on the $\sigma(\epem\to\PH\nuenuebar) \times \BR(\PH\to\mpmm)$ is $27\,\%$, corresponding to a significance of $3.7$, at $1.4\,\TeV$, and $19\,\%$, corresponding to a significance of $5.2$, at $3\,\TeV$. The corresponding uncertainties for unpolarised beams are:
\begin{align*}
\frac{\Delta[\sigma(\PH\nuenuebar)\times\BR(\PH\to\mpmm)]}{\sigma(\PH\nuenuebar)\times\BR(\PH\to\mpmm)} & = 38\,\% \ \text{at} \ 1.4\,\TeV\,, \\
\frac{\Delta[\sigma(\PH\nuenuebar)\times\BR(\PH\to\mpmm)]}{\sigma(\PH\nuenuebar)\times\BR(\PH\to\mpmm)} & = 25\,\% \ \text{at} \ 3\,\TeV\,. \\
\end{align*}

% Section 6
\section{ZZ-fusion} % Aidan
\label{sec:zz_fusion}

Higgs boson production through the $t$-channel fusion of two $\PZ$
bosons, $\epem \to \PH\epem$, is analogous to the $\WW$-fusion process
but gives access to $\gHZZ$ and $\gHbb$ using a complementary technique.  At
$\roots=1.4\,\TeV$, $\ZZ$-fusion is the sub-leading Higgs production
process, with a cross section of around $25\,\fb$, which is 10\,\% of
that for the $\WW$-fusion process. The potential for the measurement of the 
$\ZZ$-fusion process has been investigated at $\sqrt{s}=1.4\,\TeV$ 
using the \clicild detector.

The characteristic signature of the $\ZZ$-fusion process is two
scattered beam electrons reconstructed in the forward regions of the
detector, plus the Higgs boson decay products.  Here, the scattered
beam electrons are required to be fully reconstructed, and the final
state $\PH\to\bb$ is considered.

\begin{figure}[hbt]
\centering
\includegraphics[width=0.9\columnwidth]{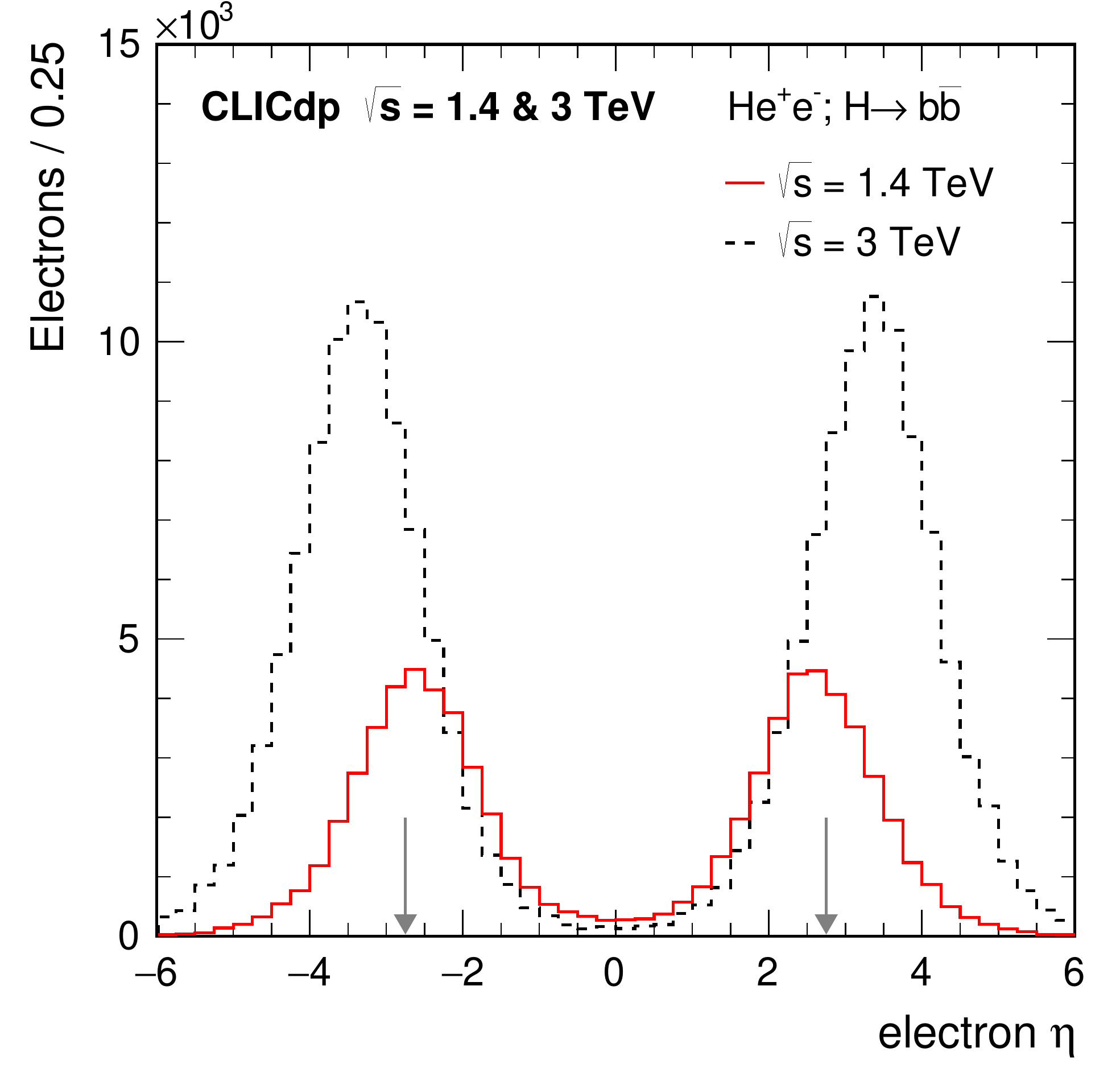}
\caption{Generated electron pseudorapidity ($\eta=-\ln\tan\frac{\theta}{2}$) distributions for $\epem \to \PH\epem$ events at $\roots=1.4\,\TeV$ and $3\,\TeV$. The distributions are normalised to~$1.5\,\abinv$ and $2\,\abinv$ respectively. The vertical arrows show the detector acceptance.\label{zzFusionFig1}}
\end{figure}

Events are clustered into a four-jet topology using a $k_{t}$ exclusive
clustering algorithm with $R=1.0$.  For a well-reconstructed signal
event, two of the resulting `jets' are expected to be the
reconstructed electrons, and the remaining two jets originate from the Higgs
decay to $\bb$.  The event selection requires two oppositely-charged
electron candidates, separated by $|\Delta\eta| > 1$, each with
$E>100\,\GeV$.  This preselection preserves 27\,\% of the
$\epem\to\PH\epem\to\bb\epem$ signal (3.6\,fb), with the lost events
almost entirely due to the scattered electrons falling outside the
detector acceptance, as shown in \autoref{zzFusionFig1}.  After the
preselection, the SM background consists mainly of events that have
two real electrons and a $\qqbar$ pair, either from the continuum or
from the decay of $\PZ$ bosons.  Although the preselection suppresses
98\,\% of the $\epem\to\qq\epem$ background, the accepted cross
section is $48\,\fb$, which is thirteen times larger than that for the
remaining signal.  A further requirement that one of the two jets
associated with the Higgs decay has a $\PQb$-tag value $> 0.4$
preserves 80\,\% of the remaining signal and rejects 80\,\% of the
remaining background.

A relative likelihood classifier $\mathcal{L}_1$, which treats
$\ZZ$-fusion events with $\PH\to\bb$ as signal and $\PH\to\WW^*$ and
$\PH\to\ZZ^*$ as background, is used to reduce contributions from
other Higgs decays.  Seven variables are used to construct the
likelihood: the jet clustering variable $y_{45}$; the invariant mass
of the two jets associated with the Higgs decay; the visible mass of
the event with the scattered beam electrons removed; the higher of the
$\PQb$-tag values of the two jets associated with the Higgs decay; the
$\PQc$-tag value corresponding to the same jet; and the b-c-separation
returned by the tagger, for both Higgs decay jets.  Requiring a high
signal likelihood, $\mathcal{L}_1 > 0.8$, reduces the $\PH\to\bb$
signal to 3000 events but leaves only 90 events from other
Higgs decays, while also reducing the non-Higgs backgrounds to
4700 events.

Finally, to separate the signal from all backgrounds, a further
relative likelihood classifier $\mathcal{L}_2$ is constructed using
four variables that provide separation power between signal and
background: the opening between the reconstructed electrons $\Delta
R$; the recoil mass of the event determined from the momenta of the
reconstructed electrons, $m_\text{rec}$; the jet clustering variable
$y_{34}$; and the invariant mass of the two jets associated with the
Higgs decay.

\begin{figure}[hbt]
\centering{
\includegraphics[width=0.9\columnwidth]{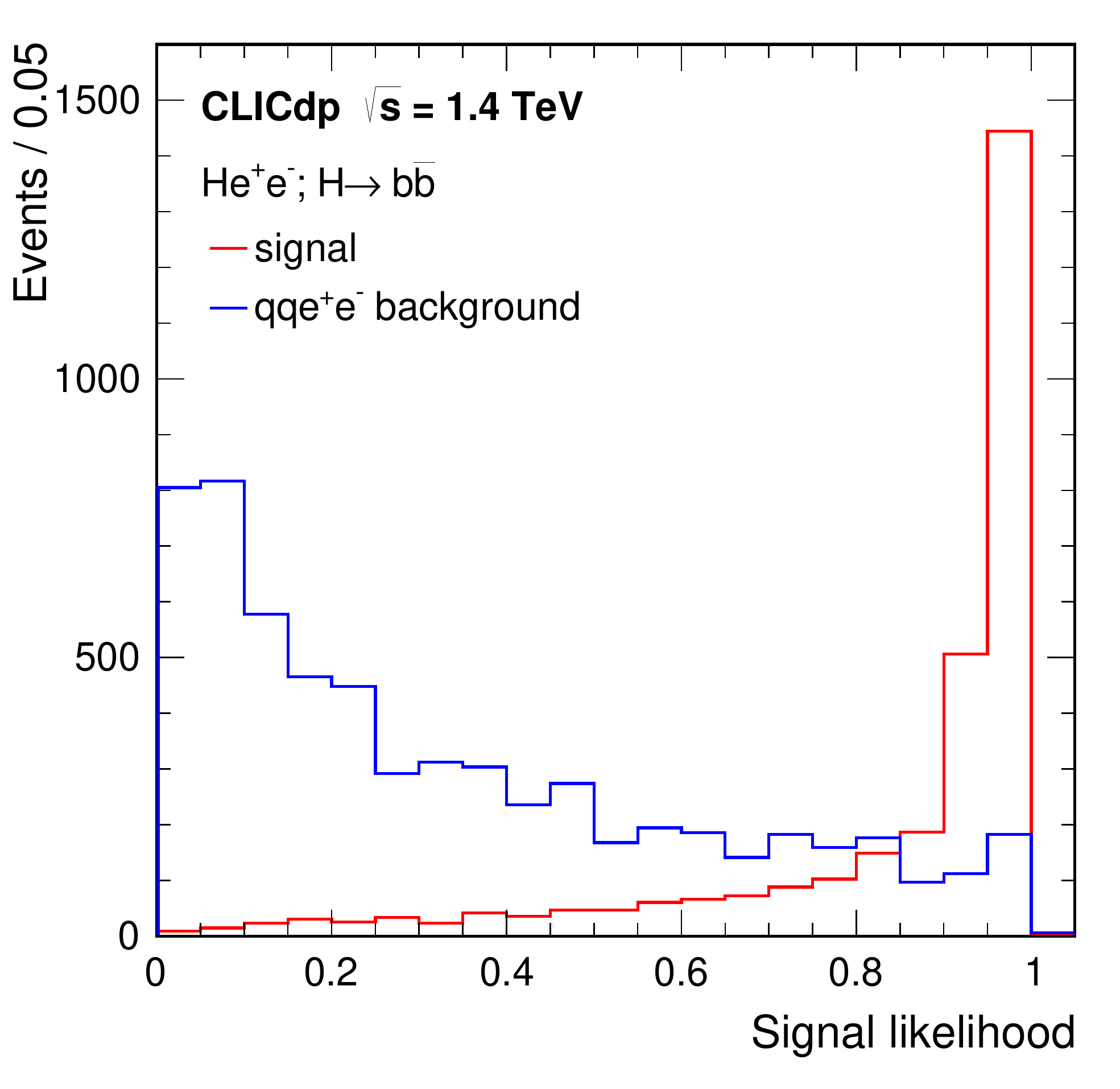}
}

\caption{Likelihood distributions for $\PH \to \bb$ events in the ZZ-fusion analysis at $\roots=1.4\,\TeV$, shown for the signal and main background. The distributions are normalised to an integrated luminosity of $1.5\,\abinv$. \label{zzFusionFig2}}
\end{figure}

The resulting likelihood is shown in \autoref{zzFusionFig2} %\autoref{zzFusionFig2}.
and gives good separation between signal and background.
The likelihood distribution is fitted by signal and background components (where the normalisation is allowed to
vary), giving:
\begin{align*}
\frac{\Delta[\sigma(\PH\epem)\times\BR{(\PH\to\PQb\PAQb)}]}{\sigma(\PH\epem)\times\BR{(\PH\to\PQb\PAQb)}} = 1.8\,\%
\end{align*}
for $1.5\,\abinv$ at $\roots=1.4\,\TeV$.

% Section 7
\section{Top Yukawa Coupling}
\label{sec:top_yukawa}

At an $\epem$ collider the top Yukawa coupling, $y_{\PQt}$, can be determined from the 
production rate in the process where a Higgs boson is produced in association with a top quark pair,
$\epem\to\PQt\PAQt\PH$. The top quarks decay almost exclusively by $\PQt\to\PQb\PW$. The 
signal event topology thus depends on the nature of the $\PW$ and Higgs boson decays. Here 
$\PH\to\PQb\PAQb$ decays have been studied for
two $\PQt \PAQt \PH$  decay channels at $\roots=1.4\,\TeV$ using the \clicsid detector model~\cite{LCD:tth_1400, LCD:tth_backgrounds_1400}:
\begin{itemize}
\item the fully-hadronic channel (where both $\PW$ bosons decay hadronically), giving a $\PQt \PAQt \PH$ final state of eight jets, including four $\PQb$ jets;
\item the semi-leptonic channel (where one $\PW$ boson decays leptonically), giving a $\PQt \PAQt \PH$ final state of six jets (four $\PQb$ jets), one lepton and one neutrino,
\end{itemize}

The two channels are distinguished by first searching for isolated leptons (muons and electrons with an energy of at least $15\,\GeV$ and tau candidates from \taufinder containing a track with $\pT > 10\,\GeV$).
If zero leptons are found, the event is classified as fully-hadronic. If one lepton is found, the event is classified as semi-leptonic. Events in which more than one lepton is found are not analysed further. 
The \kT algorithm is used to cluster the particles of each event into a specific number of jets, and remove particles arising from beam-beam interactions that are closer to the beam axis than to a hard jet as described in \autoref{sec:simreco}.
Events classified as fully-hadronic are clustered into eight jets.  In semi-leptonic events, the lepton is removed 
and the remaining particles are clustered into six jets. A semi-leptonic event is shown in \autoref{fig:tth:event}. The particles not clustered into jets by the \kT algorithm are removed from the event and the remaining particles are then re-clustered using the $\epem$ Durham algorithm in \lcfiplus, which performs flavour tagging for each jet, and prevents particles from displaced vertices being split between two or more jets.
The jets are combined to form candidate primary particles in such a way so as to minimise a $\chi^{2}$ function expressing the consistency
of the reconstructed di- and tri-jet invariant masses with the $\PQt\PAQt(\PH\to\PQb\PAQb)$ hypothesis. For example, in the case of
the semi-leptonic channel, the jet assignment with the minimum of:
\begin{equation*}
    \chi^{2} = \frac{(m_{ij}-m_{\PW})^{2}}{\sigma_{\PW}^{2}} + \frac{(m_{ijk} - m_{\PQt})^{2}}{\sigma_{\PQt}^{2}} + \frac{(m_{lm} - m_{\PH})^{2}}{\sigma_{\PH}^{2}} \,,
\end{equation*}
gives the \PW, top and Higgs candidates, where $m_{ij}$ is the invariant mass of the jet pair 
used to reconstruct the $\PW$ candidate, $m_{ijk}$ is the invariant mass of the three jets used to 
reconstruct the top quark candidate and $m_{lm}$ is the invariant mass of the jet pair used to reconstruct the 
Higgs candidate. The expected invariant mass resolutions $\sigma_{\PW, \PQt, \PH}$ were estimated from
combinations of two or three reconstructed jets matched to $\PW$, top and Higgs particles on generator level.

\begin{figure}
\begin{center}
\includegraphics[width=\columnwidth]{top_yukawa/event_tth_1400.pdf}
\caption{Event display of a $\PQt\PAQt\PH \to \PQb\PAQb\PQb\PAQb\PQq\PAQq\PGtprm\PAGnGt$ event at $\roots = 1.4\,\TeV$ in the \clicsid detector. The tau lepton decays hadronically.\label{fig:tth:event}}
\end{center}
\end{figure}

\begin{figure}
\begin{center}
\includegraphics[width=0.9\columnwidth]{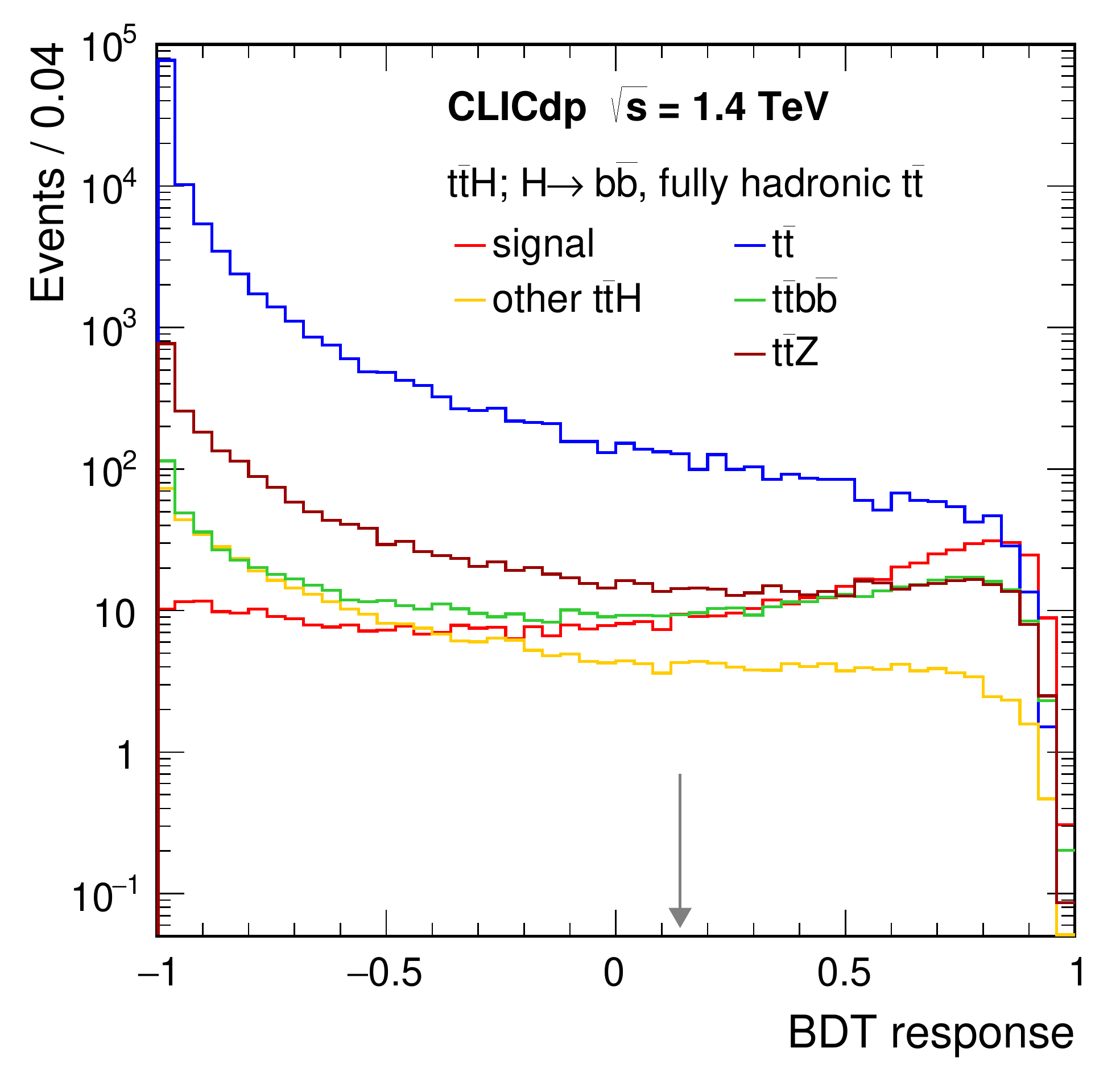}
\caption{BDT classifier distributions for fully-hadronic $\PQt\PAQt\PH$ events at $\roots=1.4\,\TeV$, shown for the $\PQt\PAQt\PH$ signal and main backgrounds. The distributions are normalised to an integrated luminosity of $1.5\,\abinv$. The vertical arrow shows the value of the cut, chosen to give the highest significance.\label{fig:tth:bdt}}
\end{center}
\end{figure}

Having forced each event into one of the two signal-like topologies, multivariate BDT classifiers (one for fully-hadronic events and one for semi-leptonic events) are used to separate signal and background. The discriminating variables include: kinematic quantities such as the reconstructed
Higgs mass, the visible energy in the jets and the missing $\pT$; angular variables such as the angles between the Higgs decay products in the rest frame of the Higgs candidate with respect to its flight direction 
and the angle between the momenta of the top and Higgs candidates; event variables such as thrust, sphericity and the number of particles in the event; and flavour tag variables for the four most likely b-jets. As an example, the BDT response distributions for the fully-hadronic channel are shown in \autoref{fig:tth:bdt}. The selection is chosen to maximise the signal significance. 
The expected numbers of selected events for $1.5\,\abinv$ of $\roots=1.4\,\TeV$ data  are listed in \autoref{tab:tth:sel_eff}. The contributions from other investigated background processes were found to be negligible.  The $\PQt \PAQt \PH$ cross section can be measured with an accuracy of $12\,\%$ in the semi-leptonic channel and $11\,\%$ in the hadronic channel. The combined precision of the two channels is $8\,\%$.

\begin{table}[htbp]
\centering
\begin{tabular}{lrrr}
\toprule
Process & Events & \multicolumn{2}{c}{Selected as} \\
              & in $1.5\,\abinv$ & HAD & SL \\
\midrule
$\epem\to\PQt \PAQt \PH$, 6 jet, $\PH \to \PQb \PAQb$ & 647 & 357 & 9  \\ % 2435
$\epem\to\PQt \PAQt \PH$, 4 jet, $\PH \to \PQb \PAQb$ & 623 & 62 & 233  \\  %2441
\midrule
$\epem\to\PQt \PAQt \PH$, 2 jet, $\PH \to \PQb \PAQb$ & 150 & 1 & 20  \\ % 2429
$\epem\to\PQt \PAQt \PH$, 6 jet, $\PH \not\to \PQb \PAQb$ & 473 & 38 & 8  \\ % 2438
$\epem\to\PQt \PAQt \PH$, 4 jet, $\PH \not\to \PQb \PAQb$ & 455 & 5 & 19  \\ % 2444
$\epem\to\PQt \PAQt \PH$, 2 jet, $\PH \not\to \PQb \PAQb$ & 110 & 0 & 1  \\ % 2432
$\epem\to\PQt \PAQt \PQb \PAQb$, 6 jet & 824 & 287 & 8  \\ % 2423
$\epem\to\PQt \PAQt \PQb \PAQb$, 4 jet & 794 & 44 & 175  \\ % 2426
$\epem\to\PQt \PAQt \PQb \PAQb$, 2 jet & 191 & 1 & 14  \\ % 2420
$\epem\to\PQt \PAQt \PZ$, 6 jet & 2,843 & 316 & 12  \\ % 2450
$\epem\to\PQt \PAQt \PZ$, 4 jet & 2,738 & 49 & 170  \\ % 2453
$\epem\to\PQt \PAQt \PZ$, 2 jet & 659 & 1 & 13  \\ % 2447
$\epem\to\PQt \PAQt$ & 203,700 & 1,399 & 523  \\ % 2417
$\epem\to\PQq\PQq\PQq\PQq\Pl\PGn (\text{non-}\PQt \PAQt)$ & 68,300 & 11 & 70  \\
$\epem\to\PQq\PQq\PQq\PQq$ & $2.0 \times 10^{6}$ & 195 & 0 \\
\bottomrule
\end{tabular}
\caption{Expected numbers of signal and background events in the fully-hadronic (HAD) and semi-leptonic (SL) channels 
for $1.5\,\abinv$ at $\roots=1.4\,\TeV$. The columns show the total numbers of events before selection and the numbers of events 
passing the fully-hadronic and semi-leptonic BDT selections. No preselection is applied in the analysis. \label{tab:tth:sel_eff}}
\end{table}

To translate the measurement of the $\PQt\PAQt\PH$ cross section into a measurement of the top Yukawa coupling, a correction is applied to take into account the contribution from the \higgsstrahlung diagram, where the Higgs boson is radiated off the intermediate $\PZ$ boson in $\epem\to\PQt\PAQt$~\cite{Djouadi:1991tk, Djouadi:1992gp}.
To evaluate the small degradation in sensitivity, the $\whizard$ program is used to
calculate the cross section for the 
inclusive process $\Pep \Pem \to \PQt \PAQt \PH$ as a function of the value of the top Yukawa coupling. 
The factor required to translate the measured cross section uncertainty into a coupling uncertainty is determined from the slope of the cross section at the SM value of the top Yukawa coupling, and is found to be:
\begin{equation*}
\frac{\Delta y_{\PQt}}{y_{\PQt}} = 0.53 \frac{\Delta \sigma}{\sigma} \, , \nonumber
\end{equation*}
which is slightly larger than the factor of 0.50 expected without the \higgsstrahlung diagram.
Thus, the expected precision on the top Yukawa coupling is: 
\begin{equation*}
          \frac{\Delta y_{\PQt}}{y_{\PQt}} =    4.2\,\% \,, 
\end{equation*}
for $1.5\,\abinv$ of data at $\roots=1.4\,\TeV$ without beam polarisation. This value is expected to improve to about $4.0\,\%$ for the same amount of 
data collected using the $P(\Pem) = -80\,\%$ polarisation configuration~\cite{Price:2014oca}. Since the cross section for the $\PQt\PAQt\PH$ cross section falls with increasing $\roots$ (see \autoref{fig:higgs:cross}), the precision with $2\,\abinv$ at $3\,\TeV$ is not expected to be better than the result presented here.

% Section 8
\section{Double Higgs Production}
\label{sec:higgs_self_coupling}

\begin{figure}[htb]
\unitlength = 1mm
\centering
\vspace{4mm}
\begin{fmffile}{triple_higgs/eevvh}
\begin{fmfgraph*}(25,20)
\put(-2,10){\large{\bf a)}}
\fmfstraight
\fmfleft{i1,i2,i3}
\fmfright{o1,o2,o3,o4}
\fmflabel{$\Pe^{-}$}{i1}
\fmflabel{$\Pe^{+}$}{i3}
\fmflabel{\PH}{o2}
\fmflabel{\PH}{o3}
\fmflabel{\PGne}{o1}
\fmflabel{\PAGne}{o4}
\fmf{fermion}{i1,v1}
\fmf{fermion}{v2,i3}
\fmf{fermion}{v1,o1}
\fmf{fermion}{o4,v2}
\fmf{photon,label.side=left,label=$\PW^{*}$}{v1,v3}
\fmf{photon,label.side=left,label=$\PW^{*}$}{v3,v2}
\fmf{dashes,tension=2.0,label=\PH}{v3,v4}
\fmfv{label.angle=90,decor.shape=circle,decor.filled=full,decor.size=2thick}{v4}
\fmf{dashes}{v4,o2}
\fmf{dashes}{v4,o3}
\fmf{phantom,tension=0.75}{i2,v3}
\end{fmfgraph*}
\end{fmffile}
\hspace{15mm}
\begin{fmffile}{triple_higgs/HHWWnunu}
\begin{fmfgraph*}(25,20)
\put(-2,10){\large{\bf b)}}
\fmfstraight
\fmfleft{i1,i2,i3}
\fmfright{o1,o2,oinv,o3,o4}
\fmflabel{$\Pe^{-}$}{i1}
\fmflabel{$\Pe^{+}$}{i3}
\fmflabel{\PH}{o2}
\fmflabel{\PH}{o3}
\fmflabel{\PGne}{o1}
\fmflabel{\PAGne}{o4}
\fmf{fermion}{i1,v1}
\fmf{fermion}{v1,o1}
\fmf{fermion}{v2,i3}
\fmf{fermion}{o4,v2}
\fmf{photon,label.side=left,label=$\PW^{*}$}{v1,v3}
\fmf{photon,label.side=left,label=$\PW^{*}$}{v3,v2}
\fmf{dashes}{v3,o2}
\fmf{dashes}{v3,o3}
\fmf{phantom,tension=1.2}{i2,v3}
\end{fmfgraph*}
\end{fmffile} \\
\vspace{15mm}
\begin{fmffile}{triple_higgs/HWHnunu}
\begin{fmfgraph*}(25,20)
\put(-2,10){\large{\bf c)}}
\fmfstraight
\fmfleft{i1,i2,i3,i4}
\fmfright{o1,o2,oinv,o3,o4}
\fmflabel{$\Pe^{-}$}{i1}
\fmflabel{$\Pe^{+}$}{i4}
\fmflabel{\PH}{o2}
\fmflabel{\PH}{o3}
\fmflabel{\PGne}{o1}
\fmflabel{\PAGne}{o4}
\fmf{fermion}{i1,v1}
\fmf{fermion}{v1,o1}
\fmf{fermion}{v2,i4}
\fmf{fermion}{o4,v2}
\fmf{photon,label.side=left}{v1,v3}
\fmf{photon,label.side=left,label=$\PW^{*}$}{v3,v4}
\fmf{photon,label.side=left}{v4,v2}
\fmf{dashes}{v3,o2}
\fmf{dashes}{v4,o3}
\fmf{phantom,tension=0.75}{i2,v3}
\fmf{phantom,tension=0.75}{i3,v4}
\end{fmfgraph*}
\end{fmffile}
\hspace{15mm}
\begin{fmffile}{triple_higgs/HWHnunuXed}
\begin{fmfgraph*}(25,20)
\put(-2,10){\large{\bf d)}}
\fmfstraight
\fmfleft{i1,i2,i3,i4}
\fmfright{o1,o2,oinv,o3,o4}
\fmflabel{$\Pe^{-}$}{i1}
\fmflabel{$\Pe^{+}$}{i4}
\fmflabel{\PH}{o2}
\fmflabel{\PH}{o3}
\fmflabel{\PGne}{o1}
\fmflabel{\PAGne}{o4}
\fmf{fermion}{i1,v1}
\fmf{fermion}{v1,o1}
\fmf{fermion}{v2,i4}
\fmf{fermion}{o4,v2}
\fmf{photon,label.side=left}{v1,v3}
\fmf{photon,label.side=left,label=$\PW^{*}$}{v3,v4}
\fmf{photon,label.side=left}{v4,v2}
\fmf{phantom}{v3,o2}
\fmf{phantom}{v4,o3}
\fmf{phantom,tension=0.75}{i2,v3}
\fmf{phantom,tension=0.75}{i3,v4}
\fmffreeze
\fmf{dashes}{v3,o3}
\fmf{dashes}{v4,o2}
\end{fmfgraph*}
\end{fmffile}

\vspace{3mm}
\caption{Feynman diagrams of leading-order processes that produce two Higgs bosons and missing energy at CLIC at $\roots=1.4\,\TeV$ and $3\,\TeV$. The diagram (a) is sensitive to the trilinear Higgs self-coupling $\lambda$. The diagram (b) is sensitive to the quartic coupling $g_{\PH\PH\PW\PW}$. All four diagrams are included in the generated $\epem\to\PH\PH\nuenuebar$ signal samples.}
\label{fig:higgs_diagram}
\end{figure}
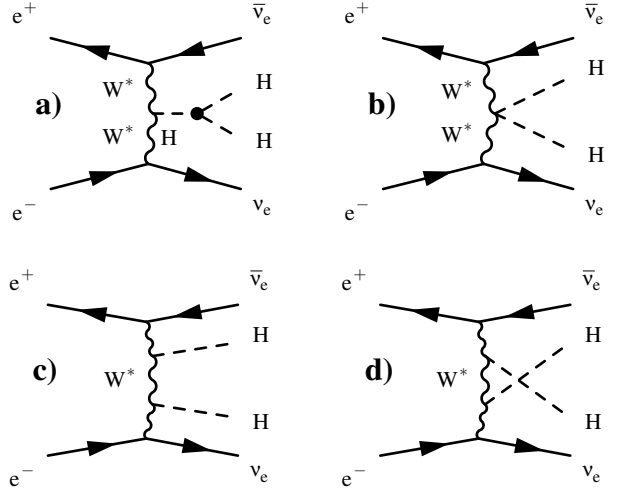

In $\epem$ collisions at high energy, double Higgs production,
$\epem\to\PH\PH\PGne\PAGne$, can occur through the processes shown
in \autoref{fig:higgs_diagram}. Despite the small cross section
(0.15\,fb and 0.59\,fb for CLIC operated at $\roots=1.4\,\TeV$ and
$3\,\TeV$, respectively), measurements of the double Higgs production
rate can be used to extract the Higgs boson trilinear self-coupling
parameter $\lambda$, that determines the shape of the fundamental
Higgs potential. BSM physics scenarios can introduce deviations of
$\lambda$ from its SM value of up to tens of
percent~\cite{Gupta:2013zza}. 
The physics potential for the measurement of this coupling has been studied 
using the \clicild detector model for $1.5\,\abinv$ of data at $\roots=1.4\,\TeV$ and 
for $2\,\abinv$ of data at $\roots=3\,\TeV$.
The process $\epem\to\PH\PH\epem$ has not been included as its cross section is 
about an order of magnitude smaller compared to $\epem\to\PH\PH\PGne\PAGne$.

Two signatures for $\epem\to\PH\PH\nuenuebar$ production are considered in the following: $\PH\PH\to\bb\bb$ and $\PH\PH\to\bb\PW\PW^*\to\bb\qqbar\qqbar$. All events without isolated leptons are considered for the analysis. These events are clustered into four jets using the \kT algorithm. Flavour tagging information is obtained from the \lcfiplus package. Events where the sum of the b-tag values of the four jets is smaller than 2.3 and the hadronic system fulfills the requirement $-\log_{10}(y_{34}) < 3.7 (3.6)$ at 1.4\,\TeV (3\,\TeV) are considered as $\bb\PW\PW^*$ candidates, while all other events are considered as $\bb\bb$ candidates. The following steps of the analysis are performed separately for the two final states.

At 1.4\,\TeV, a cut on the sum of the four b-tag values of at least 1.5 is imposed for $\bb\bb$ candidate events. Those events with a sum of the four b-tag values less than 2.3 are required to have a sum of the jet energies of at least 150\,\GeV and a second highest jet transverse momentum of at least 25\,\GeV. A cut on the sum of the four b-tag values of at least 2.3 is imposed for all events at 3 TeV. The jets are grouped into two Higgs boson candidates by minimising $|m_{ij}-m_{kl}|$, where $m_{ij}$ and $m_{kl}$ are the invariant masses of the jet pairs used to reconstruct the Higgs candidates. For events passing the preselection cuts, at both energies BDT classifiers with the same 10 input variables are used to suppress the backgrounds further.

For the $\bb\PW\PW^*$ final state, the events are re-clustered into six jets. These jets are then grouped into $\PW$ and $\PH$ candidates by minimising:
\begin{align*}
\chi^2=\frac{(m_{ij}-m_{\PH})^{2}}{\sigma_{\PH\to\bb}^2}+\frac{(m_{klmn}-m_{\PH})^{2}}{\sigma_{\PH\to\PW\PW^*}^2}+\frac{(m_{kl}-m_{\PW})^{2}}{\sigma_{\PW}^2},
\end{align*}
where $m_{ij}$ and $m_{klmn}$ are the jet combinations used to reconstruct the Higgs candidates, $m_{kl}$ is the invariant mass of the jet pair used to reconstruct the \PW candidate and $\sigma_{\PH\to\bb}$, $\sigma_{\PH\to\PW\PW^*}$, $\sigma_{\PW}$ are the estimated invariant mass resolutions for the reconstruction of $\PH\to\bb$, $\PH\to\PW\PW^*$ and $\PW$ decays. Events with an invariant mass of the two \PH boson candidates above 150\,\GeV are considered further. At 3\,\TeV a highest b-tag value of at least 0.7 is required while at 1.4\,\TeV the second highest b-tag values has to be larger than 0.2 and the visible transverse momentum has to be larger than 30 GeV. After this preselection, BDT classifiers using 32 input variables are used to suppress the backgrounds further.

The event selections for both studies at 1.4\,\TeV and 3\,\TeV are summarised in \autoref{tab:double_higgs_1400} and \autoref{tab:double_higgs_3000}, respectively. Combining the expected precisions on the cross sections for both signatures leads to:
\begin{align*}
\frac{\Delta[\sigma(\PH\PH\nuenuebar)]}{\sigma(\PH\PH\nuenuebar)} & = 44\,\% \ \text{at} \ 1.4\,\TeV\,, \\
\frac{\Delta[\sigma(\PH\PH\nuenuebar)]}{\sigma(\PH\PH\nuenuebar)} & = 20\,\% \ \text{at} \ 3\,\TeV\,. \\
\end{align*}

\begin{table}[tb]
  \centering
  \begin{tabular}{lrrrr}
      \toprule 
      Process  & \tabt{$\sigma/\text{fb}$}   & \tabt{$\varepsilon_\text{presel}$}   & \tabt{$\varepsilon_{\text{BDT}}$}  & \tabt{$N_{\text{BDT}}$} \\ 
      \midrule
      $\PH\PH\nuenuebar;\,\PH\PH\to\bb\bb$    &  0.047  & 94\,\%     & 24\,\%      & 16  \\
      \midrule
      $\PH\PH\nuenuebar;\,\PH\PH\to\textnormal{other}$    &  0.102  & 29\,\%     & 0.77\,\%      & 0.3  \\
%       $\epem \to \qqbar\qqbar$                              &  xx      &    xx\,\%   & xx\,\%   & 0 \\                                                                               
       $\epem \to \qqbar\qqbar\nunubar$                              &   23     &    6.2\,\%   & 0.38\,\%   & 8 \\
%       $\epem \to \qqbar\qqbar\lplm$                              &  xx      &    xx\,\%   & 0\,\%   & 0 \\                                                                           
       $\epem \to \qqbar\qqbar\Pl\PGn$                              &  110      &    16\,\%   & 0.03\,\%   & 7 \\
%       $\epem \to \qqbar\nunubar$                              &  xx      &    xx\,\%   & 0\,\%   & 0 \\                                                                              
%       $\epem \to \qqbar\Pl\PGn$                              &  xx      &    xx\,\%   & 0\,\%   & 0 \\                                                                               
%       $\epem \to \qqbar\lplm$                              &  xx      &    xx\,\%   & 0\,\%   & 0 \\                                                                                 
%       $\epem \to \qqbar$                              &  xx      &    xx\,\%   & 0\,\%   & 0 \\                                                                                      
       $\epem \to \qqbar\PH\nunubar$            &  1.5      &    39\,\%   & 2.0\,\%   & 18 \\
%       $\egam \to \Pepm\qqbar\qqbar$            &  xx   &    xx\,\%   & 0\,\%   & 0 \\                                                                                                
       $\egam \to \PGn\qqbar\qqbar$            &  154   &    13\,\%   & 0.01\,\%   & 3 \\
       $\egam \to \PQq\PQq\PH\nu$            &  30   &    28\,\%   & 0.01\,\%   & 1 \\
      \midrule
      $\PH\PH\nuenuebar;\,\PH\PH\to\bb\PW\PW^*$;      &  0.018       &    60\,\% & 8.2\,\% & 1.3 \\   
      $\ww\to\qqqq$ & & & & \\
      \midrule
      $\PH\PH\nuenuebar;\,\PH\PH\to\bb\bb$       &  0.047              &    15\,\% & 0.5\,\% & 0.1  \\   
      $\PH\PH\nuenuebar;\,\PH\PH\to\textnormal{other}$      &  0.085              &    20\,\% & 1.7\,\% & 0.5 \\   
%       $\epem \to \qqbar\qqbar$                 &  1245                &    0.2\,\% & 0.2\,\% & 3 \\   
       $\epem \to \qqbar\qqbar\nunubar$         &  23                 &    17\,\% & 0.002\,\% & 0.1 \\   
%       $\epem \to \qqbar\qqbar\lplm$           &  62                &    xx\,\%   & 0\,\%   & 0 \\   
       $\epem \to \qqbar\qqbar\Pl\PGn$          &  110                &    10\,\% & 0.01\,\%   & 2 \\   
%       $\epem \to \qqbar\nunubar$              &  788               &    xx\,\%   & 0\,\%   & 0 \\   
%       $\epem \to \qqbar\Pl\PGn$               &  4310               &    xx\,\%   & 0\,\%   & 0 \\   
%       $\epem \to \qqbar\lplm$                 &  2726               &    xx\,\%   & 0\,\%   & 0 \\   
%       $\epem \to \qqbar$                      &  4010               &    xx\,\%   & 0\,\%   & 0 \\   
       $\epem \to \qqbar\PH\nunubar$            &  1.5               &    35\,\%   & 0.1\,\%   & 0.8 \\   
%       $\egam \to \Pepm\qqbar\qqbar$           &  2891               &    xx\,\%   & 0\,\%   & 0 \\   
       $\egam \to \PGn\qqbar\qqbar$             &  154                &    22\,\%   & 0.0045\,\%   & 2 \\   
       $\egam \to \PQq\PQq\PH\nu$          &  30                &    27\,\%   & 0.02\,\%   & 3 \\   
%      \midrule
%       $\PGg\PGg \to \qqbar\qqbar$             &  30212              &    xx\,\%   & 0\,\%   & 0 \\   
    \bottomrule

  \end{tabular}
\caption{Preselection and selection efficiencies for the double Higgs signal and most important background processes in both considered decay channels at $\roots=1.4\,\TeV$. The numbers of events correspond to $1.5\,\abinv$. Contributions from all other backgrounds are found to be negligibly small. \label{tab:double_higgs_1400}}
\end{table}

\begin{table}[tb]
  \centering
  \begin{tabular}{lrrrr}
      \toprule 
      Process  & \tabt{$\sigma/\text{fb}$}   & \tabt{$\varepsilon_\text{presel}$}   & \tabt{$\varepsilon_{\text{BDT}}$}  & \tabt{$N_{\text{BDT}}$} \\ 
      \midrule
      $\PH\PH\nuenuebar;\,\PH\PH\to\bb\bb$    &  0.19  & 66\,\%     & 24\,\%      & 61  \\   
      \midrule
      $\PH\PH\nuenuebar;\,\PH\PH\to\textnormal{other}$    &  0.40  & 5.4\,\%     & 3.2\,\%      & 1  \\   
       $\epem \to \qqbar\qqbar$                              &  547      &    0.16\,\%   & 0.16\,\%   & 3 \\   
       $\epem \to \qqbar\qqbar\nunubar$                              &  72      &    1.8\,\%   & 0.68\,\%   & 17 \\   
       $\epem \to \qqbar\qqbar\Pl\PGn$                              &  107      &    1.8\,\%   & 0.15\,\%   & 6 \\   
       $\epem \to \qqbar\PH\nunubar$            &  4.7      &    18\,\%   & 3.0\,\%   & 50 \\   
       $\egam \to \PGn\qqbar\qqbar$            &  523   &    1.2\,\%   & 0.09\,\%   & 11 \\   
       $\egam \to \PQq\PQq\PH\nu$            &  116   &    2.7\,\%   & 0.14\,\%   & 9 \\   
      \midrule
      $\PH\PH\nuenuebar;\,\PH\PH\to\bb\PW\PW^*$;      &  0.07       &    62\,\% & 12\,\% & 10 \\   
      $\ww\to\qqqq$ & & & & \\
      \midrule
      $\PH\PH\nuenuebar;\,\PH\PH\to\bb\bb$       &  0.19              &    19\,\% & 1.5\,\% & 1  \\   
      $\PH\PH\nuenuebar;\,\PH\PH\to\textnormal{other}$      &  0.34              &    20\,\% & 3.6\,\% & 5 \\   
       $\epem \to \qqbar\qqbar$                 &  547                &    1.4\,\% & 0.01\,\% & 1 \\   
       $\epem \to \qqbar\qqbar\nunubar$         &  72                 &    9.0\,\% & 0.05\,\% & 6 \\   
%       $\epem \to \qqbar\qqbar\lplm$           &  169                &    xx\,\%   & 0\,\%   & 0 \\   
       $\epem \to \qqbar\qqbar\Pl\PGn$          &  107                &    7.3\,\% & 0.05\,\%   & 8 \\   
%       $\epem \to \qqbar\nunubar$              &  1318               &    xx\,\%   & 0\,\%   & 0 \\   
%       $\epem \to \qqbar\Pl\PGn$               &  5561               &    xx\,\%   & 0\,\%   & 0 \\   
%       $\epem \to \qqbar\lplm$                 &  3320               &    xx\,\%   & 0\,\%   & 0 \\   
%       $\epem \to \qqbar$                      &  2950               &    xx\,\%   & 0\,\%   & 0 \\   
       $\epem \to \qqbar\PH\nunubar$            &  4.8                &    32\,\%   & 0.6\,\%   & 19 \\   
%       $\egam \to \Pepm\qqbar\qqbar$           &  3112               &    xx\,\%   & 0\,\%   & 0 \\   
       $\egam \to \PGn\qqbar\qqbar$             &  523                &    15\,\%   & 0.04\,\%   & 67 \\   
       $\egam \to \PQq\PQq\PH\nu$          &  116                &    27\,\%   & 0.2\,\%   & 140 \\   
%      \midrule
%       $\PGg\PGg \to \qqbar\qqbar$             &  18297              &    xx\,\%   & 0\,\%   & 0 \\     
    \bottomrule

  \end{tabular}
\caption{Preselection and selection efficiencies for the double Higgs signal and most important background processes in both considered decay channels at $\roots=3\,\TeV$. The numbers of events correspond to $2\,\abinv$. Contributions from all other backgrounds are found to be negligibly small. \label{tab:double_higgs_3000}}
\end{table}

\begin{figure}[hbp]
    \centering
    \includegraphics[width=0.9\columnwidth]{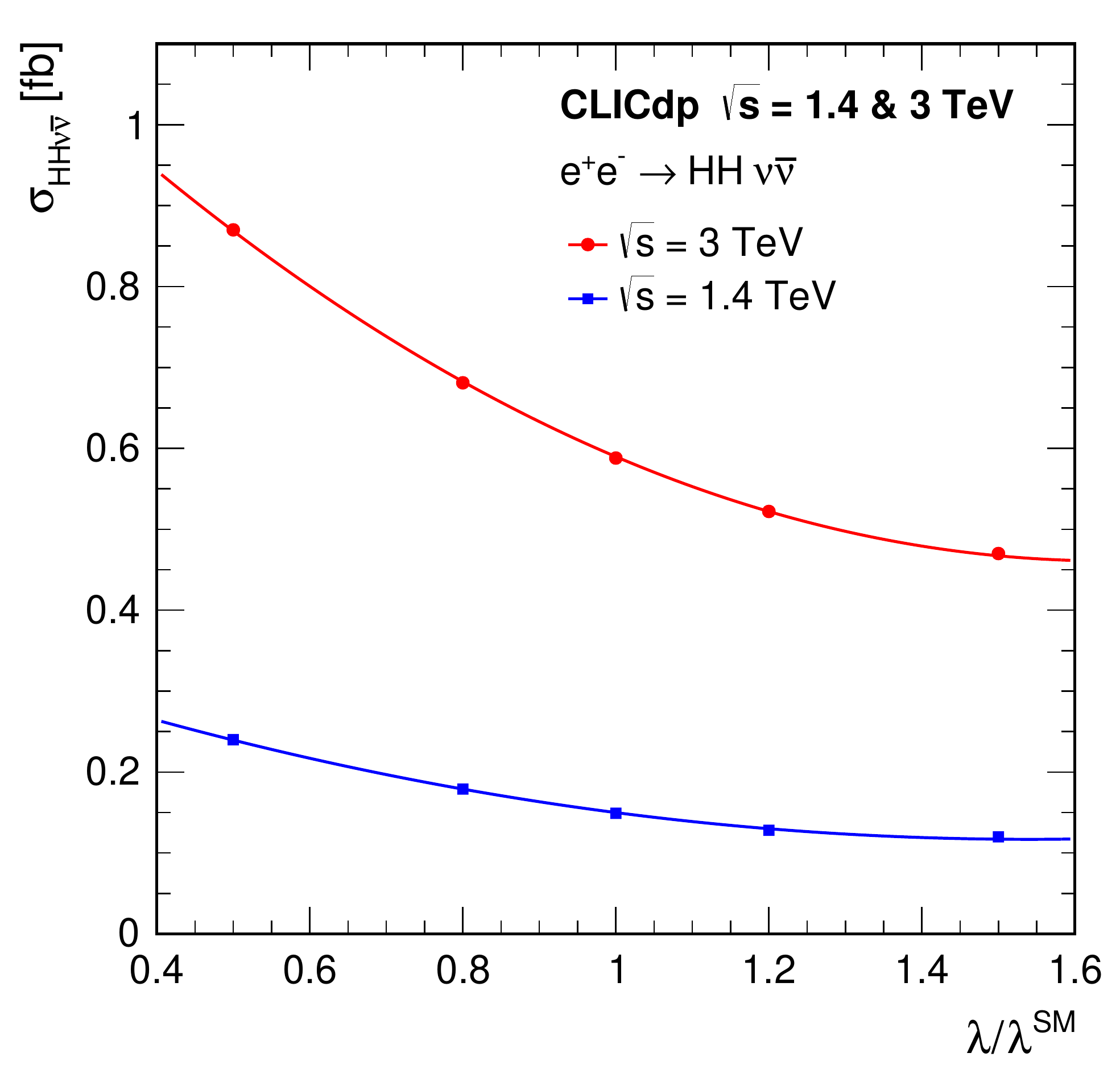}
    \caption{Cross section for the $\epem\to\PH\PH\PGne\PAGne$ process as a function of the ratio $\ghThree/\ghThree^\text{SM}$ at $\roots=1.4\,\TeV$ and $3\,\TeV$.}
    \label{fig:self_coupling_ratio}
\end{figure}

The double Higgs production cross section is sensitive to the trilinear Higgs self-coupling $\ghThree$. Since diagrams not
involving $\ghThree$ also contribute to the $\epem\to\PH\PH\nuenuebar$ process, their effect must be taken into account.
The relation between the relative uncertainty on the cross section and the relative uncertainty of the Higgs trilinear coupling
can be approximated as:
\begin{equation*}
\frac{\Delta \ghThree}{\ghThree} \approx \kappa \cdot \frac{\Delta[\sigma(\PH\PH\nuenuebar)]}{\sigma(\PH\PH\nuenuebar)}\,.
\end{equation*}
The value of $\kappa$ can be determined from the $\whizard$ generator by parameterising the $\epem\to\PH\PH\nuenuebar$ cross section
as a function of the input value for $\ghThree$, as indicated in \autoref{fig:self_coupling_ratio}. The fact that the slope is negative indicates that the main dependence on $\ghThree$ enters through interference with other SM diagrams.
The value of 
$\kappa$ is determined from the derivative of the cross section dependence as a function of 
$\ghThree$, evaluated at its SM value, giving $\kappa= 1.22$ and $\kappa = 1.47$ at 1.4\,TeV and 3\,TeV, respectively. However, this 
method does not account for the possibility that the event selection might preferentially favour some diagrams over others, and hence change the analysis 
sensitivity to $\ghThree$. 

In the case of zero beam polarisation, the combined cross sections for double Higgs production give:
\begin{align*}
  \Delta\ghThree/\ghThree & = 54\,\% \ \ \text{at} \ \roots=1.4\,\TeV\,,\\
  \Delta\ghThree/\ghThree & = 29\,\% \ \ \text{at} \ \roots=3\,\TeV\,.
\end{align*}
Because the process involving the trilinear Higgs coupling involves $t$-channel $\PW\PW$-fusion, it can be enhanced by operating with
polarised beams. For the case of $P(\Pem) = -80\,\%$, this yields:
\begin{align*}
  \Delta\ghThree/\ghThree & = 40\,\% \ \ \text{at} \ \roots=1.4\,\TeV\,,\\
  \Delta\ghThree/\ghThree & = 22\,\% \ \ \text{at} \ \roots=3\,\TeV\,.
\end{align*}
The statistical precision on $\ghThree$ improves to $26\,\%$ for unpolarised beams and to $19\,\%$ for $P(\Pem) = -80\,\%$ when combining both energy stages. These results will be improved further using template fits to the BDT output distributions as the different diagrams contributing to double Higgs production lead to different event topologies.

% Section 9
\section{Higgs Mass}
\label{sec:higgs_mass}

At a centre-of-mass energy of $\roots=350\,\GeV$, the Higgs boson mass can be measured in the $\epem\to\PZ\PH$ process. The Higgs boson mass can be extracted from the four-momentum recoiling against in \PZ boson using $\PZ\to\epem$ or $\PZ\to\mpmm$ events as described in \autoref{sec:higgsstrahlung}. Due to the small branching ratios for leptonic \PZ boson decay channels and the impact of the CLIC beamstrahlung spectrum, the achievable precision is limited to $110\,\MeV$.

In a different approach, the Higgs mass is reconstructed from the measured four-vectors of its decay products. The best precision is expected using $\PH\to\PQb\PAQb$ decays in $\epem\to\PH\PGne\PAGne$ events at high energy. For this purpose, the analysis described in \autoref{sec:ww_fusion_bbccgg} has been modified. After the preselection, a single BDT is used at each energy to select $\PH\to\PQb\PAQb$ decays. In contrast to the coupling measurement, the flavour tagging information is included in the BDT classifier.

\begin{figure}
\centering
\includegraphics[width=0.9\columnwidth]{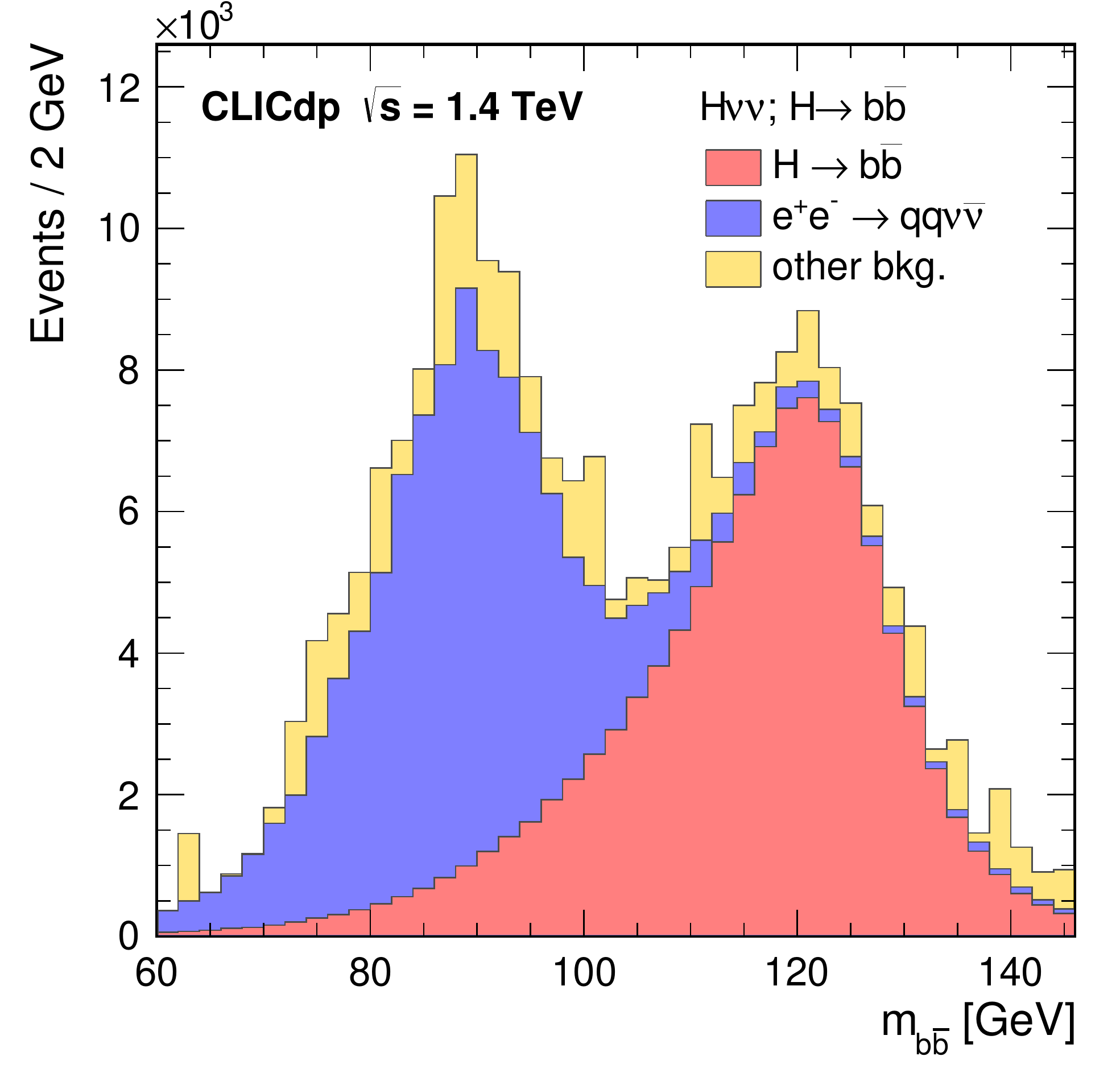}
\caption{\label{fig:bbvv_events_selected_1400} Reconstructed di-jet invariant mass distribution of selected $\PH\to\PQb\PAQb$ events at $\roots=1.4\,\TeV$, showing the signal and backgrounds as stacked histograms. The distributions are normalised to an integrated luminosity of $1.5\,\abinv$.}
\end{figure}

The invariant mass distribution for selected events at $\roots=1.4\,\TeV$ is shown in \autoref{fig:bbvv_events_selected_1400}. The Higgs mass is extracted in the range $105\,\GeV < m_{\PQb\PAQb} < 145\,\GeV$ where good purity of the signal channel is achieved. At the nominal \PZ boson mass, a second peak from $\epem\to\PZ\PGne\PAGne; \PZ\to\PQb\PAQb$ events is visible. These events can be used to calibrate the jet energy scale for the precision measurement of the Higgs boson mass.

A template fit using $\epem\to\PH\PGne\PAGne; \PH\to\PQb\PAQb$ event samples generated using slightly shifted values for the Higgs mass parameter is performed. The Higgs mass and production cross section are extracted simultaneously. The following statistical precisions on the Higgs mass are achieved:
\begin{align*}
    \Delta(m_{\PH}) & = 47\,\MeV \ \text{at} \ 1.4\,\TeV\,, \\
    \Delta(m_{\PH}) & = 44\,\MeV \ \text{at} \ 3\,\TeV\,. \\
\end{align*}
A combination of both energy stages would lead to a precision of $32\,\MeV$.

\section{Systematic Uncertainties}
\label{sec:systematics}

The complete Higgs physics potential of a CLIC collider implemented in three energy stages is described in this paper. The expected statistical uncertainties given in the previous sections do not include potential sources of systematic uncertainty. The obtained results therefore illustrate the level of precision desirable for the control of systematic effects. This is crucial input for the choice of detector technologies and the development of calibration procedures in the coming years.

% A trivial example for this is the total luminosity which should be known to better precision than the statistical precision of the cross section and cross section times branching ratio measurements described in the previous sections of this document.

A comprehensive study of systematic uncertainties requires more knowledge on the technical implementation of the detector. This is beyond the scope of this paper. At this stage, the impact of potentially relevant sources of systematic uncertainty is discussed. The measurements of $\sigma(\PH\PGne\PAGne)\times \BR(\PH\to\PQb\PAQb)$ and the Higgs mass at $\roots=3$\,\TeV, described in \autoref{sec:ww_fusion_bbccgg} and \autoref{sec:higgs_mass}, are used as examples. These measurements are the most challenging test cases for many systematic effects due to the very small expected statistical uncertainties of 0.3\,\% and 44\,\MeV, respectively. In addition, the experimental conditions are most challenging at 3\,\TeV.

The impact of theoretical uncertainties on the Higgs branching fractions is discussed in \autoref{sec:combined_fits} in the context of a combined fit.

\begin{itemize}

\item \textbf{Luminosity spectrum:} A good knowledge of the luminosity spectrum is mandatory for precision Higgs physics at CLIC. The reconstruction of the CLIC luminosity spectrum from Bhabha scattering events is described in~\cite{Poss:2013oea}. A model of the CLIC luminosity spectrum with 19 free parameters is assumed. The expected uncertainties of these parameters and their correlations are propagated to the measurement of $\sigma(\PH\PGne\PAGne)\times \BR(\PH\to\PQb\PAQb)$ and lead to a systematic uncertainty of 0.15\,\%. The luminosity spectrum affects the event rate more than the observed invariant mass of the two jets. Concerning the Higgs mass extraction, the luminosity spectrum is not expected to represent a dominant source of systematic uncertainty since the cross section is a free parameter in the template fit.

\item \textbf{Total luminosity:} The expected statistical precision of the $\sigma(\PH\PGne\PAGne)\times \BR(\PH\to\PQb\PAQb)$ measurement indicates the desired precision for the knowledge of the total luminosity. It is expected that an accuracy of a few permille can be achieved using the luminometer envisaged for CLIC~\cite{Lukic:2013fw, Bozovic-Jelisavcic:2013aca}.

\item \textbf{Beam polarisation:} The knowledge of the beam polarisation at the interaction point is most important for the measurement of $\PW\PW$-fusion events at high energy. The beam polarisation can be controlled to a level of 0.2\,\% using single $\PW$, $\PZ$ and $\PGg$ events with missing energy~\cite{Wilson:lcws2012}. The resulting systematic uncertainty on $\sigma(\PH\PGne\PAGne)\times \BR(\PH\to\PQb\PAQb)$ is 0.1\,\%. For the Higgs mass measurement, the effect of the estimated beam polarisation uncertainty is negligible.

\item \textbf{Jet energy scale:} The measurement of the Higgs boson mass using $\PH\to\PQb\PAQb$ decays requires a precise knowledge of the energy scale correction for b-jets. An uncertainty on the jet energy scale of $3.5\times10^{-4}$ leads to a systematic uncertainty on the Higgs mass similar to the statistical error at 3 TeV. The same jet energy scale uncertainty would have negligible impact on $\sigma(\PH\PGne\PAGne)\times \BR(\PH\to\PQb\PAQb)$. A suitable process for the calibration is $\epem\to\PZ\PGne\PAGne; \PZ\to\PQb\PAQb$ which is kinematically similar to Higgs production in $\PW\PW$-fusion. $\sigma(\PZ\PGne\PAGne)\times \BR(\PZ\to\PQb\PAQb) = 276$\,fb leads to an expected number of events for calibration which is slightly larger than the signal event sample. To improve the precision further, additional high-statistics $\PZ$ boson samples would be needed. Generator-level studies show that $\Pepm\PGg\to\PZ\Pepm; \PZ\to\PQb\PAQb$ with a cross section about one order of magnitude larger compared to the signal process is a promising channel for this purpose.

\item \textbf{Flavour tagging:} Several of the precision measurements discussed in this paper rely on b-tagging information. The calibration of the flavour tagging at CLIC is a topic for future study. To illustrate the impact of a non-perfect understanding of the mistag rate for charm and light quark jets, an ad hoc variation of the b-tag distributions for jets in background events is performed. Even after the BDT selection, the background contains only very few b-jets in the $\sigma(\PH\PGne\PAGne)\times \BR(\PH\to\PQb\PAQb)$ analysis. First, the b-tag distributions for both jets were decreased (increased) by 0.5\,\% using event reweighting for values below (above) the median keeping the overall number of background events constant. The opposite variation is applied in a second step. These variations lead to a $\pm0.25\,\%$ change of the result. \\ As the flavour tagging efficiency mostly affects the event rate, it is not expected to be a dominant source of systematic uncertainty for the Higgs mass measurement.

\end{itemize}

In summary, it seems possible to control the systematic uncertainties discussed above with similar or better precision compared to the statistical uncertainty for the measurement of $\sigma(\PH\PGne\PAGne)\times \BR(\PH\to\PQb\PAQb)$. An excellent understanding of the b-jet energy scale is necessary for a competitive Higgs mass measurement at CLIC.

Many of the analyses described in this paper, especially where hadronisation is relevant, will require a careful tuning of the Monte Carlo models using other high-precision processes. Such an investigation is beyond the scope of this first study of Higgs physics at CLIC presented here.

% This is a temporary fix
% \newpage

% Section 10
\section{Combined Fits}
\label{sec:combined_fits}

The results discussed in the preceding sections are summarised in
\autoref{tab:GlobalFit:Input350} and
\autoref{tab:GlobalFit:Input143}. From the $\sigma$ and $\sigma \times
\BR$ measurements given in the tables the Higgs coupling parameters and
total width are extracted by a global fit as described below. Here, a
$-80\,\%$ electron polarisation is assumed for the $1.4\,\TeV$ and the
$3\,\TeV$ stages. The increase in cross section is taken into account
by multiplying the event rates with a factor of 1.8 for all WW-fusion measurements (see
\autoref{tab:higgs:polarisation}), resulting in a reduction of the
uncertainties by a factor of $\sqrt{1.8}$. This approach is
conservative since it assumes that all backgrounds including those
from $s$-channel processes, which do not receive the same enhancement
by polarisation, scale with the same factor.

\begin{table*}[htp]\centering
  \begin{tabular}{lllc}\toprule
                        &                                                           &                              & \tabt{Statistical precision}                        \\\cmidrule(l){4-4}
        \tabt{Channel}  & \tabt{Measurement}                                        & \tabt{Observable}            & $350\,\GeV$       \\ 
                        &                                                           &                              & $500\,\fbinv$        \\ \midrule
    $\PZ\PH$            & Recoil mass distribution                                  & $\mH$                        & $110\,\MeV$  \\
    $\PZ\PH$            & $\sigma(\PZ\PH)\times \BR(\PH\to\text{invisible})$         & $\Gamma_\text{inv}$          & $0.6\,\%$  \\ \midrule
    $\PZ\PH$            & $\sigma(\PZ\PH)\times \BR(\PZ\to\Plp\Plm)$             & $\gHZZ^{2}$                  & $3.8\,\%$  \\
    $\PZ\PH$            & $\sigma(\PZ\PH)\times \BR(\PZ\to\PQq\PAQq)$                  & $\gHZZ^{2}$                  & $1.8\,\%$  \\
    $\PZ\PH$            & $\sigma(\PZ\PH)\times \BR(\PH\to\PQb\PAQb)$                & $\gHZZ^{2}\gHbb^{2}/\GH$     & $0.86\,\%$ \\
    $\PZ\PH$            & $\sigma(\PZ\PH)\times \BR(\PH\to\PQc\PAQc)$                & $\gHZZ^{2}\gHcc^2/\GH$       & $14\,\%$ \\
    $\PZ\PH$            & $\sigma(\PZ\PH)\times \BR(\PH\to\Pg\Pg)$                   &                              & $6.1\,\%$ \\
    $\PZ\PH$            & $\sigma(\PZ\PH)\times \BR(\PH\to\tptm)$               & $\gHZZ^{2}\gHTauTau^{2}/\GH$ & $6.2\,\%$ \\
    $\PZ\PH$            & $\sigma(\PZ\PH)\times \BR(\PH\to\PW\PW^*)$                 & $\gHZZ^{2}\gHWW^{2}/\GH$     & $5.1\,\%$ \\
    $\PH\PGne\PAGne$    & $\sigma(\PH\PGne\PAGne)\times \BR(\PH\to\PQb\PAQb)$        & $\gHWW^{2}\gHbb^{2}/\GH$     & $1.9\,\%$ \\
    $\PH\PGne\PAGne$    & $\sigma(\PH\PGne\PAGne)\times \BR(\PH\to\PQc\PAQc)$        & $\gHWW^{2}\gHcc^{2}/\GH$     & $26\,\%$ \\
    $\PH\PGne\PAGne$    & $\sigma(\PH\PGne\PAGne)\times \BR(\PH\to\Pg\Pg)$        &     & $10\,\%$ \\    
    \bottomrule
  \end{tabular}
    \caption{Summary of the precisions obtainable for the Higgs
      observables in the first stage of CLIC for an integrated
      luminosity of $500\,\fbinv$ at $\roots=350\,\GeV$, assuming
      unpolarised beams. For the branching ratios, the measurement
      precision refers to the expected statistical uncertainty on the
      product of the relevant cross section and branching ratio; this
      is equivalent to the expected statistical uncertainty of the
      product of couplings divided by $\Gamma_{\PH}$ as indicated in
      the third column. \label{tab:GlobalFit:Input350}}
\end{table*}

\begin{table*}[htp]\centering
    \begin{tabular}{lllcc}\toprule
                        &                                                           &                              & \tabtt{Statistical precision}                        \\\cmidrule(l){4-5}
        \tabt{Channel}  & \tabt{Measurement}                                        & \tabt{Observable}  & $1.4\,\TeV$         & $3\,\TeV$           \\ 
                        &                                                           &                           & $1.5\,\abinv$      & $2.0\,\abinv$        \\ \midrule
   $\PH\PGne\PAGne$    & $\PH\to\PQb\PAQb$ mass distribution                       & $\mH$                & $47\,\MeV$     & $44\,\MeV$       \\ \midrule
   $\PH\PGne\PAGne$    & $\sigma(\PH\PGne\PAGne)\times \BR(\PH\to\PQb\PAQb)$        & $\gHWW^{2}\gHbb^{2}/\GH$   & $0.4\,\%$         & $0.3\,\%$           \\
   $\PH\PGne\PAGne$    & $\sigma(\PH\PGne\PAGne)\times \BR(\PH\to\PQc\PAQc)$        & $\gHWW^{2}\gHcc^{2}/\GH$  & $6.1\,\%$         & $6.9\,\%$           \\
   $\PH\PGne\PAGne$    & $\sigma(\PH\PGne\PAGne)\times \BR(\PH\to\Pg\Pg)$           &                     & $5.0\,\%$         & $4.3\,\%$           \\
   $\PH\PGne\PAGne$    & $\sigma(\PH\PGne\PAGne)\times \BR(\PH\to\tptm)$       & $\gHWW^{2}\gHTauTau^{2}/\GH$ & $4.2\,\%$         & $4.4\,\%$               \\
   $\PH\PGne\PAGne$    & $\sigma(\PH\PGne\PAGne)\times \BR(\PH\to\mpmm)$       & $\gHWW^{2}\gHMuMu^{2}/\GH$   & $38\,\%$        & $25\,\%$            \\
   $\PH\PGne\PAGne$    & $\sigma(\PH\PGne\PAGne)\times \BR(\PH\to\upgamma\upgamma)$ &                          & $15\,\%$          & $10\,\%^*$               \\
   $\PH\PGne\PAGne$    & $\sigma(\PH\PGne\PAGne)\times \BR(\PH\to\PZ\upgamma)$      &                             & $42\,\%$           & $30\,\%^*$               \\
   $\PH\PGne\PAGne$    & $\sigma(\PH\PGne\PAGne)\times \BR(\PH\to\PW\PW^*)$         & $\gHWW^{4}/\GH$            & $1.0\,\%$         & $0.7\,\%^*$         \\
   $\PH\PGne\PAGne$    & $\sigma(\PH\PGne\PAGne)\times \BR(\PH\to\PZ\PZ^*)$         & $\gHWW^{2}\gHZZ^{2}/\GH$  & $5.6\,\%$ & $3.9\,\%^*$   \\
   $\PH\epem$       & $\sigma(\PH\epem)\times \BR(\PH\to\PQb\PAQb)$           & $\gHZZ^{2}\gHbb^{2}/\GH$    & $1.8\,\%$ & $2.3\,\%^*$ \\ \midrule
   $\PQt\PAQt\PH$      & $\sigma(\PQt\PAQt\PH)\times \BR(\PH\to\PQb\PAQb)$          & $\gHtt^{2}\gHbb^{2}/\GH$  & $8\,\%$         & $-$             \\
   $\PH\PH\PGne\PAGne$ & $\sigma(\PH\PH\PGne\PAGne)$                               & $\lambda$                   & $54\,\%$          & $29\,\%$            \\
   $\PH\PH\PGne\PAGne$ & with $-80\,\%$ $\Pem$ polarisation                             & $\lambda$                  & $40\,\%$          & $22\,\%$            \\ \bottomrule
  \end{tabular}
  \caption{Summary of the precisions obtainable for the Higgs
    observables in the higher-energy CLIC stages for integrated
    luminosities of $1.5\,\abinv$ at $\roots=1.4\,\TeV$, and
    $2.0\,\abinv$ at $\roots=3\,\TeV$. In both cases unpolarised beams
    have been assumed. 
    For $\gHtt$, the $3\,\TeV$ case has not yet been studied,
    but is not expected to result in substantial improvement due to the
    significantly reduced cross section at high energy.
    Numbers marked with $*$ are extrapolated from $\roots=1.4\,\TeV$
    to $\roots=3\,\TeV$ as explained in the text.
    For the branching ratios, the measurement precision refers to the expected
    statistical uncertainty on the product of the relevant cross
    section and branching ratio; this is equivalent to the expected
    statistical uncertainty of the product of couplings divided by
    $\Gamma_{\PH}$, as indicated in the third column. For the
    measurements from the $\PH\PH\PGne\PAGne$
    process, the measurement precisions give the expected
    statistical uncertainties on the self-coupling parameter $\lambda$. \label{tab:GlobalFit:Input143}}
\end{table*}

A few of the observables listed in \autoref{tab:GlobalFit:Input143} were 
studied only at $\roots=1.4\,\TeV$, but not at $\roots=3\,\TeV$. In cases where 
those observables have a significant impact on the combined fits described in this 
section, the precisions obtained at $\roots=1.4\,\TeV$ were extrapolated to $\roots=3\,\TeV$. 
The extrapolation is based on the number of signal events within the detector acceptance 
at $1.4\,\TeV$ and $3\,\TeV$. It is assumed that the background processes scale 
in the same way with $\roots$ as the signal events.  However, in fact the 
signal Higgs bosons are produced in vector boson fusion which increases with increasing $\roots$, while several backgrounds 
are dominated by $s$-channel diagrams which decrease with increasing $\roots$.

Since the physical observables ($\sigma$ or $\sigma \times \BR$)
typically depend on several coupling parameters and on the total
width, these parameters are extracted with a combined fit of all
measurements. To provide a first indication of the overall impact of
the CLIC physics programme, simple fits considering only the
statistical uncertainties of the measurements are performed. Two types
of fits are used: A model-independent fit making minimal theoretical
assumptions, and a model-dependent fit following the strategies used
for the interpretation of LHC Higgs results.

Both fits are based on a $\chi^2$ minimisation using the \minuit
package~\cite{James1975343}. The measurements which serve as input to the fit,
presented in detail in the preceding sections, are either a total
cross section $\sigma$ in the case of the measurement of
$\epem\to\PZ\PH$ via the recoil mass technique, or a cross section
$\times$ branching ratio $\sigma \times \BR$ for specific Higgs
production modes and decays. To obtain the expected sensitivity for
CLIC it is assumed that for all measurements the value expected in the
SM has been measured, so only the statistical uncertainties of each
measurement are used in the $\chi^2$ calculation. In the absence of correlations, the contribution of a single measurement is given by
\begin{equation*}
\chi^2_{i} =\frac{(C_i/C_i^{\text{SM}} - 1)^2}{\Delta F_i^2},
\end{equation*}
where $C_i$ is the fitted value of the relevant combination of relevant
Higgs couplings (and total width) describing the particular
measurement, $C_i^{\text{SM}}$ is the SM expectation, and $\Delta F_i$
is the statistical uncertainty of the measurement of the considered
process. Since this simplified description does not allow the accurate treatment of correlations between measurements, nor the inclusion of correlated theory systematics in the model-dependent fit, the global $\chi^2$ of the fit is constructed from the covariance matrix of all measurements. It is given by
\begin{equation*}
\chi^2 = {\zeta}^T {\bf{V}}^{-1} {\zeta},
\end{equation*}
where ${\bf{V}}$ is the covariance matrix and $\zeta$ is the vector of deviations of fitted values of the relevant combination of Higgs couplings and total width describing the particular measurement deviation from the SM expectation as introduced above, 
$\zeta_i = C_i/C_i^{\text{SM}} - 1$. 

The $C_i$'s depend on the particular measurements and on the type of
fit (model-independent or model-dependent), given in detail below. In the absence of systematic uncertainties, the diagonal elements of ${\bf{V}}$ are given by the statistical uncertainty of the measurement, 
\begin{equation*}
{\bf{V}}_{ii} = \Delta F_i^2,
\end{equation*}
while the off-diagonal elements represent the correlations between measurements. In the fit, correlations are  taken into account in
cases where they are expected to be large. This applies to the
measurements of $\sigma \times \BR$ for $\PH \to \bb, \cc, \Pg\Pg$ in
\higgsstrahlung and $\PW\PW$-fusion events at 350\,\GeV and in $\PW\PW$-fusion events only at 1.4\,\TeV and 3\,\TeV, which are extracted in a
combined fitting procedure at each energy. These measurements show correlation coefficients with absolute values as large as 0.32. 

In signal channels with substantial contaminations from other Higgs decays, penalty terms were added to the $\chi^2$ to take into account the normalisation of the other channels. These additional uncertainties, which are also of a statistical nature, are derived from the statistical uncertainties of the respective Higgs final state analysis, taking the level of contamination into account. The channels where this results in non-negligible effects are the $\PH\to\PW\PW^*$ analyses at all energies, in particular in the all-hadronic decay modes, with corrections to the statistical uncertainties as large as 8\% at 350 GeV.

\subsection{Model-independent Fit}

The model-independent fit uses the zero-width approximation to
describe the individual measurements in terms of 
Higgs couplings and the total width, $\Gamma_{\PH}$. Here, the
total cross section of $\epem\to\PZ\PH$ depends on:
\begin{equation*}
C_{\PZ\PH} = g_{\PH\PZ\PZ}^2,
\end{equation*}
while for specific final states such as $\epem\to\PZ\PH$; $\PH\to\PQb\PAQb$ and $\epem\to\PH\nuenuebar$; $\PH\to \PQb\PAQb$:
\begin{equation*}
   C_{{\PZ\PH},\,\PH\to\PQb\PAQb} = \frac{\gHZZ^2 \gHbb^2}{\Gamma_{\PH}}
\end{equation*}
and:
\begin{equation*}
C_{\PH\nuenuebar,\,\PH\to \PQb\PAQb} = \frac{\gHWW^2 \gHbb^2}{\Gamma_{\PH}},
\end{equation*}
respectively. 

The fit is performed with 11 free parameters: $\gHZZ$,
$\gHWW$, $\gHbb$, $\gHcc$, $\gHTauTau$, $\gHMuMu$, $\gHtt$ and
$\Gamma_{\PH}$, as well as the three effective couplings
$g^\dagger_\mathrm{\PH\Pg\Pg}$, $g^\dagger_{\PH\PGg\PGg}$ and $g^\dagger_{\PH\PZ\PGg}$. The
latter three parameters are treated in the same way as the physical
Higgs couplings in the fit.

\begin{table}[htp]\centering
\begin{tabular}{lccc}
\toprule
Parameter & \multicolumn{3}{c}{Relative precision}\\
\midrule
& $350\,\GeV$ & + $1.4\,\TeV$ & + $3\,\TeV$\\
&$500\,\fbinv$& + $1.5\,\abinv$& + $2\,\abinv$\\
\midrule
$\gHZZ$ & 0.8\,\% & 0.8\,\% & 0.8\,\% \\
$\gHWW$ & 1.4\,\% & 0.9\,\% & 0.9\,\% \\
$\gHbb$ & 3.0\,\% & 1.0\,\% & 0.9\,\% \\
$\gHcc$ & 6.2\,\% & 2.3\,\% & 1.9\,\% \\
$\gHTauTau$ & 4.3\,\% & 1.7\,\% & 1.4\,\% \\
$\gHMuMu$ & $-$ & 14.1\,\% & 7.8\,\% \\
$\gHtt$ & $-$ & 4.2\,\% & 4.2\,\% \\
\midrule
$g^\dagger_{\PH\Pg\Pg}$ & 3.7\,\% & 1.8\,\% & 1.4\,\% \\
$g^\dagger_{\PH\PGg\PGg}$ & $-$ & 5.7\,\% & 3.2\,\% \\
$g^\dagger_{\PH\PZ\PGg}$ & $-$ & 15.6\,\% & 9.1\,\% \\
\midrule
$\Gamma_{\PH}$ & 6.7\,\% & 3.7\,\% & 3.5\,\% \\
\bottomrule
\end{tabular}
\caption{Results of the model-independent fit. Values marked "$-$" can
  not be measured with sufficient precision at the given energy.
  For $\gHtt$, the $3\,\TeV$ case has not yet been studied,
  but is not expected to result in substantial improvement due to the
  significantly reduced cross section at high energy. The three
  effective couplings $g^\dagger_{\PH\Pg\Pg}$, 
  $g^\dagger_{\PH\PGg\PGg}$ and $g^\dagger_{\PH\PZ\PGg}$ are also included in the fit. Operation with $-80\,\%$ electron beam polarisation is assumed above 1\,\TeV. \label{tab:MIResults}}
\end{table}

\begin{figure}
  \centering
  \includegraphics[width=0.9\columnwidth]{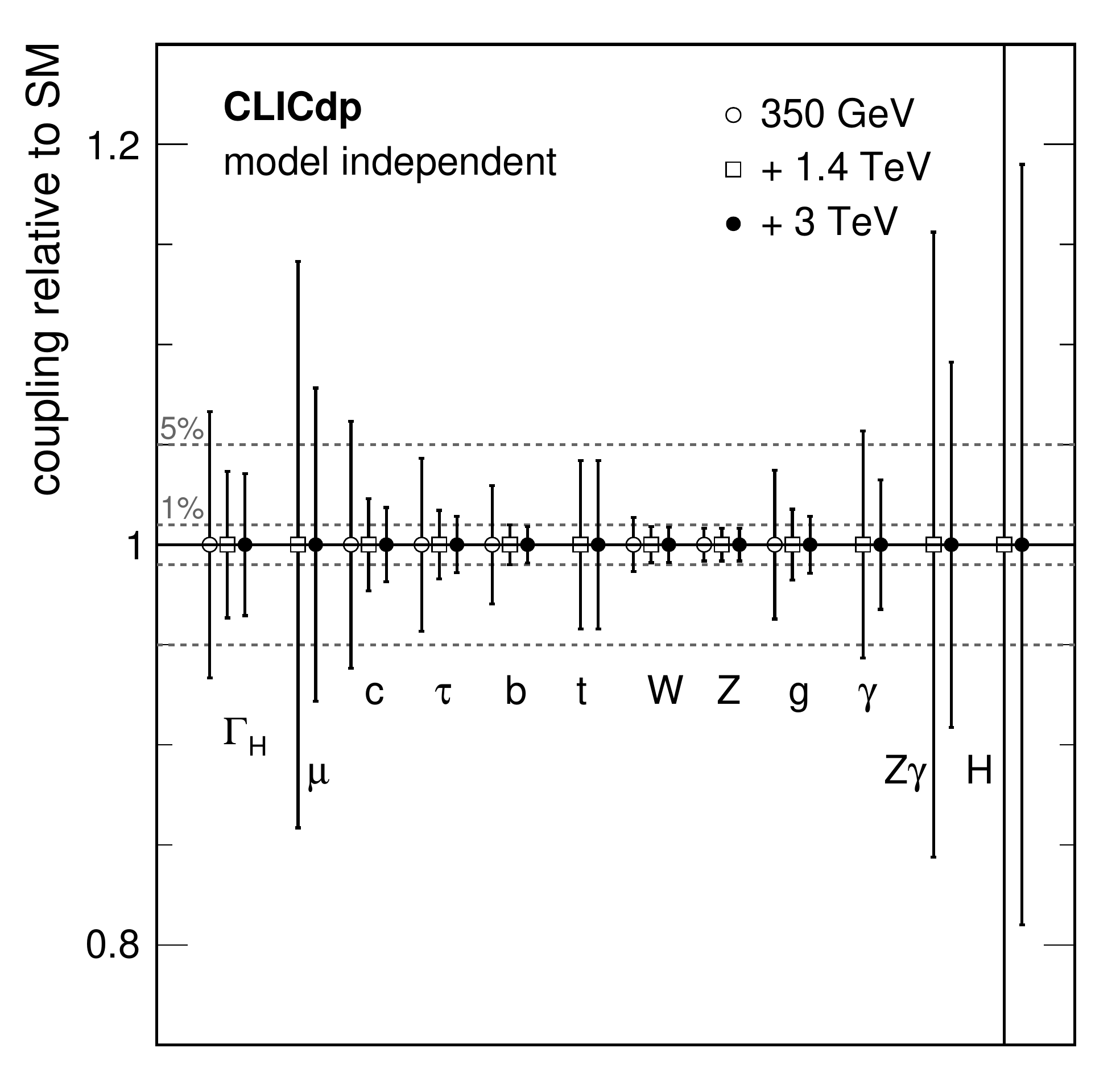}
  \caption{\label{fig:combinedFit:MI} Illustration of the precision of
    the Higgs couplings of the three-stage CLIC programme determined
    in a model-independent fit without systematic or theoretical uncertainties. 
    The dotted lines show the relative precisions of 1\,\% and 5\,\%.}
\end{figure}

The fit is performed in three stages, taking the statistical
uncertainties obtainable from CLIC at the three considered energy
stages ($350\,\GeV$, $1.4\,\TeV$, $3\,\TeV$) successively into
account. Each new stage also includes all measurements of the previous
stages. \autoref{tab:MIResults} summarises the results. They are
graphically illustrated in \autoref{fig:combinedFit:MI}. Since the
model-independence of the analysis hinges on the absolute measurement
of $\sigma(\PZ\PH)$ at $350\,\GeV$, which provides the coupling
$\gHZZ$, the precision of all other couplings is ultimately limited by
this uncertainty.

\subsection{Model-dependent Fit}

For the model-dependent fit, it is assumed that the Higgs decay
properties can be described by ten independent parameters
$\kappa_{\PH\PZ\PZ}$, $\kappa_{\PH\PW\PW}$, $\kappa_{\PH\PQb\PQb}$,
$\kappa_{\PH\PQc\PQc}$, $\kappa_{\PH\PGt\PGt}$,
$\kappa_{\PH\PGm\PGm}$, $\kappa_{\PH\PQt\PQt}$, $\kappa_{{\PH\Pg\Pg}}$,
 $\kappa_{\PH\PGg\PGg}$ and $\kappa_{\PH\PZ\PGg}$. These factors are defined by the ratio of
the Higgs partial width divided by the partial width expected in the
Standard Model as:
\begin{equation*}
\kappa_i^2 = \Gamma_i/\Gamma_i^{\text{SM}}\,.
\end{equation*}

In this scenario, the total width is given by the sum of the ten
partial widths considered, which is equivalent to assuming no non-Standard-Model Higgs decays such as decays into new invisible particles.
The ratio of the total width to its SM value is thus given by:
\begin{equation}
\frac{\Gamma_{\PH,\text{md}}}{\Gamma_{\PH}^{\text{SM}}} = \sum_i \kappa_i^2 \ \BR_i, \label{eq:KappaWidth}
\end{equation}
where $\BR_i$ is the SM branching fraction for the respective final
state and the subscript ``md'' stands for ``model-dependent''. To
obtain these branching fractions, a fixed value for the Higgs mass has
to be imposed. For the purpose of this study, $126\,\GeV$ is
assumed. The branching ratios are taken from the LHC Higgs cross
section working group~\cite{Dittmaier:2012vm}. To exclude
effects from numerical rounding errors, the total sum of $\BR$'s is
normalised to unity.

With these definitions, the $C_i$'s in the $\chi^2$ take the following
forms: for the total $\epem\to\PZ\PH$ cross section:
\begin{equation*}
C_{\PZ\PH} = \kappa_{\PH\PZ\PZ}^2;
\end{equation*}
while for specific final states such as $\epem\to\PZ\PH$; $\PH\to \PQb\PAQb$ and $\epem\to\PH\nuenuebar$; $\PH\to \PQb\PAQb$:
\begin{equation*}
C_{{\PZ\PH},\,\PH\to\PQb\PAQb} = \frac{\kappa_{\PH\PZ\PZ}^2\kappa_{\PH\PQb\PQb}^2}{\left(\Gamma_{\PH,\text{md}} / \Gamma_{\PH}^{\text{SM}}\right)}
\end{equation*}
and:
\begin{equation*}
C_{\PH\nuenuebar,\,\PH\to \PQb\PAQb} = \frac{\kappa_{\PH\PW\PW}^2\kappa_{\PH\PQb\PQb}^2}{\left(\Gamma_{\PH,\text{md}} / \Gamma_{\PH}^{\text{SM}}\right)},
\end{equation*}
respectively.  

Since at the first energy stage of CLIC no significant measurements of
the $\PH\to\mpmm$, $\PH\to\PGg\PGg$ and $\PH\to\PZ\PGg$ decays are possible, the fit is
reduced to six free parameters (the coupling to top is also not
constrained, but this is without effect on the total width) 
by setting  $\PH\to\mpmm$, $\PH\to\PGg\PGg$ and $\PH\to\PZ\PGg$ 
to zero.  These branching ratios are much smaller than the derived 
uncertainty on the total width.

Two versions of the model-dependent fit are performed, one ignoring theoretical uncertainties to illustrate the full potential of the constrained fit, and one taking the present theoretical uncertainties of the branching fractions into account~\cite{Dittmaier:2012vm}. To avoid systematic biases in the fit results, the uncertainties are symmetrised, preserving the overall size of the uncertainties. Theoretical uncertainties on the production are assumed to be substantially smaller than in the decay, and are ignored in the present study. Depending on the concrete Higgs decay, multiple measurements may enter in the fit, originating from different centre-of-mass energies, different production channels or different signal final states. To account for this, the theoretical uncertainties are treated as fully correlated for each given Higgs decay. 

\begin{table}[htp]\centering
\begin{tabular}{lccc}
\toprule
Parameter & \multicolumn{3}{c}{Relative precision}\\
\midrule
& $350\,\GeV$ & + $1.4\,\TeV$& + $3\,\TeV$\\
&$500\,\fbinv$& + $1.5\,\abinv$& + $2\,\abinv$\\
\midrule
$\kappa_{\PH\PZ\PZ}$ & 0.6\,\% & 0.4\,\% & 0.3\,\% \\
$\kappa_{\PH\PW\PW}$ & 1.1\,\% & 0.2\,\% & 0.1\,\% \\
$\kappa_{\PH\PQb\PQb}$ & 1.8\,\% & 0.4\,\% & 0.2\,\% \\
$\kappa_{\PH\PQc\PQc}$ & 5.8\,\% & 2.1\,\% & 1.7\,\% \\
$\kappa_{\PH\PGt\PGt}$ & 3.9\,\% & 1.5\,\% & 1.1\,\% \\
$\kappa_{\PH\PGm\PGm}$ & $-$ & 14.1\,\% & 7.8\,\% \\
$\kappa_{\PH\PQt\PQt}$ & $-$ & 4.1\,\% & 4.1\,\% \\
$\kappa_{\PH\Pg\Pg}$ & 3.0\,\% & 1.5\,\% & 1.1\,\% \\
$\kappa_{\PH\PGg\PGg}$ & $-$ & 5.6\,\% & 3.1\,\% \\
$\kappa_{\PH\PZ\PGg}$ & $-$ & 15.6\,\% & 9.1\,\% \\
\midrule
$\Gamma_{\PH,\text{md,\,derived}}$ & 1.4\,\% & 0.4\,\% & 0.3\,\% \\
\bottomrule
\end{tabular}
\caption{Results of the model-dependent fit without theoretical uncertainties. Values marked "$-$" can
  not be measured with sufficient precision at the given energy.
  For $g_{\PH\PQt\PQt}$, the $3\,\TeV$ case has not yet been
  studied, but is not expected to result in substantial improvement
  due to the significantly reduced cross section at high energy. The
  uncertainty of the total width is calculated from the fit results
  following \autoref{eq:KappaWidth}, taking the parameter correlations
  into account. Operation with $-80\,\%$ electron beam polarisation is assumed above 1\,\TeV. \label{tab:MDResults}}
\end{table}

\begin{figure}
  \centering
  \includegraphics[width=0.9\columnwidth]{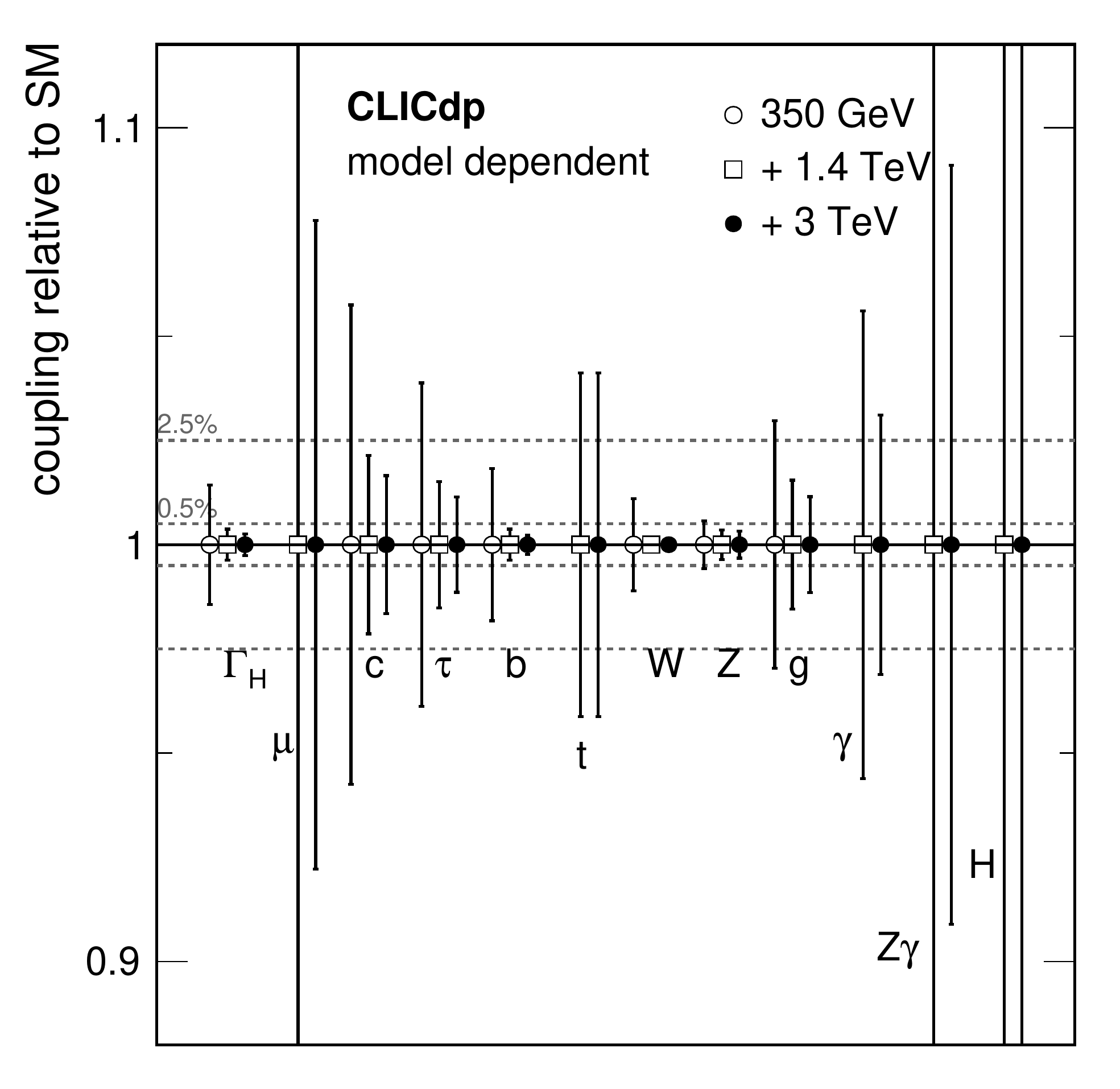}
  \caption{\label{fig:combinedFit:MD} Illustration of the precision of
    the Higgs couplings of the three-stage CLIC programme determined
    in a model-dependent fit without systematic or theoretical uncertainties. The dotted lines show the relative precisions of 0.5\,\% and 2.5\,\%.}
\end{figure}

\begin{table}[htp]\centering
\begin{tabular}{lccc}
\toprule
Parameter & \multicolumn{3}{c}{Relative precision}\\
\midrule
& $350\,\GeV$ & + $1.4\,\TeV$& + $3\,\TeV$\\
&$500\,\fbinv$& + $1.5\,\abinv$& + $2\,\abinv$\\
\midrule
$\kappa_{\PH\PZ\PZ}$ & 0.6\,\% & 0.5\,\% & 0.5\,\% \\
$\kappa_{\PH\PW\PW}$ & 1.2\,\% & 0.5\,\% & 0.5\,\% \\
$\kappa_{\PH\PQb\PQb}$ & 2.6\,\% & 1.5\,\% & 1.4\,\% \\
$\kappa_{\PH\PQc\PQc}$ & 6.3\,\% & 3.2\,\% & 2.9\,\% \\
$\kappa_{\PH\PGt\PGt}$ & 4.2\,\% & 2.1\,\% & 1.8\,\% \\
$\kappa_{\PH\PGm\PGm}$ & $-$ & 14.2\,\% & 7.9\,\% \\
$\kappa_{\PH\PQt\PQt}$ & $-$ & 4.2\,\% & 4.1\,\% \\
$\kappa_{\PH\Pg\Pg}$ & 5.1\,\% & 4.0\,\% & 3.9\,\% \\
$\kappa_{\PH\PGg\PGg}$ & $-$ & 5.9\,\% & 3.5\,\% \\
$\kappa_{\PH\PZ\PGg}$ & $-$ & 16.0\,\% & 9.8\,\% \\
\midrule
$\Gamma_{\PH,\text{md,\,derived}}$ & 2.0\,\% & 1.1\,\% & 1.1\,\% \\
\bottomrule

\end{tabular}
\caption{Results of the model-dependent fit with the current theoretical uncertainties on the decay branching fractions. Values marked "$-$" can
  not be measured with sufficient precision at the given energy.
  For $g_{\PH\PQt\PQt}$, the $3\,\TeV$ case has not yet been
  studied, but is not expected to result in substantial improvement
  due to the significantly reduced cross section at high energy. The
  uncertainty of the total width is calculated from the fit results
  following \autoref{eq:KappaWidth}, taking the parameter correlations
  into account. Operation with $-80\,\%$ electron beam polarisation is assumed above 1\,\TeV. \label{tab:MDResultsTheory}}
\end{table}

As in the model-independent case the fit is performed in three stages,
taking the statistical errors of CLIC at the three considered energy
stages ($350\,\GeV$, $1.4\,\TeV$, $3\,\TeV$) successively into
account. Each new stage also includes all measurements of the previous
stages. The total width is not a free parameter of the fit. Instead,
its uncertainty, based on the assumption given in
\autoref{eq:KappaWidth}, is calculated from the fit results, taking
the full correlation of all parameters into
account. \autoref{tab:MDResults} summarises the results of the fit without taking theoretical uncertainties into account,
and \autoref{fig:combinedFit:MD} illustrates the evolution of the
precision over the full CLIC programme. \autoref{tab:MDResultsTheory} summarises the results of the model-dependent fit with theoretical uncertainties of the branching fractions. 

\subsection{Discussion of Fit Results}

\begin{figure}
  \centering
  \includegraphics[width=0.9\columnwidth]{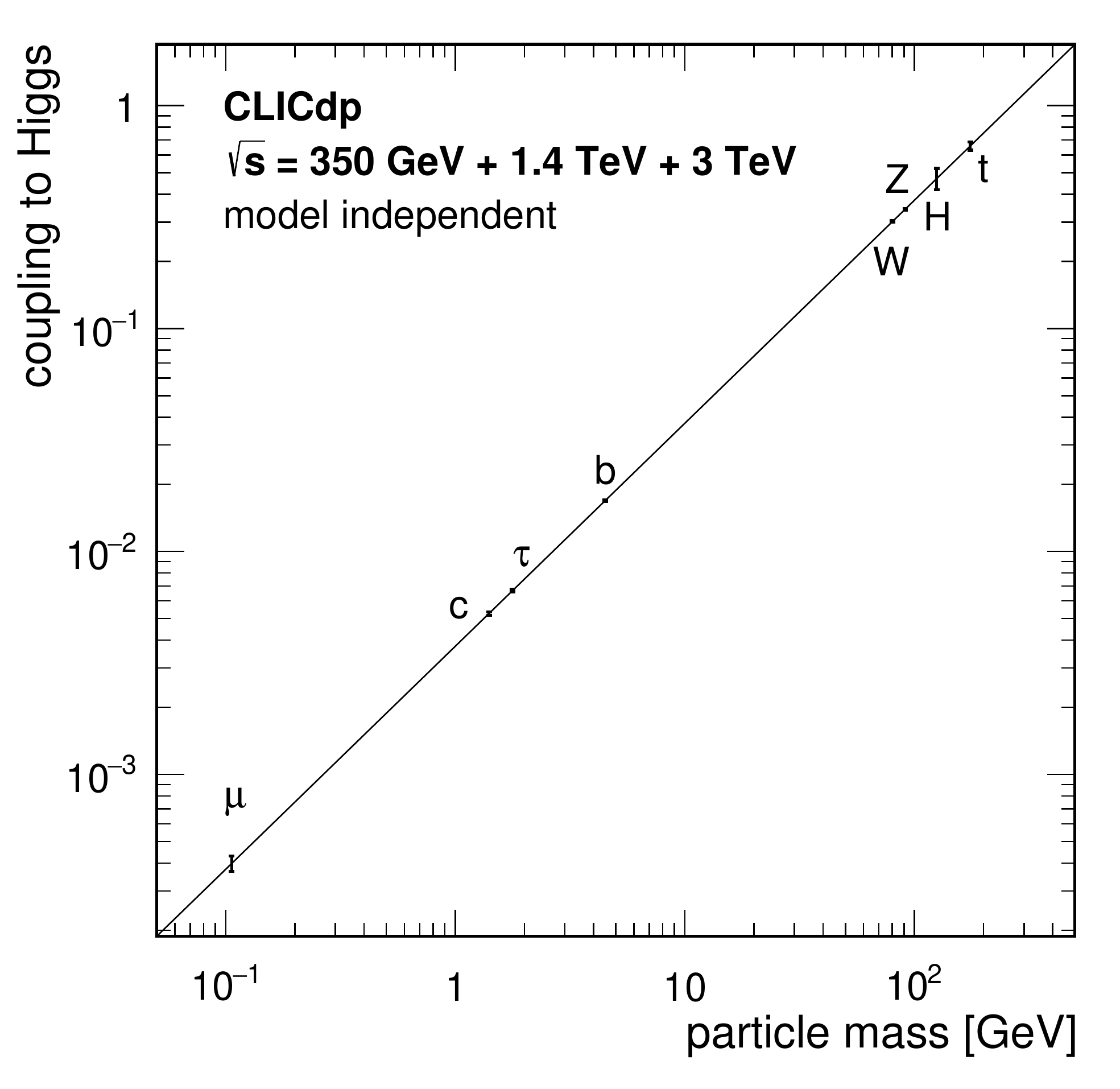}
  \caption{\label{fig:combinedFit:CouplingvsMass} Illustration of the
    precision of the model-independent Higgs couplings and of the
    self-coupling as a function of particle mass. The line shows the 
    SM prediction that the Higgs coupling of each particle is proportional to
    its mass.}
\end{figure}

The full Higgs physics programme of CLIC, interpreted with a combined
fit of the couplings to fermions and gauge bosons as well as the total
width, and combined with the measurement of the self-coupling, will
provide a comprehensive picture of the properties of this recently 
discovered particle. \autoref{fig:combinedFit:CouplingvsMass}
illustrates the expected uncertainties of the various couplings
determined in the model-independent fit as well as the self-coupling
as a function of the particle mass. Combined with the quasi
model-independent measurement of the total width with a precision of
$3.5\,\%$, this illustrates the power of the three-stage CLIC
programme. Each of the stages contributes significantly to the total
precision, with the first stage at $350\,\GeV$ providing the
model-independent "anchor" of the coupling to the $\PZ$ boson, as well
as a first measurement of the total width and coupling measurements to
most fermions and bosons. The higher-energy stages add direct
measurements of the coupling to top quarks, to muons and photons as
well as overall improvements of the branching ratio measurements and
with that of the total width and all couplings except the one to the
$\PZ$ already measured in the first stage. They also provide a
measurement of the self-coupling of the Higgs boson. In a
model-dependent analysis, the improvement with increasing energy is
even more significant than in the model-independent fit, since the
overall limit of all couplings imposed by the model-independent
measurement of the $\PZ\PH$ recoil process is removed.

% Section 11
\section{Summary and Conclusions}
\label{sec:summary-conclusions}

A detailed study of the Higgs physics reach of CLIC has been presented
in the context of CLIC operating in three energy stages, $\roots =
350\,\GeV$, $1.4\,\TeV$ and $3\,\TeV$. The initial stage of operation,
$500\,\fbinv$ at $\roots=350\,\GeV$,  allows the study of Higgs
production from both the $\epem\to\PZ\PH$ and the $\PW\PW$-fusion
process. These data yield precise model-independent measurements
of the Higgs boson couplings, in particular $\Delta(g_{\PH\PZ\PZ}) = 
0.8\,\%$, $\Delta(g_{\PH\PW\PW}) = 1.4\,\%$ and
$\Delta(g_{\PH\PQb\PQb}) = 3.0\,\%$. In addition, the branching ratio
to invisible decay modes is constrained to
$\Gamma_{\text{invis}}/\Gamma_{\PH} < 0.01$ at $90\,\%$ C.L. and the
total Higgs width is measured to $\Delta(\Gamma_{\PH}) = 
6.7\,\%$. Operation of CLIC at $\roots > 1\,\TeV$ provides
high-statistics samples of Higgs bosons produced through the
$\PW\PW$-fusion process and give access to rarer processes such as
$\epem\to\PQt\PAQt\PH$ and $\epem\to\PH\PH\nuenuebar$. Studies of
these rare processes provide measurements of the top Yukawa
coupling to $4.2\,\%$ and the Higgs boson self-coupling to about 
$20\,\%$. Furthermore, the full data sample leads to very strong
constraints on the Higgs couplings to vector bosons and fermions. 
In a model-independent treatment, many of the
accessible couplings are measured to better than $2\,\%$, and the
model-dependent $\kappa$ parameters are determined with a precision of
between $0.1\,\%$ and $1\,\%$.

\begin{acknowledgements}
This work benefited from services provided by the ILC Virtual Organisation, supported by the national resource pro\-viders of the EGI Federation. This research was done using resources provided by the Open Science Grid, which is supported by the National Science Foundation and the U.S.\ Department of Energy's Office of Science. The authors would like to acknowledge the use of the Oxford Particle Physics Computing Cluster.
This work was supported by
the Comisi\'{o}n Nacional de Investigaci\'{o}n Cient\'{i}fica y Tecnol\'{o}gica (CONICYT), Chile;
the Ministry of Education, Youth and Sports, Czech Republic, under Grant INGO II-LG 14033;
the DFG cluster of excellence ``Origin and Structure of the Universe'', Germany;
the EC HIGGSTOOLS project, under contract PITN-GA-2012-316704;
the European Union's Horizon 2020 Research and Innovation programme under Grant Agreement no.\ 654168;
the German - Israel Foundation (GIF);
the Israel Science Foundation (ISF);
the I-CORE programme of VATAT, ISF and the Israel Academy of Sciences, Israel;
the Research Council of Norway;
the Ministry of Education, Science and Technological Development of the Republic of Serbia through the national project OI171012;
the Polish Ministry of Science and Higher Education under contract nr 3501/H2020/2016/2;
the National Science Centre, Poland, HARMONIA project, under contracts 2013/10/M/ST2/00629 and UMO-2015/18/M/ST2/00518;
the Romanian agencies UEFISCDI and ROSA;
the Secretary of State of Research, Development and Innovation of Spain, under project FPA2011-15330-E, FPA2015-71956-REDT;
the Gates Foundation, United Kingdom;
the UK Science and Technology Facilities Council (STFC), United Kingdom;
and the U.S.\ Department of Energy, Office of Science, Office of Basic Energy Sciences and Office of High Energy Physics under contract DE-AC02-06CH11357.
\end{acknowledgements}

\bibliographystyle{Main/spphys_mod}
\bibliography{Bibliography/higgspaper}

\end{document}